\documentclass
	[
		10 pt
	]
	{article}

\usepackage
	[
		a4paper,
		left = 2.500 cm,
		right = 2.500 cm,
		top = 2.500 cm,
		bottom = 3.000 cm
	]
	{geometry}
\usepackage{tikz, pgf, pgfplots}
\usetikzlibrary
	{
		3d,
		arrows,											
		arrows.meta,
		automata,										
		backgrounds,
		bending,
		calc,
		chains,
		datavisualization.formats.functions,			
		decorations.pathmorphing,
		decorations.pathreplacing,
		decorations.text,
		fadings,
		fit,
		graphs,
		graphs.standard,
		matrix,
		positioning,									
		quantikz2,
		quotes,
		shadings,
		shadows,
		shadows.blur,
		trees
	}
\pgfplotsset{compat = newest}
\usepgflibrary
	{
		shapes.misc
	}
\usepgfplotslibrary
	{
		dateplot
	}
\usepackage{tikzpeople}
\usepackage{fontawesome5}
\usepackage{smartdiagram}

\allowdisplaybreaks
\usepackage{amsthm}
\usepackage{amssymb}
\usepackage[bb = boondox]{mathalfa}
\usepackage{dsfont}
\usepackage{newtxmath}
\usepackage{mathtools}
\usepackage{multicol}
\usepackage{physics2}
\usephysicsmodule{braket}
\usepackage{empheq}
\usepackage[ compat = newest ]{yquant}

\usepackage{sectsty}
\allsectionsfont{\color{WordBlue}}
\usepackage{xcolor}
\usepackage{varwidth}
\usepackage{graphicx}
\usepackage{wrapfig}
\usepackage{subcaption}
\usepackage{xurl}                                                                                  %
\usepackage{hyperref}
\hypersetup
	{
		colorlinks,
		linkcolor = {WordRed!75!black},
		citecolor = {WordBlueDarker25},
		urlcolor = {WordAquaDarker50}
	}
\usepackage{tabularray}
\usepackage{colortbl}
\usepackage{float}
\usepackage{nicematrix}
\usepackage{cellspace}
\usepackage{makecell}
\usepackage{multirow}
\usepackage{caption}
\usepackage[ linesnumbered, tworuled, vlined ] {algorithm2e}
\usepackage{appendix}
\usepackage{enumitem}
\usepackage{siunitx}
\usepackage{xintexpr}
\usepackage{spreadtab}
\usepackage{framed}
\usepackage{svg}
\usepackage{lipsum}
\usepackage{listings}
\definecolor{MyLightRed}{RGB}{244, 213, 245}
\definecolor{Purple}{HTML}{911146}
\definecolor{PurpleDark}{RGB}{102, 0, 102}
\definecolor{RedDarkLight}{HTML}{ea005f}
\definecolor{RedDarkLightest}{HTML}{ff0088}
\definecolor{RedPurple}{HTML}{AA007F}
\definecolor{WordRed}{RGB}{255, 0, 102}
\definecolor{WordRedAccent5Lighter60}{HTML}{F5B5A7}
\definecolor{WordRedAccent5Darker25}{HTML}{B23214}
\definecolor{GreenDark}{HTML}{225522}
\definecolor{GreenLighter1}{HTML}{00B383}
\definecolor{GreenLighter2}{HTML}{00AA7F}
\definecolor{GreenLightest}{HTML}{00FFA0}
\definecolor{GreenTeal}{HTML}{008080}
\definecolor{WordLightGreen}{RGB}{140, 214, 192}
\definecolor{WordGreen}{RGB}{0, 176, 80}
\definecolor{BlueVeryDark}{HTML}{222255}
\definecolor{MyBlue}{RGB}{0, 64, 128}
\definecolor{MyDarkBlue}{RGB}{0, 51, 102}
\definecolor{MyVeryLightBlue}{RGB}{211, 245, 247}
\definecolor{WordBlue}{RGB}{19, 65, 99}
\definecolor{WordBlueDark}{RGB}{46, 116, 181}
\definecolor{WordBlueDarker}{RGB}{31, 78, 121}
\definecolor{WordBlueDarker25}{RGB}{54, 96, 146}
\definecolor{WordBlueDarker50}{RGB}{36, 64, 98}
\definecolor{WordBlueDarkest}{RGB}{0, 32, 96}
\definecolor{WordBlueLight}{RGB}{0, 112, 192}
\definecolor{WordBlueVeryLight}{HTML}{00B0F0}
\definecolor{WordIceBlue}{RGB}{223, 227, 229}
\definecolor{MagentaDark}{RGB}{106, 65, 152}
\definecolor{MagentaLight}{RGB}{128, 100, 162}
\definecolor{MagentaLighter}{RGB}{161, 106, 221}
\definecolor{MagentaVeryDark}{RGB}{97, 75, 128}
\definecolor{MagentaVeryLight}{RGB}{178, 162, 201}
\definecolor{WordAquaAccent2Darker25}{HTML}{398E98}
\definecolor{WordAquaDarker25}{HTML}{31869B}
\definecolor{WordAquaDarker50}{HTML}{215967}
\definecolor{WordAquaLighter40}{HTML}{92CDDC}
\definecolor{WordAquaLighter60}{HTML}{B7DEE8}
\definecolor{WordAquaLighter80}{HTML}{DAEEF3}
\definecolor{WordDarkerTeal}{RGB}{48, 82, 80}
\definecolor{WordDarkTeal}{RGB}{72, 123, 119}
\definecolor{WordDarkTealLighter80}{RGB}{207, 223, 234}
\definecolor{WordLightTeal}{RGB}{160, 199, 197}
\definecolor{WordVeryLightTeal}{RGB}{223, 236, 235}
\definecolor{WordTurquoiseLighter80}{RGB}{209, 238, 249}
\definecolor{Brown}{HTML}{666633}
\definecolor{WordGoldAccent1Darker25}{HTML}{C49A00}
\definecolor{WordGoldAccent1Lighter40}{HTML}{FFDF6A}
\definecolor{WordOrangeAccent2Lighter60}{HTML}{FCD3A4}
\definecolor{WordOrangeAccent4Lighter60}{HTML}{F7C5A1}
\definecolor{LavenderBlush}{RGB}{255, 240, 245}
\definecolor{MediumTurquoise}{RGB}{72, 209, 204}
\definecolor{PowderBlue}{RGB}{176, 224, 230}
\definecolor{SkyBlue}{RGB}{135, 206, 235}
\definecolor{Azure2}{RGB}{224, 238, 238}
\definecolor{Azure3}{RGB}{193, 205, 205}
\definecolor{CadetBlue4}{RGB}{83, 134, 139}
\definecolor{DarkSeaGreen1}{RGB}{193, 255, 193}
\definecolor{DeepPink4}{RGB}{139, 10, 80}
\definecolor{Honeydew2}{RGB}{224, 238, 224}
\definecolor{LightSkyBlue1}{RGB}{176, 226, 255}
\definecolor{LightSkyBlue3}{RGB}{141, 182, 205}
\definecolor{LightSkyBlue4}{RGB}{96, 123, 139}
\definecolor{LightSteelBlue1}{RGB}{202, 225, 255}
\definecolor{LightSteelBlue4}{RGB}{110, 123, 139}
\definecolor{MediumPurple1}{RGB}{171, 130, 255}
\definecolor{PaleTurquoise3}{RGB}{150, 205, 205}
\definecolor{PaleVioletRed3}{RGB}{205, 104, 137}
\definecolor{Purple1}{RGB}{155, 48, 255}
\definecolor{SeaGreen1}{RGB}{84, 255, 159}
\definecolor{SeaGreen2}{RGB}{78, 238, 148}
\definecolor{SeaGreen3}{RGB}{67, 205, 128}
\definecolor{SkyBlue1}{HTML}{87CEFF}
\definecolor{SkyBlue4}{RGB}{74, 112, 139}
\definecolor{SteelBlue1}{RGB}{99, 184, 255}
\definecolor{Thistle3}{RGB}{205, 181, 205}
\definecolor{Turquoise4}{RGB}{0, 134, 139}
\definecolor{VioletRed1}{RGB}{255, 62, 150}
\definecolor{VioletRed2}{RGB}{208, 32, 144}
\definecolor{VioletRed3}{RGB}{199, 21, 133}
\definecolor{VioletRed4}{RGB}{139, 10, 80}
\usepackage
	[
		skins,
		breakable,
		listings,
		raster,
		theorems
	]
	{tcolorbox}
\NewTcbTheorem
	[
		number within = section
	]
	{definition}
	{Definition}
	{
		breakable,
		skin = enhanced,
		enhanced jigsaw,			
		attach boxed title to top left = { xshift = -5.000 mm, yshift = 0.000 mm },
		boxed title style = { boxrule = 0.000 pt, sharp corners = all },
		varwidth boxed title,
		fonttitle = \bfseries,
		coltitle = SkyBlue!30!black,
		colbacktitle = SkyBlue!70!white,
		colframe = SkyBlue,
		colback = SkyBlue!20!white,
		sharp corners = all,
		toprule = 0.000 mm,
		bottomrule = 0.500 mm,
		leftrule = 0.500 mm,
		rightrule = 0.000 mm,
	}
	{def}
\NewTcbTheorem
	[
		number within = section
	]
	{example}
	{Example}
	{
		breakable,
		skin = enhanced,
		enhanced jigsaw,			
		attach boxed title to top left = { xshift = -5.000 mm, yshift = 0.000 mm },
		boxed title style = { boxrule = 0.000 pt, sharp corners = all },
		varwidth boxed title,
		fonttitle = \bfseries,
		coltitle = PowderBlue!30!black,
		colbacktitle = PowderBlue!70!white,
		colframe = PowderBlue,
		colback = white,
		sharp corners = all,
		toprule = 0.000 mm,
		bottomrule = 0.500 mm,
		leftrule = 0.500 mm,
		rightrule = 0.000 mm,
	}
	{xmp}
\NewTcbTheorem
	[
		number within = section
	]
	{theorem}
	{Theorem}
	{
		breakable,
		skin = enhanced,
		enhanced jigsaw,			
		attach boxed title to top left = { xshift = -5.000 mm, yshift = 0.000 mm },
		boxed title style = { boxrule = 0.000 pt, sharp corners = all },
		varwidth boxed title,
		fonttitle = \bfseries,
		coltitle = VioletRed3!20!black,
		colbacktitle = VioletRed2!30!white,
		colframe = VioletRed4,
		colback = VioletRed1!03,
		sharp corners = all,
		toprule = 0.000 mm,
		bottomrule = 0.500 mm,
		leftrule = 0.500 mm,
		rightrule = 0.000 mm,
	}
	{thr}
\NewTcbTheorem
	[
		number within = section
	]
	{proposition}
	{Proposition}
	{
		breakable,
		skin = enhanced,
		enhanced jigsaw,			
		attach boxed title to top left = { xshift = -5.000 mm, yshift = 0.000 mm },
		boxed title style = { boxrule = 0.000 pt, sharp corners = all },
		varwidth boxed title,
		fonttitle = \bfseries,
		coltitle = cyan5!30!black,
		colbacktitle = cyan7!50!white,
		colframe = cyan5,
		colback = cyan9!07,
		sharp corners = all,
		toprule = 0.000 mm,
		bottomrule = 0.500 mm,
		leftrule = 0.500 mm,
		rightrule = 0.000 mm,
	}
	{prp}

\newcounter{mathseed}
\pgfmathsetseed{\arabic{mathseed}}
\pgfdeclarelayer{background}
\pgfsetlayers{background, main}
\pgfdeclaredecoration{irregular fractal line}{init}
{
	\state{init}[width=\pgfdecoratedinputsegmentremainingdistance]
	{
		\pgfpathlineto{\pgfpoint{random*\pgfdecoratedinputsegmentremainingdistance}{(random*\pgfdecorationsegmentamplitude-0.02)*\pgfdecoratedinputsegmentremainingdistance}}
		\pgfpathlineto{\pgfpoint{\pgfdecoratedinputsegmentremainingdistance}{0pt}}
	}
}
\def\tornpaper#1{%
	\ifthenelse{\isodd{\value{mathseed}}}
	{%
		\tikz
		{
			\node[inner sep = 1em] (A) {#1};		
			\begin{pgfonlayer}{background}			
				\fill[paper]						
				\pgfextra{\pgfmathsetseed{\arabic{mathseed}}\addtocounter{mathseed}{1}}%
				{decorate[irregular cloudy border]{decorate{decorate{decorate{decorate[ragged border]{
										(A.north west) -- (A.north east)
				}}}}}}
				-- (A.south east)
				\pgfextra{\pgfmathsetseed{\arabic{mathseed}}}%
				{decorate[irregular spiky border]{decorate{decorate{decorate{decorate[ragged border]{
										-- (A.south west)
				}}}}}}
				-- (A.north west);
			\end{pgfonlayer}
		}
	}
	{%
		\tikz{
			\node[inner sep=1em] (A) {#1};  
			\begin{pgfonlayer}{background}  
				\fill[paper] 
				\pgfextra{\pgfmathsetseed{\arabic{mathseed}}\addtocounter{mathseed}{1}}%
				{decorate[irregular spiky border]{decorate{decorate{decorate{decorate[ragged border]{
										(A.north east) -- (A.north west)
				}}}}}}
				-- (A.south west)
				\pgfextra{\pgfmathsetseed{\arabic{mathseed}}}%
				{decorate[irregular cloudy border]{decorate{decorate{decorate{decorate[ragged border]{
										-- (A.south east)
				}}}}}}
				-- (A.north east);
		\end{pgfonlayer}}
	}
}
\title
	{
		Quantum Shadows: The Dining Information Brokers
	}

\usepackage{orcidlink}
\newcommand{\orcidicon}[1]{\href{https://orcid.org/#1}{\includegraphics[height=\fontcharht\font`\B]{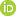}}}

\author
{
	Theodore Andronikos$^1$\orcidicon{0000-0002-3741-1271},
	Constantinos Bitsakos$^2$\orcidicon{0009-0003-3669-0453},
	Konstantinos Nikas$^2$\orcidicon{0000-0003-4424-9951},
	\\
	Georgios I. Goumas$^2$\orcidicon{0000-0001-7811-4831}
	and
	Nectarios Koziris$^2$\orcidicon{0000-0002-4890-8427}
	\\
	$^1$ \
	Department of Informatics, Ionian University, \\
	7 Tsirigoti Square, 49100 Corfu, Greece; \\
	andronikos@ionio.gr
	\\
	$^2$ \ 
	Computing Systems Laboratory, \\
	National Technical University of Athens, Greece; \\
	\{kbitsak, knikas, goumas, nkoziris\}@cslab.ece.ntua.gr
}
\begin{document}

\maketitle

\begin{abstract}
	This article introduces the innovative Quantum Dining Information Brokers Problem, presenting a novel entanglement-based quantum protocol to address it. The scenario involves $n$ information brokers, all located in distinct geographical regions, engaging in a metaphorical virtual dinner. The objective is for each broker to share a unique piece of information with all others simultaneously. Unlike previous approaches, this protocol enables a fully parallel, single-step communication exchange among all brokers, regardless of their physical locations. A key feature of this protocol is its ability to ensure both the anonymity and privacy of all participants are preserved, meaning no broker can discern the identity of the sender behind any received information. At its core, the Quantum Dining Information Brokers Problem serves as a conceptual framework for achieving anonymous, untraceable, and massively parallel information exchange in a distributed system. The proposed protocol introduces three significant advancements. First, while quantum protocols for one-to-many simultaneous information transmission have been developed, this is, to the best of our knowledge, one of the first quantum protocols to facilitate many-to-many simultaneous information exchange. Second, it guarantees complete anonymity and untraceability for all senders, a critical improvement over sequential applications of one-to-many protocols, which fail to ensure such robust anonymity. Third, leveraging quantum entanglement, the protocol operates in a fully distributed manner, accommodating brokers in diverse spatial locations. This approach marks a substantial advancement in secure, scalable, and anonymous communication, with potential applications in distributed environments where privacy and parallelism are paramount.
\end{abstract}
\section{Introduction} \label{sec: Introduction}

In the dynamic landscape of the modern digital age, technology has become an integral part of daily life, making robust cybersecurity measures more essential than ever. As we navigate the intricate web of digital interactions, we encounter a complex environment where the open exchange of information coexists with sophisticated and unpredictable threats. The concept of privacy has undergone a profound transformation, now encompassing not only individual autonomy but also the security of personal data in an era defined by rapid technological advancements and interconnected digital ecosystems. Privacy, in this context, refers to an individual’s ability to control their personal information, determining how it is collected, utilized, and shared. The scope of privacy concerns has expanded dramatically, covering areas such as personal communications meet, financial transactions, health records, and even behavioral data generated by online activities. The pervasive integration of digital platforms, social media, and smart technologies has revolutionized convenience and connectivity, but it has also amplified concerns about personal privacy, as these systems often collect and process vast amounts of sensitive data.

The role of cybersecurity is to protect this interconnected digital world, safeguarding data, privacy, and the trust that underpins our networked society. Cybersecurity encompasses the protection of digital systems, networks, and devices against a wide array of threats, including ransomware, sophisticated cyberespionage, malware, and data breaches. These threats pose not only significant economic risks but also undermine the foundational trust in digital infrastructure. The proliferation of the Internet and the widespread adoption of smart devices have fundamentally altered how we communicate, work, shop, and engage in leisure activities. However, this digital transformation has also expanded the attack surface, providing cybercriminals with new opportunities to exploit vulnerabilities. As a result, cybersecurity is a dynamic and ever-evolving field, striving to anticipate and counter emerging threats. This requires a multifaceted approach, combining cutting-edge technological innovations, robust regulatory frameworks, and skilled human expertise.

The field of quantum computing is progressing at an unprecedented pace, with recent advancements indicating that transformative quantum systems are on the horizon, poised to challenge classical computing paradigms, even though current quantum computers have not yet fully surpassed their classical counterparts. Leading industry and research entities have achieved remarkable milestones, pushing the boundaries of quantum technology.
IBM has made significant strides with the introduction of the 1,121-qubit Condor and the high-performance R2 Heron, building upon the foundations laid by the 127-qubit Eagle \cite{IBMEagle2021} and the 433-qubit Osprey \cite{IBMOsprey2022} \cite{IBMCondor2023, IBMHeron2024}.
Google has showcased the superior performance of its quantum computers, demonstrating their ability to outperform advanced supercomputers in specific tasks \cite{GoogleWillow2024, NatureGoogle2024}.
Microsoft has advanced the field with the development of the Majorana 1 quantum chip, leveraging topological qubits to enhance stability and scalability \cite{Aasen2025, Aghaee2025, MicrosoftMajorana2025}.
D-Wave has contributed significantly by utilizing its quantum annealer to solve a scientifically significant problem more efficiently than classical computers, marking a notable achievement in practical quantum computing applications \cite{King2025, NatureD-Wave2025}.
In parallel, China’s 105-qubit Zuchongzhi 3.0 processor has emerged as a technological breakthrough, further solidifying the global race in quantum innovation \cite{Gao2025, APSNewsZuchongzhi3.0-2025}.
Beyond these achievements, the quantum computing landscape is enriched by innovative design concepts \cite{Cacciapuoti2024, Illiano2024} and hardware advancements, such as those in photonic quantum systems \cite{Photonic2024, NuQuantum2024}. A particularly promising development is in distributed quantum computing, where two quantum processors were interconnected via a photonic network to operate as a unified system \cite{Main2025, OxfordNewsEvents2025}. These advancements highlight the increasing viability of distributed quantum architectures and their potential to revolutionize fields such as quantum cryptography, secure communication, and complex computational problem-solving. Collectively, these efforts underscore the rapid maturation of quantum computing technologies and their transformative potential for future applications.

In the rapidly evolving domain of cryptographic protocols, the Dining Cryptographers Problem, introduced by David Chaum in 1988 \cite{Chaum1988}, stands as a pioneering framework for exploring anonymous communication within a social context. This thought experiment was designed to illustrate the potential for secure and private message exchange, prioritizing the anonymity and privacy of each participant. The protocol employs cryptographic techniques to ensure that only a pre-agreed binary outcome ($0$ or $1$) is revealed, effectively concealing individual contributions. Inspired by real-world scenarios where individuals seek to share information while preserving confidentiality, this problem has significantly influenced classical cryptography, particularly in applications focused on obscuring the identities of senders and receivers \cite{Chaum1981, Ahn2003}. The emphasis on anonymity as a core cryptographic primitive has catalyzed extensive research, transitioning from classical to quantum cryptography, where novel approaches leverage quantum mechanics to enhance security and privacy.

The advent of quantum cryptography has spurred significant advancements in anonymous communication protocols. In 2002, Boykin proposed a quantum protocol utilizing pairs of entangled qubits, known as EPR pairs, to generate cryptographic keys for anonymous transmission of classical information via quantum teleportation \cite{Boykin2002}. An EPR pair, consisting of two qubits in a maximally entangled state, serves as a cornerstone for quantum communication and computation tasks, such as teleportation. Subsequently, Christandl and Wehner developed a protocol for anonymously distributing qubits using EPR pairs, enabling the transmission of a quantum coin without requiring all honest participants to share the same qubit, thus adhering to the no-cloning theorem \cite{Christandl2005}. Bouda and Sprojcar further advanced the field by achieving quantum communication without relying on a pre-shared trusted state among participants \cite{Bouda2007}. Brassard and Tapp et al. introduced information-theoretically secure protocols for anonymous quantum communication, incorporating fail-safe teleportation to ensure precise and secure message delivery despite potential errors or malicious actors \cite{Brassard2007, Broadbent2007}.

Further innovations include a quantum communication scheme based on non-maximally entangled qubit pairs \cite{Shimizu2009}, and Wang’s protocol for anonymous entanglement using single photons and CNOT operations \cite{Wang2010}. Shi et al. proposed a quantum anonymous communication method in a public receiver model, leveraging DC-Nets and non-maximally entangled channels \cite{Shi2011}. Wang and Zhang identified vulnerabilities in these protocols, particularly risks to sender anonymity in the presence of malicious participants, and suggested improvements \cite{Wang2014}. In 2015, Rahaman and Kar introduced quantum protocols for the Dining Cryptographers Problem and the Anonymous Veto (AV) problem, utilizing GHZ correlations and the GHZ paradox to ensure anonymity \cite{Rahaman2015}. Hameedi et al. advanced this work with a one-way sequential protocol using a single qubit and GHZ states, extending its application to the Anonymous Veto problem \cite{Hameedi2017}. In 2021, Li et al. proposed an anonymous transmission protocol using single-particle states with collective detection \cite{Li2021}, followed by Mishra et al.’s series of Quantum Anonymous Veto (QAV) protocols in 2022 \cite{Mishra2022}. Most recently, an innovative entanglement-based protocol for the Dining Cryptographers Problem was introduced, further advancing the field by leveraging quantum entanglement for enhanced anonymity and security \cite{Karananou2024}. These developments collectively underscore the growing sophistication of quantum cryptographic protocols in addressing anonymity and privacy challenges in distributed communication systems.

In this research, we introduce the innovative Quantum Dining Information Brokers Problem, a significant extension of the classic Dining Cryptographers Problem. Unlike the traditional setting, which implies a localized gathering of participants around a shared table, our framework removes this constraint, embracing a fully distributed environment where $n$ information brokers are situated in diverse geographic locations. The ``dining'' scenario is reimagined as a virtual, metaphorical interaction, reflecting the distributed nature of modern communication networks. Each broker aims to share a piece of information with all others, moving beyond the original problem’s limitation of exchanging a single bit (indicating whether a cryptographer paid for the meal) to allow for the transmission of arbitrarily large volumes of data. To tackle this challenge, we propose a novel quantum protocol that leverages entanglement to enable secure, anonymous, and parallel information exchange across distributed nodes.

To elucidate this complex protocol, we present it as a quantum game featuring signature players like Alice, Bob, Charlie, etc. harnessing the engaging and intuitive nature of games to demystify intricate quantum concepts. Quantum games, a concept popularized since 1999 \cite{Meyer1999, Eisert1999}, have demonstrated superior performance over classical strategies in various contexts \cite{Andronikos2018,Andronikos2021,Andronikos2022a}, such as the Prisoners’ Dilemma \cite{Eisert1999} and other abstract strategic scenarios \cite{Giannakis2015a,Koh2024}. Beyond their entertainment value, quantum games have proven effective in addressing serious challenges, including cryptographic protocols \cite{Bennett1984,Ampatzis2021, Ampatzis2022, Ampatzis2023, Andronikos2023, Andronikos2023a, Andronikos2023b, Karananou2024, Andronikos2024, Andronikos2024a, Andronikos2024b, Andronikos2025}, quantum classification of Boolean functions \cite{Andronikos2025a,Andronikos2025b}. Our game-based approach provides a powerful tool for advancing the design of quantum protocols. Moreover, the transformation of classical systems into quantum frameworks, as explored in recent studies on political structures \cite{Andronikos2022},
underscores the versatility of quantum approaches. Concerning games that take place in unusual settings, we mention that games that feature biological systems have attracted a lot of attention \cite{Theocharopoulou2019,Kastampolidou2020a,Kostadimas2021}. The fact that biosystems can produce biostrategies that might perform better than conventional strategies—even in the well-known Prisoners' Dilemma game—is especially fascinating \cite{Kastampolidou2020,Kastampolidou2021,Papalitsas2021,Kastampolidou2023,Adam2023}.
Therefore, it is easy to see that the game-theoretic framework not only facilitates a deeper understanding of the Quantum Dining Information Brokers Problem but also highlights its potential to revolutionize secure, distributed communication systems.

\textbf{Contribution}. In this work, we build upon the strengths of prior research, such as \cite{Andronikos2023b,Karananou2024}, preserving their key advantages: scalability, which supports both an increasing number of participants and the transmission of arbitrary volumes of anonymous
information; streamlined implementation, where all participants utilize identical quantum circuits for consistency and efficiency; and robust privacy and anonymity without any compromise. Our approach not only retains these strengths but also introduces three groundbreaking advancements that significantly enhance the Quantum Dining Information Brokers Problem.

\begin{itemize}
	\item	
	\textbf{Many-to-Many Simultaneous Information Exchange.} A key innovation of our protocol is its ability to facilitate communication among all participants, regardless of their geographical dispersion, in a single, fully parallel operation. While previous quantum protocols have achieved one-to-many simultaneous information transmission \cite{Andronikos2023b}, the current protocol is, as far as we are aware, one of the very first to enable many-to-many simultaneous exchange. This advancement ensures efficient, large-scale information sharing without sequential delays, marking a significant leap forward in distributed quantum communication.
	\item	
	\textbf{Enhanced Anonymity.} Leveraging the unique properties of quantum entanglement, our protocol encodes information into the relative phase of a distributed entangled system, rendering it untraceable and fully anonymous. This ensures that the identities of all senders remain completely protected, a critical improvement over sequential applications of one-to-many protocols, which cannot guarantee such robust anonymity. Unlike approaches that repeat one-to-many transmission $n - 1$ times, our protocol achieves the coveted goal of complete anonymity in a single operation, providing a transformative solution for secure communication.
	\item	
	\textbf{Fully Distributed Framework.} Traditional formulations of the Dining Cryptographers Problem often assume a localized setting where participants are physically co-located. Our protocol transcends this limitation by addressing a fully distributed scenario, where information brokers are situated in diverse geographical locations. By exploiting quantum entanglement, it ensures seamless and secure communication across vast distances. Notably, this protocol remains applicable to localized settings, as they represent a special case of the distributed framework, thus offering unparalleled flexibility for various real-world applications.
\end{itemize}

\subsection*{Organization} \label{subsec: Organization}

This article is structured to provide a comprehensive exploration of the Quantum Dining Information Brokers Problem and its associated protocol. Section \ref{sec: Introduction} presents an overview of the topic within the context of existing research and including citations to pertinent literature. Section \ref{sec: Preliminary Concepts} offers a concise introduction to essential concepts, laying the groundwork for understanding the technical intricacies of the protocol. Section \ref{sec: Introducing The Quantum Dining Information Brokers Problem} rigorously defines the Quantum Dining Information Brokers Problem, articulating its scope and significance. Section \ref{sec: Protagonists And Hypotheses} details the configuration and assumptions underlying the proposed quantum protocol, setting the stage for its implementation. Section \ref{sec: Detailed Analysis Of The QDIBP} provides a thorough examination of the protocol’s mechanics, offering a step-by-step analysis of its execution. Section \ref{sec: A Small Scale Realization Of The QDIBP} illustrates the protocol’s functionality through a practical, small-scale example, designed to enhance reader comprehension. Finally, Section \ref{sec: Discussion and Conclusions} summarizes the findings, discusses the protocol’s implications providing a holistic conclusion to the study.

\section{Preliminary concepts} \label{sec: Preliminary Concepts}

\subsection{GHZ states} \label{subsec: GHZ States}

Quantum entanglement, one of the most profound and defining features of quantum mechanics, serves as the foundation for a wide array of quantum protocols, enabling phenomena that defy classical intuition. Unlike separable states, entangled states of composite quantum systems cannot be represented by a single product state; rather, they require a superposition of multiple product states of their subsystems to capture their correlated nature. For multipartite systems with $r \geq 3$ qubits, the most well-known example of maximal entanglement is the $\ket{ GHZ_{ r } }$ state—named after researchers Greenberger, Horne, and Zeilinger. This state entangles $r$ distinct qubits, each treated as a spatially separated subsystem, into a highly correlated quantum state. The mathematical formulation of the $\ket{ GHZ_{ r } }$ state is detailed in equation \eqref{eq: Extended General GHZ_r State}, providing a precise description of its structure.

\begin{align}
	\label{eq: Extended General GHZ_r State}
	\ket{ GHZ_{ r } }
	=
	\frac
	{
		\ket{ 0 }_{ r - 1 } \ket{ 0 }_{ r - 2 } \dots \ket{ 0 }_{ 0 }
		+
		\ket{ 1 }_{ r - 1 } \ket{ 1 }_{ r - 2 } \dots \ket{ 1 }_{ 0 }
	}
	{ \sqrt{ 2 } }
	\ .
\end{align}

To clearly denote the entanglement of $r$ distinct qubits, we employ indices $i$, where $0 \leq i \leq r - 1$, to represent the $i^{ th }$ qubit, maintaining this convention throughout the paper. Qubits assigned to specific participants, such as Alice, Bob, and others, are denoted as $\ket{ \cdot }_A $, $\ket{ \cdot }_B $, and so forth. Modern quantum computers, including IBM’s advanced systems \cite{IBMOsprey2022, IBMCondor2023, IBMHeron2024}, are capable of preparing $\ket{ GHZ_{ r } }$  states using fundamental quantum operations like Hadamard and CNOT gates. Remarkably, the preparation of these states is highly efficient, requiring only $\lg r$ steps \cite{Cruz2019}. For a deeper exploration of entanglement, readers are directed to comprehensive resources such as \cite{Nielsen2010, Yanofsky2013, Wong2022}. For the purposes of our proposed protocol, a single $\ket{ GHZ_{ r } }$ tuple is insufficient; instead, we utilize a compound system comprising $p$ such tuples, as described in equation \eqref{eq: p-Fold Extended General GHZ_r State} and further elaborated in \cite{Ampatzis2023}. This configuration enhances the protocol’s capacity to handle complex, distributed quantum communication tasks.

\begin{align}
	\label{eq: p-Fold Extended General GHZ_r State}
	\ket{ GHZ_{ r } }^{ \otimes p }
	&=
	2^{ - \frac { p } { 2 } }
	\sum_{ \mathbf{ x } \in \mathbb{ B }^{ p } }
	\ket{ \mathbf{ x } }_{ r - 1 } \dots \ket{ \mathbf{ x } }_{ 0 }
	\ .
\end{align}

The notation used in formulating equation \eqref{eq: p-Fold Extended General GHZ_r State} is as follows:

\begin{itemize}
	\item	
	Subscripts are extensively used to clearly indicate the subsystem to which each qubit belongs, ensuring unambiguous identification.
	\item	
	The binary set $\mathbb{ B } = \{ 0, 1 \}$ represents the possible states of a single bit.
	\item	
	Bit vectors $\mathbf{ x } \in \mathbb{ B }^{ p }$ are denoted in boldface to distinguish them from single bits $x \in \mathbb{ B } $, which are written in regular typeface.
	\item	
	A bit vector $\mathbf{ x } = x_{ p - 1 } \dots x_{ 0 }$ is a sequence of $p$ bits. The zero bit vector, denoted $\mathbf{ 0 }$, consists of all zero bits, i.e., $\mathbf{ 0 } = 0 \dots 0$. Whenever we want to precisely specify the length of the zero bit vector, we use the notation $\mathbf{ 0 }_{ p }$ to designate the zero vector of length $p$.
	\item	
	Each bit vector $\mathbf{ x } \in \mathbb{ B }^{ p }$ corresponds to one of the $2^{ p }$ basis kets in the computational basis of the $2^{ p }$-dimensional Hilbert space, facilitating the representation of complex quantum states.
\end{itemize}

The proposed protocol also requires two other well-known states, $\ket{ + }$ and $\ket{ - }$, which are defined as

\begin{tcolorbox}
	[
		enhanced,
		breakable,
		grow to left by = 0.000 cm,
		grow to right by = 0.000 cm,
		colback = white,
		enhanced jigsaw,			
		frame hidden,
		sharp corners,
	]
	\begin{minipage}[b]{0.45 \textwidth}
		\begin{align}
			\label{eq: Ket +}
			\ket{ + }
			=
			H \ket{ 0 } = \frac { \ket{ 0 } + \ket{ 1 } } { \sqrt{ 2 } }
		\end{align}
	\end{minipage} 
	\hfill
	\begin{minipage}[b]{0.45 \textwidth}
		\begin{align}
			\label{eq: Ket -}
			\ket{ - }
			=
			H \ket{ 1 } = \frac { \ket{ 0 } - \ket{ 1 } } { \sqrt{ 2 } }
		\end{align}
	\end{minipage}
\end{tcolorbox}

\subsection{Inner product modulo $2$ operation} \label{subsec: Inner Product Modulo $2$ Operation}

In this work, we leverage the inner product modulo $2$ operation, which takes two bit vectors $\mathbf{ x }, \mathbf{ y } \in \mathbb{ B }^{ p }$ and computes their inner product $\mathbf{ x \bullet y }$. For bit vectors defined as $\mathbf{ x } = x_{ p - 1 } \dots x_{ 0 }$  and $\mathbf{ y } = y_{ p - 1 } \dots y_{ 0 }$, the inner product is expressed as:

\begin{align}
	\label{eq: Inner Product Modulo $2$}
	\mathbf{ x }
	\bullet
	\mathbf{ y }
	\coloneq
	x_{ p - 1 } y_{ p - 1 }
	\oplus \dots \oplus
	x_{ 0 } y_{ 0 }
	\ ,
\end{align}

where $\coloneq$ denotes ``is defined as'' and $\oplus$ represents addition modulo $2$. This operation is pivotal in quantum information theory, particularly in the context of the $p$-fold Hadamard transform applied to a basis ket $\ket{ \mathbf{ x } }$, as described below. Its proof is available in most standard textbooks, e.g., \cite{Mermin2007, Nielsen2010}.

\begin{align}
	\label{eq: Hadamard p-Fold Ket x}
	H^{ \otimes p } \ket{ \mathbf{ x } }
	=
	2^{ - \frac { p } { 2 } }
	\sum_{ \mathbf{ z } \in \mathbb{ B }^{ p } }
	\
	( - 1 )^{ \mathbf{ z \bullet x } } \ket{ \mathbf{ z } }
	\ .
\end{align}

Our protocol exploits a critical property of the inner product modulo $2$, referred to as the Characteristic Inner Product (CIP) property \cite{Andronikos2023b}. Specifically, for any non-zero bit vector $\mathbf{ c }$ of $\mathbb{ B }^{ p }$, exactly half of the $2^{ p }$ bit vectors $\mathbf{ x } \in \mathbb{ B }^{ p }$ satisfy $\mathbf{ c } \bullet \mathbf{ x } = 0$, while the other half satisfy $\mathbf{ c } \bullet \mathbf{ x } = 0$. In contrast, for the zero bit vector $\mathbf{ 0 }$, the inner product $\mathbf{ 0 } \bullet \mathbf{ x } = 0$ holds for all $\mathbf{ x } \in \mathbb{ B }^{ p }$. This balanced distribution of outcomes for non-zero $\mathbf{ c }$ enhances the protocol’s ability to encode and process information securely and anonymously in quantum systems.

\begin{tcolorbox}
	[
		enhanced,
		breakable,
		grow to left by = 0.500 cm,
		grow to right by = 0.500 cm,
		colback = white,
		enhanced jigsaw,			
		frame hidden,
		sharp corners,
	]
	\begin{minipage}[c]{0.490 \textwidth}
		{\small
			\begin{align}
				\label{eq: Inner Product Modulo $2$ Property For Zero}
				\mathbf{ c } = \mathbf{ 0 }
				&\Rightarrow
				\text{for all } 2^{ p } \text{ bit vectors } \mathbf{ x } \in \mathbb{ B }^{ p },
				\text{ } \mathbf{ c } \bullet \mathbf{ x } = 0
			\end{align}
		}
	\end{minipage} 
	\hfill
	\begin{minipage}[c]{0.510 \textwidth}
		{\small
			\begin{align}
				\label{eq: Inner Product Modulo $2$ Property For NonZero}
				\mathbf{ c } \neq \mathbf{ 0 }
				&\Rightarrow
				\left\{
				\
				\begin{matrix*}[l]
					\text{for } 2^{ p - 1 } \text{ bit vectors } \mathbf{ x } \in \mathbb{ B }^{ p }, \ \mathbf{ c } \bullet \mathbf{ x } = 0
					\\
					\text{for } 2^{ p - 1 } \text{ bit vectors } \mathbf{ x } \in \mathbb{ B }^{ p }, \ \mathbf{ c } \bullet \mathbf{ x } = 1
				\end{matrix*}
				\
				\right\}
			\end{align}
		}
	\end{minipage}
\end{tcolorbox}
\section{Introducing the Quantum Dining Information Brokers Problem} \label{sec: Introducing The Quantum Dining Information Brokers Problem}

In this section, we present a comprehensive examination of the Quantum Dining Information Brokers Problem, beginning with an exploration of its conceptual origins and inspirations. We then elaborate on how this problem extends and generalizes prior frameworks, highlighting its key advantages and novel contributions. The quantum protocol designed to address this problem, referred to as the Quantum Dining Information Brokers Protocol (QDIBP for short), is thoroughly detailed in Sections \ref{sec: Protagonists And Hypotheses} and \ref{sec: Detailed Analysis Of The QDIBP}.

\subsection{Inspirational Foundations} \label{subsec: Inspirational Foundations}

The QDIBP draws significant inspiration from the Dining Cryptographers Problem, a seminal cryptographic protocol introduced by David Chaum in his groundbreaking 1988 paper \cite{Chaum1988}. The Dining Cryptographers Problem is a thought experiment that illustrates the feasibility of anonymous communication within a social context, emphasizing the preservation of participants’ privacy and anonymity during message exchanges. In Chaum’s scenario, cryptographers aim to determine whether one of them paid for a shared dinner without revealing individual contributions, using cryptographic techniques to ensure that only the pre-agreed outcome (a binary $0$ or $1$) is disclosed. This setup mirrors real-world situations where individuals seek to share sensitive information while safeguarding their privacy and the confidentiality of their messages. The Dining Cryptographers Problem has significantly influenced classical cryptography, particularly in applications focused on obfuscating the identities of senders and receivers, as evidenced by works such as \cite{Chaum1981, Ahn2003}.

Further inspiration for the QDIBP stems from a recent advancement in quantum cryptography presented in \cite{Karananou2024}. This work introduced a scalable, quantum entanglement-based protocol to address the Dining Cryptographers Problem, utilizing maximally entangled $\ket{ GHZ_{ n } }$ states as its cornerstone. The protocol’s primary innovation lies in its scalability, accommodating an arbitrary number of cryptographers ($n$) and enabling the transmission of a variable amount of anonymous information, represented by $m$ qubits per quantum register. Unlike the original Dining Cryptographers Problem, which is limited to conveying a single bit of information (e.g., whether a cryptographer paid for the dinner), this quantum protocol allows $m$ to be any arbitrarily large positive integer. This flexibility facilitates the transmission of complex data, such as the cost of the dinner, the timing of arrangements, or other multifaceted information, significantly enhancing the protocol’s practical utility.

\subsection{Extending the scope} \label{subsec: Extending The Scope}

The Quantum Dining Information Brokers Problem (QDIBP) establishes a framework for secure, anonymous, and scalable information exchange among multiple participants in a distributed quantum environment. Below, we outline the key components of this setting, emphasizing its innovations and extensions over prior work.

\begin{itemize}
	\item	
	\textbf{Multiple Participants.} There are $n$ information brokers, denoted by $IB_{ 0 }$, \dots, $IB_{ n - 1 }$, where $n$ is an arbitrarily large positive integer, enabling the protocol to accommodate a scalable number of participants.
	\item	
	\textbf{Fully Distributed Environment.} Although the word ``Dining'' evokes images of a local gathering of the players around a table, something that was assumed in previous works, the QDIBP operates in a fully distributed setting. Here, the $n$ information brokers are geographically dispersed, and the concept of a ``dinner'' is metaphorical, representing a virtual interaction rather than a physical meeting.
	\item	
	\textbf{Secret Information Sharing.} Every information broker $IB_{ i }$, $0 \leq i \leq n - 1$, aims to transmit a piece of secret information to all other brokers $IB_{ j }$, $j \neq i$, ensuring secure and anonymous communication across the network.
	\item	
	\textbf{Arbitrary Information Volume.} In contrast to the original Dining Cryptographers Problem, which is limited to a single bit of information (e.g., whether a cryptographer paid for the dinner), the QDIBP supports the transmission of $m$ qubits, where $m$ is an arbitrarily large positive integer. This allows for the encoding and exchange of complex, multi-dimensional information.
	\item	
	\textbf{Parallel Many-to-Many Exchange.} A defining feature of the QDIBP is its ability to facilitate simultaneous many-to-many information exchange among all participants in a single operation. Unlike prior quantum protocols that support one-to-many transmission \cite{Andronikos2023b}, this is, as far as we are aware, the first quantum protocol to achieve fully parallel many-to-many communication.
	\item	
	\textbf{Uncompromised Anonymity and Privacy.} The protocol ensures that information is exchanged without compromising the anonymity or privacy of any participant. Each broker receives the information transmitted by others without discerning the sender’s identity, embodying the essence of the QDIBP as a paradigm for anonymous and untraceable information transmission in a massively parallel and distributed manner.
\end{itemize}

\noindent Compared to prior works, such as \cite{Andronikos2023b, Karananou2024}, the QDIBP retains and enhances their strengths, delivering a robust framework for quantum-based anonymous communication.

\begin{itemize}
	\item	
	\textbf{Scalability.} The QDIBP is designed for scalability in both the number of participants ($n$) and the volume of information transmitted ($m$ qubits). This dual scalability ensures the protocol can handle large networks and complex data exchanges seamlessly.
	\item	
	\textbf{Robust Anonymity.} The QDIBP guarantees that the anonymity and privacy of all participants are preserved. Information is exchanged such that no participant can trace the origin of any received message, reinforcing the protocol’s focus on secure and anonymous communication.
	\item	
	\textbf{Modular and Streamlined Implementation.} The protocol employs identical quantum circuits for all participants, ensuring modularity and ease of implementation. These circuits rely solely on standard quantum gates, such as Hadamard and CNOT, making them compatible with contemporary quantum computing platforms.
\end{itemize}

\noindent The present setup extends previous advantages and brings additional novelties to the table in three fundamental ways.

\begin{enumerate}
	[ left = 0.600 cm, labelsep = 0.500 cm, start = 1 ]
	\renewcommand \labelenumi { $($\textbf{E}$_{ \theenumi }$$)$ }
	\item
	\textbf{Simultaneous Many-to-Many Communication.} The QDIBP enables all participants to exchange information concurrently in a single, fully parallel operation, regardless of their geographical locations. While earlier quantum protocols achieved one-to-many simultaneous transmission (see for instance \cite{Andronikos2023b}), the QDIBP is the first to realize many-to-many communication in one step. This eliminates the inefficiencies of sequential transmissions and ensures robust anonymity, unlike repeated one-to-many protocols that may fail to guarantee complete anonymity after $n - 1$ iterations.
	\item
	\textbf{Enhanced Anonymity through Quantum Entanglement.} By leveraging quantum entanglement, the QDIBP encodes information into the relative phases of a distributed entangled system, ensuring that messages are untraceable and sender identities remain fully protected. This quantum approach provides a higher degree of anonymity compared to classical or sequential quantum protocols, marking a transformative advancement in secure communication.
	\item
	\textbf{Fully Distributed and Flexible Framework.} The QDIBP transcends the localized assumptions of earlier works, such as the Dining Cryptographers Problem, by supporting a fully distributed network where participants are geographically dispersed. Quantum entanglement facilitates secure communication across vast distances, while the protocol remains adaptable to localized settings as a special case, offering greater versatility for diverse applications.
\end{enumerate}

\noindent These advancements are enabled through the integrated use of ideal pairwise quantum channels, complemented by pairwise authenticated classical channels, ensuring secure and efficient communication across the distributed network.

\section{Protagonists and hypotheses} \label{sec: Protagonists And Hypotheses}

In this work, we present the Quantum Dining Information Brokers Protocol (QDIBP), a novel quantum protocol designed to address the Quantum Dining Information Brokers Problem. For brevity, we refer to this protocol as QDIBP throughout the rest of the text. This section outlines the setup and hypotheses essential for the correct implementation of the QDIBP, with a comprehensive explanation of its execution provided in Section \ref{sec: Detailed Analysis Of The QDIBP}.

\subsection{Protagonists \& rules} \label{subsec: Protagonists & Rules}

To enhance clarity and engagement, we adopt the format of a quantum game, a common approach in cryptographic protocol literature to make complex concepts more accessible. Before introducing the participants, we first clarify the critical concept of a \emph{semi-honest} player, which is pivotal to the protocol’s security model. Security concerns in quantum computation, particularly in two-party scenarios \cite{Lo1997}, necessitate additional assumptions to ensure robust protection. One widely recognized assumption is the involvement of a semi-honest third party, defined as follows.

\begin{definition} {Semi-honest player} { Semi-Honest Player}
	A \emph{semi-honest} player is characterized by the following properties:
	\begin{itemize}
		\item	Faithfully executes the protocol as specified.
		\item	Does not collude with any other player.
		\item	Cannot be corrupted by an outside entity.
		\item	Records all intermediate computations and may attempt to extract information from these records.
	\end{itemize}
\end{definition}

In essence, a semi-honest player adheres strictly to the protocol’s rules to facilitate the intended outcomes but may seek to gain unauthorized insights from the data processed during execution.

The QDIBP protocol evolves as a game played by $n + 1$ players, where $n$ is an arbitrarily large positive integer. So, without further ado, we list the protagonists and the rules governing their behavior below.

\begin{tcolorbox}
	[
		enhanced,
		breakable,
		center title,
		fonttitle = \bfseries,
		colbacktitle = SkyBlue4,
		title = Players \& Rules,
		grow to left by = 0.000 cm,
		grow to right by = 0.000 cm,
		colframe = SkyBlue4,
		colback = SkyBlue1!20,
		enhanced jigsaw,			
		sharp corners,
		boxrule = 0.500 pt,
	]
	\begin{enumerate}
		[ left = 0.250 cm, labelsep = 0.500 cm, start = 1 ]
		\renewcommand\labelenumi{(\textbf{R}$_{ \theenumi }$)}
		\item
		\textbf{Participants.} The protocol includes $n$ primary participants, referred to as information brokers, denoted $IB_{ 0 }$, \dots, $IB_{ n - 1 }$, where $n$ is an arbitrarily large positive integer. Each broker $IB_{ i }$, $0 \leq i \leq n - 1$, aims to transmit her secret information to all other brokers $IB_{ j }$, $0 \leq j \neq i \leq n - 1$, while ensuring the sender’s identity remains concealed from all in-game participants. The secret information of each broker $IB_{ i }$ is represented by the secret bit vector $\mathbf{ s }_i$. To enhance security, all bit vectors $\mathbf{ s }_i$ are assumed to be unique.
		\item
		\textbf{Semi-Honest Third Party.} A single semi-honest third party, named Trent, is integral to the protocol’s execution. Thus, the total number of participants is $n + 1$, each located in distinct geographical regions.
		\item
		\textbf{Trent’s Role.} Trent plays a pivotal role by generating and distributing $\ket{ GHZ_{ n + 1 } }$ tuples to all $n + 1$ participants, following the entanglement distribution scheme outlined in Definition \ref{def: The $r$-Uniform Entanglement Distribution Scheme}. In the protocol’s second phase, Trent applies a random permutation to the contents of his quantum register, as detailed in Section \ref{sec: Detailed Analysis Of The QDIBP}. This permutation ensures that the secret information is transmitted anonymously, with Trent strictly prohibited from disclosing the permutation used.
		\item
		\textbf{Information Encoding.} All participants agree in advance on the number $m$ of qubits required to encode their information bit vectors $\mathbf{ s }_{ i }$ (for $0 \leq i \leq n - 1$), allowing for flexible and scalable information transmission.
		\item
		\textbf{Restricted Communication.} The $n$ information brokers and Trent are prohibited from communicating outside the protocol’s designated channels and scope, ensuring all interactions occur within the game’s framework.
		\item
		\textbf{Unique Information Vectors.} For technical robustness, all $n$ information bit vectors are assumed to be unique and non-zero, preventing ambiguities in the protocol’s execution.
	\end{enumerate}
\end{tcolorbox}

In illustrative small-scale examples, the information brokers are represented by named actors—Alice, Bob, Charlie, and Dave—to make the scenarios more relatable. Consistent with standard practices in theoretical quantum cryptography, we assume ideal quantum channels, which are free from noise, particle loss, decoherence, and environmental challenges such as those encountered in free-space or optical fiber transmissions. While we acknowledge the critical importance of practical issues like noise, channel loss, entanglement control, and scalability, these are beyond the scope of this work, which focuses on establishing the theoretical foundations of the QDIBP. Additionally, classical channels used in the protocol are assumed to be authenticated, ensuring that messages are publicly accessible but protected against tampering by adversaries.

\subsection{Blocks \& segments} \label{subsec: Blocks & Segments}

In this subsection, we elucidate the methodology employed by the QDIBP for securely managing and exchanging sensitive information. The QDIBP is designed to enable simultaneous, many-to-many communication among multiple parties while ensuring robust anonymity and confidentiality. To achieve this, the protocol employs a sophisticated framework for structuring, encoding, and transmitting information, leveraging quantum channels to maintain security against potential adversaries, including semi-honest third parties.

\begin{definition} {Secret Vectors} { Secret Vectors}
	Every information broker $IB_{ i }$ encodes her confidential information as the secret bit vector $\mathbf{ s }_{ i }$, $0 \leq i \leq n - 1$.
\end{definition}

Each secret vector $\mathbf{ s }_{ i }$ encapsulates the sensitive data that $IB_{ i }$ aims to share anonymously with all other information brokers $IB_{ j }$, $0 \leq j \neq i \leq n - 1$. The protocol ensures that the sender’s identity remains concealed throughout the communication process, while simultaneously preventing a semi-honest third party, Trent, from obtaining $\mathbf{ s }_{ i }$. As stipulated in subsection \ref{subsec: Protagonists & Rules}, point (\textbf{R}$_{ 4 }$), all $n$ secret vectors share a uniform length of $m$ bits. Additionally, point (\textbf{R}$_{ 6 }$) mandates that these vectors are unique and distinct from the zero bit vector, thereby eliminating potential ambiguities and safeguarding the integrity of the information exchange.

The confidential information itself is represented by the set of secret vectors $\mathbf{ s }_{ 0 }$, $\dots$, $\mathbf{ s }_{ n - 1 }$, where each vector has a length of $m$ bits. To facilitate fully parallel many-to-many communication while preserving anonymity, each information broker, denoted $IB_{ i }$ transforms the data intended for transmission into a larger, structured bit vector known as the extended secret vector, represented as \emph{extended} secret vector, and denoted by $\widetilde { \mathbf{ s } }_{ i }$, $0 \leq j \neq i \leq n - 1$. This extended vector is designed with a specialized hierarchical configuration to support secure and efficient data exchange across a quantum communication framework. The detailed structure of this hierarchical schema is formalized in Definition \ref{def: Blocks & Segments}.

\begin{definition} {Blocks \& segments} { Blocks & Segments}
	The extended secret vector $\widetilde { \mathbf{ s } }_{ i }$ comprises $n^{ 2 } m$ bits and is systematically organized into a hierarchical structure consisting of $n$ segments, each containing $n m$ bits. Each segment is further subdivided into $n$ blocks, with every block consisting of $m$ bits.

	This structured organization is directly reflected in the corresponding quantum registers. Each quantum register is composed of $n^{ 2 } m$ qubits and is similarly partitioned into $n$ segments, each containing $n m$ qubits. These segments are further divided into $n$ blocks, with each block comprising $m$ qubits. This establishes a one-to-one correspondence between the segments and blocks of the extended secret vectors and those of the quantum registers, ensuring seamless alignment between classical and quantum data representations.
\end{definition}

This hierarchical segmentation, illustrated in Figure \ref{fig: Hierarchical Organization Into Segments & Blocks}, significantly enhances the protocol’s flexibility, scalability, and robustness. By organizing data into segments and blocks, the system can efficiently handle complex data structures while maintaining the anonymity of participants. This structure also ensures reliable and secure information transfer across the quantum channel, as the segmented organization allows for precise error detection and correction mechanisms. Furthermore, the one-to-one correspondence between the classical extended secret vectors and the quantum registers enables the protocol to leverage quantum properties, such as superposition and entanglement, to enhance security and anonymity. The design supports a broad range of applications, from secure multi-party computation to distributed quantum networks, while upholding the integrity and confidentiality of the transmitted information.

\begin{tcolorbox}
	[
		enhanced,
		breakable,
		grow to left by = 0.000 cm,
		grow to right by = 0.000 cm,
		colback = white,
		enhanced jigsaw,			
		frame hidden,
		sharp corners,
	]
	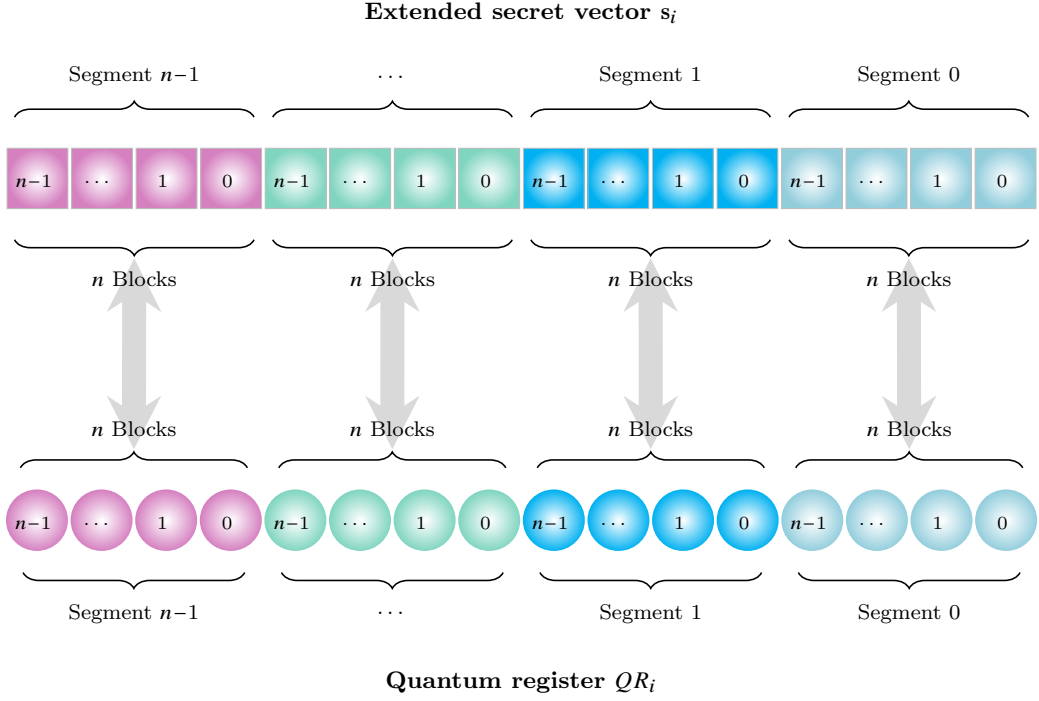
\begin{figure}[H]
		\centering
		\vspace*{ - 0.500 cm }
		\begin{tikzpicture} [ scale = 1.000, transform shape,
			bitcell/.style = {
				draw = gray!50, rectangle, minimum size = 0.800 cm, inner sep = 0pt, align = center, font = \scriptsize, line width = 0.400 pt, text height = 1.800 ex, text depth = 0.300 ex
			},
			circlebit/.style = {
				draw = none, circle, minimum size = 0.815 cm, inner sep = 0.000pt, align = center, font = \scriptsize, line width = 0.000 pt, text height = 1.800 ex, text depth = 0.300 ex
			},
			title/.style = {
				font = \small\bfseries, above = 0.000 cm, yshift = 1.500cm
			},
			circletitle/.style = {
				font = \small\bfseries, below = 1.500 cm
			},
			seglabel/.style = {
				font = \footnotesize, above = 0.400 cm
			},
			blocklabel/.style = {
				font = \footnotesize, below = 0.400 cm
			},
			dotlabel/.style = {
				font = \footnotesize, above = 0.400 cm, yshift = 0.100 cm
			},
			dotlabelbelow/.style = {
				font = \footnotesize, below = 0.400 cm, yshift = -0.100 cm
			},
			brace/.style = {
				line width = 0.600 pt, decorate, decoration = { brace, amplitude = 6.000 pt, raise = 3.000 pt }
			},
			mirrorbrace/.style = {
				line width = 0.600 pt, decorate, decoration = { brace, amplitude = 6.000 pt, mirror, raise = 3.000 pt}
			},
			connectarrow/.style = {
				line width = 9.000 pt,
				>= { Stealth [ length = 20.000pt, width = 20.000pt ] },
				<->,
				draw = gray!30,
				preaction = { draw = white, line width = 4.000 pt },
				shorten >= 10.000 pt,
				shorten <= 10.000 pt,
			}
			]
			\matrix (reg) [
				matrix of nodes,
				nodes = bitcell,
				column sep = 1.000 pt,
				row sep = 0.000 pt
			]
			{
				| [ bitcell, shade, outer color = RedPurple!50, inner color = white ] | $n{-}1$ &
				| [ bitcell, shade, outer color = RedPurple!50, inner color = white ] | $\cdots$ &
				| [ bitcell, shade, outer color = RedPurple!50, inner color = white ] | $1$ &
				| [ bitcell, shade, outer color = RedPurple!50, inner color = white ] | $0$ &
				| [ bitcell, shade, outer color = GreenLighter2!50, inner color = white ] | $n{-}1$ &
				| [ bitcell, shade, outer color = GreenLighter2!50, inner color = white ] | $\cdots$ &
				| [ bitcell, shade, outer color = GreenLighter2!50, inner color = white ] | $1$ &
				| [ bitcell, shade, outer color = GreenLighter2!50, inner color = white ] | $0$ &
				| [ bitcell, shade, outer color = WordBlueVeryLight, inner color = white ] | $n{-}1$ &
				| [ bitcell, shade, outer color = WordBlueVeryLight, inner color = white ] | $\cdots$ &
				| [ bitcell, shade, outer color = WordBlueVeryLight, inner color = white ] | $1$ &
				| [ bitcell, shade, outer color = WordBlueVeryLight, inner color = white ] | $0$ &
				| [ bitcell, shade, outer color = WordAquaLighter40, inner color = white ] | $n{-}1$ &
				| [ bitcell, shade, outer color = WordAquaLighter40, inner color = white ] | $\cdots$ &
				| [ bitcell, shade, outer color = WordAquaLighter40, inner color = white ] | $1$ &
				| [ bitcell, shade, outer color = WordAquaLighter40, inner color = white ] | $0$ \\
			};
			\draw [ brace ] ([ xshift = 3.000 pt, yshift = 0.300 cm ]reg-1-13.north west) -- ([ xshift = - 3.000 pt, yshift = 0.300 cm ]reg-1-16.north east) node [ seglabel, pos = 0.500 ] { Segment $0$ };
			\draw [ brace ] ([ xshift = 3.000 pt, yshift = 0.300 cm ]reg-1-9.north west) -- ([ xshift = - 3.000 pt, yshift = 0.300 cm ]reg-1-12.north east) node [ seglabel, pos = 0.500 ] { Segment $1$ };
			\draw [ brace ] ([ xshift = 3.000 pt, yshift = 0.300 cm ]reg-1-5.north west) -- ([ xshift = - 3.000 pt, yshift = 0.300 cm ]reg-1-8.north east) node [ dotlabel, pos = 0.500 ] { \dots };
			\draw [ brace ] ([ xshift = 3.000 pt, yshift = 0.300 cm ]reg-1-1.north west) -- ([ xshift = - 3.000pt, yshift = 0.300 cm ]reg-1-4.north east) node [ seglabel, pos = 0.500 ] { Segment $n{-}1$ };
			\foreach \start/\end/\pos in {13/16/0, 9/12/1, 5/8/2, 1/4/3} {
				\draw [ mirrorbrace ] ([ xshift = 3.000 pt, yshift = - 0.300 cm ]reg-1-\start.south west) -- ([ xshift = - 3.000 pt, yshift = - 0.300 cm ]reg-1-\end.south east) coordinate [ pos = 0.500 ] (regblock\pos) node [ blocklabel, pos = 0.5 ] { $n$ Blocks };
			}
			\node [ title ] at ([yshift = - 0.100 cm ]reg.north) { Extended secret vector $\mathbf{ s }_{ i }$ };
			\matrix (circ) [
				matrix of nodes,
				nodes = circlebit,
				column sep = 1.000 pt,
				row sep = 0.000 pt,
				yshift = - 4.000 cm
			]
			at (reg.south)
			{
				| [ circlebit, shade, outer color = RedPurple!50, inner color = white ] | $n{-}1$ &
				| [ circlebit, shade, outer color = RedPurple!50, inner color = white ] | $\cdots$ &
				| [ circlebit, shade, outer color = RedPurple!50, inner color = white ] | $1$ &
				| [ circlebit, shade, outer color = RedPurple!50, inner color = white ] | $0$ &
				| [ circlebit, shade, outer color = GreenLighter2!50, inner color = white ] | $n{-}1$ &
				| [ circlebit, shade, outer color = GreenLighter2!50, inner color = white ] | $\cdots$ &
				| [ circlebit, shade, outer color = GreenLighter2!50, inner color = white ] | $1$ &
				| [ circlebit, shade, outer color = GreenLighter2!50, inner color = white ] | $0$ &
				| [ circlebit, shade, outer color = WordBlueVeryLight, inner color = white ] | $n{-}1$ &
				| [ circlebit, shade, outer color = WordBlueVeryLight, inner color = white ] | $\cdots$ &
				| [ circlebit, shade, outer color = WordBlueVeryLight, inner color = white ] | $1$ &
				| [ circlebit, shade, outer color = WordBlueVeryLight, inner color = white ] | $0$ &
				| [ circlebit, shade, outer color = WordAquaLighter40, inner color = white ] | $n{-}1$ &
				| [ circlebit, shade, outer color = WordAquaLighter40, inner color = white ] | $\cdots$ &
				| [ circlebit, shade, outer color = WordAquaLighter40, inner color = white ] | $1$ &
				| [ circlebit, shade, outer color = WordAquaLighter40, inner color = white ] | $0$ \\
			};
			\foreach \start/\end/\pos in {13/16/0, 9/12/1, 5/8/2, 1/4/3} {
				\draw[ brace ] ([ xshift = 2.000 pt, yshift = 0.300 cm ]circ-1-\start.north west) -- ([ xshift = - 2.000 pt, yshift = 0.300 cm ]circ-1-\end.north east) coordinate [ pos = 0.500 ] (circblock\pos) node [ seglabel, pos = 0.500 ] { $n$ Blocks };
			}
			\draw [ mirrorbrace ] ([ xshift = 3.000 pt, yshift = - 0.300 cm ]circ-1-13.south west) -- ([ xshift = - 3.000 pt, yshift = - 0.300 cm ]circ-1-16.south east) node [ blocklabel, pos = 0.500 ] { Segment $0$ };
			\draw [ mirrorbrace ] ([ xshift = 3.000 pt, yshift = - 0.300 cm ]circ-1-9.south west) -- ([ xshift = - 3.000 pt, yshift = - 0.300 cm ]circ-1-12.south east) node [ blocklabel, pos = 0.500 ] { Segment $1$ };
			\draw [ mirrorbrace ] ([ xshift = 3.000 pt, yshift = - 0.300 cm ]circ-1-5.south west) -- ([ xshift = - 3.000 pt, yshift = - 0.300 cm ]circ-1-8.south east) node [ dotlabelbelow, pos = 0.500] { \dots };
			\draw [ mirrorbrace ] ([ xshift = 3.000 pt, yshift = - 0.300 cm]circ-1-1.south west) -- ([ xshift = - 3.000 pt, yshift = - 0.300 cm]circ-1-4.south east) node [ blocklabel, pos = 0.500 ] { Segment $n{-}1$ };
			\node [ circletitle ] at ([yshift = 0.150 cm ]circ.south) { Quantum register $QR_{ i }$ };
			\begin{scope} [ on background layer ]
				\foreach \i in {0,1,2,3} {
					\draw [ connectarrow ] (regblock\i) -- (circblock\i);
				}
			\end{scope}
		\end{tikzpicture}
		\vspace*{ 0.050 cm }
		\caption{This figure gives a pictorial representation of the structure of the extended
			secret information vectors $\widetilde { \mathbf{ s } }_{ 0 }$, $\dots$, $\widetilde { \mathbf{ s } }_{ n - 1 }$.}
		\label{fig: Hierarchical Organization Into Segments & Blocks}
	\end{figure}
\end{tcolorbox}

Prior to introducing Definition \ref{def: Primary & Auxiliary Segments}, which delineates the concepts of primary and auxiliary segments, we establish the notation $\mathbf{ 0 }_{ m }$ to represent the zero bit vector of length $m$. This notation provides clarity for the subsequent discussion of segment and blocks.

\begin{definition} {Primary \& Auxiliary Segments} { Primary & Auxiliary Segments}
	Each information broker $IB_{ i }$, $0 \leq i \leq n - 1$, constructs the \emph{primary} and \emph{auxiliary segments}, denoted by $\mathbf{ p }_{ i }$ and $\mathbf{ a }_{ i }$ respectively, as illustrated in Figures \ref{fig: The Primary Segments of the Information Brokers} and \ref{fig: The Auxiliary Segments of the Information Brokers}.
\end{definition}

\begin{tcolorbox}
	[
		enhanced,
		breakable,
		grow to left by = 0.000 cm,
		grow to right by = 0.000 cm,
		colback = white,
		enhanced jigsaw,			
		frame hidden,
		sharp corners,
	]
	\begin{minipage} [ b ] { 0.480 \textwidth }
		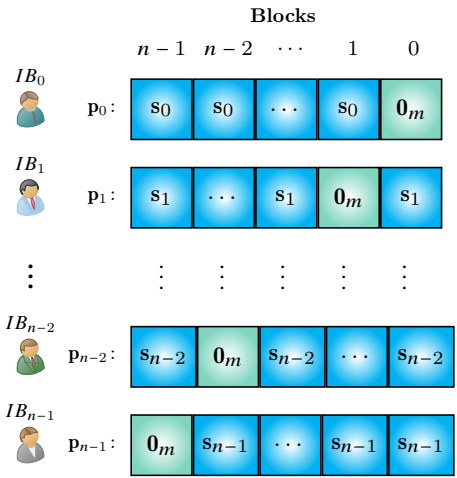
\begin{figure}[H]
			\centering
			\vspace*{ - 0.500 cm }
			\begin{tikzpicture} [ scale = 0.750, transform shape ]
				\node
				[
				charlie,
				scale = 1.250,
				anchor = center,
				label = { [ label distance = 0.000 cm ] north: $IB_{ 0 }$ }
				]
				(Charlie)
				{ };
				\matrix
				[
				column sep = 0.000 mm,
				right = 1.350 cm of Charlie,
				nodes = { draw = black, fill = white, minimum size = 8 mm, semithick, font = \small },
				label = { [ label distance = - 0.150 cm ] west: $\mathbf{ p }_{ 0 } \colon$ },
				]
				(IR_0)
				{
					\node [ shade, outer color = WordBlueVeryLight, inner color = white ] { $\mathbf{ s }_{ 0 }$ }; &
					\node [ shade, outer color = WordBlueVeryLight, inner color = white ] { $\mathbf{ s }_{ 0 }$ }; &
					\node [ shade, outer color = WordBlueVeryLight, inner color = white ] { \dots }; &
					\node [ shade, outer color = WordBlueVeryLight, inner color = white ] { $\mathbf{ s }_{ 0 }$ }; &
					\node [ shade, outer color = GreenLighter2!50, inner color = white ] { $\mathbf{ 0 }_{ m }$ }; \\
				};
				\matrix
				[
				column sep = 0.000 mm,
				right = 1.350 cm of Charlie,
				above = -0.350 cm of IR_0, 
				nodes = { minimum size = 8 mm, font = \footnotesize, semithick }, 
				label = { [ label distance = 0.000 cm, align = center, yshift = -0.350 cm ] north: \textbf{Blocks} }, 
				]
				(LabelMatrix)
				{
					\node { $n - 1$ }; &
					\node { $n - 2$ }; &
					\node { \dots }; &
					\node { $1$ }; &
					\node { $0$ }; \\
				};
				\node
				[
				dave,
				below = 1.250 cm of Charlie,
				scale = 1.250,
				anchor = center,
				label = { [ label distance = 0.000 cm ] north: $IB_{ 1 }$ }
				]
				(Dave)
				{ };
				\matrix
				[
				column sep = 0.000 mm,
				right = 1.350 cm of Dave,
				nodes = { draw = black, fill = white, minimum size = 8 mm, semithick, font = \small },
				label = { [ label distance = - 0.150 cm ] west: $\mathbf{ p }_{ 1 } \colon$ },
				]
				(IR_1)
				{
					\node [ shade, outer color = WordBlueVeryLight, inner color = white ] { $\mathbf{ s }_{ 1 }$ }; &
					\node [ shade, outer color = WordBlueVeryLight, inner color = white ] { \dots }; &
					\node [ shade, outer color = WordBlueVeryLight, inner color = white ] { $\mathbf{ s }_{ 1 }$ }; &
					\node [ shade, outer color = GreenLighter2!50, inner color = white ] { $\mathbf{ 0 }_{ m }$ }; &
					\node [ shade, outer color = WordBlueVeryLight, inner color = white ] { $\mathbf{ s }_{ 1 }$ }; \\
				};
				\node
				[
				below = 0.950 cm of Dave,
				scale = 1.250,
				anchor = center,
				]
				(VDots)
				{ \Large \vdots };
				\matrix
				[
				column sep = 0.000 mm,
				right = 1.350 cm of VDots,
				nodes = { minimum size = 8 mm, font = \small, semithick }, 
				]
				(DotsMatrix)
				{
					\node { \vdots }; &
					\node { \vdots }; &
					\node { \vdots }; &
					\node { \vdots }; &
					\node { \vdots }; \\
				};
				\node
				[
				businessman,
				below = 2.500 cm of Dave,
				scale = 1.250,
				anchor = center,
				label = { [ label distance = 0.000 cm ] north: $IB_{ n - 2 }$ }
				]
				(Businessman)
				{ };
				\matrix
				[
				column sep = 0.000 mm,
				right = 1.350 cm of Businessman,
				nodes = { draw = black, fill = white, minimum size = 8 mm, semithick, font = \small },
				label = { [ label distance = - 0.150 cm ] west: $\mathbf{ p }_{ n - 2 } \colon$ },
				]
				(IR_n_2)
				{
					\node [ shade, outer color = WordBlueVeryLight, inner color = white ] { $\mathbf{ s }_{ n - 2 }$ }; &
					\node [ shade, outer color = GreenLighter2!50, inner color = white ] { $\mathbf{ 0 }_{ m }$ }; &
					\node [ shade, outer color = WordBlueVeryLight, inner color = white ] { $\mathbf{ s }_{ n - 2 }$ }; &
					\node [ shade, outer color = WordBlueVeryLight, inner color = white ] { \dots }; &
					\node [ shade, outer color = WordBlueVeryLight, inner color = white ] { $\mathbf{ s }_{ n - 2 }$ }; \\
				};
				\node
				[
				bob,
				below = 1.250 cm of Businessman,
				scale = 1.250,
				anchor = center,
				label = { [ label distance = 0.000 cm ] north: $IB_{ n - 1 }$ }
				]
				(Bob)
				{ };
				\matrix
				[
				column sep = 0.000 mm,
				right = 1.350 cm of Bob,
				nodes = { draw = black, fill = white, minimum size = 8 mm, semithick, font = \small },
				label = { [ label distance = - 0.150 cm ] west: $\mathbf{ p }_{ n - 1 } \colon$ },
				]
				(IR_n_1)
				{
					\node [ shade, outer color = GreenLighter2!50, inner color = white ] { $\mathbf{ 0 }_{ m }$ }; &
					\node [ shade, outer color = WordBlueVeryLight, inner color = white ] { $\mathbf{ s }_{ n - 1 }$ }; &
					\node [ shade, outer color = WordBlueVeryLight, inner color = white ] { \dots }; &
					\node [ shade, outer color = WordBlueVeryLight, inner color = white ] { $\mathbf{ s }_{ n - 1 }$ }; &
					\node [ shade, outer color = WordBlueVeryLight, inner color = white ] { $\mathbf{ s }_{ n - 1 }$ }; \\
				};
			\end{tikzpicture}
			\vspace*{ 0.250 cm }
			\caption{This figure shows the construction of the primary segments $\mathbf{ p }_{ 0 }$, $\dots$, $\mathbf{ p }_{ n - 1 }$.}
			\label{fig: The Primary Segments of the Information Brokers}
		\end{figure}
	\end{minipage}
	\hspace{ 0.500 cm }
	\begin{minipage} [ b ] { 0.480 \textwidth }
		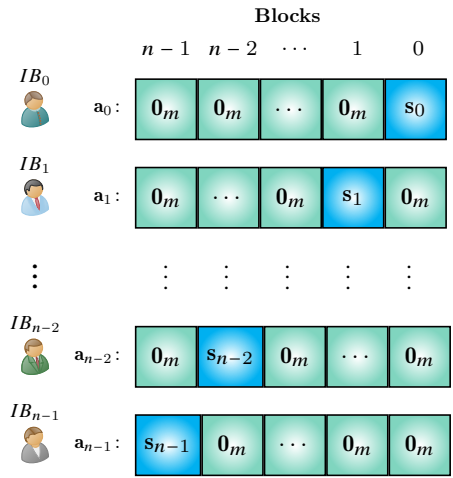
\begin{figure}[H]
			\centering
			\vspace*{ - 0.500 cm }
			\begin{tikzpicture} [ scale = 0.750, transform shape ]
				\node
				[
				charlie,
				scale = 1.250,
				anchor = center,
				label = { [ label distance = 0.000 cm ] north: $IB_{ 0 }$ }
				]
				(Charlie)
				{ };
				\matrix
				[
				column sep = 0.000 mm,
				right = 1.350 cm of Charlie,
				nodes = { draw = black, fill = white, minimum size = 8 mm, semithick, font = \small },
				label = { [ label distance = - 0.150 cm ] west: $\mathbf{ a }_{ 0 } \colon$ },
				]
				(IR_0)
				{
					\node [ shade, outer color = GreenLighter2!50, inner color = white ] { $\mathbf{ 0 }_{ m }$ }; &
					\node [ shade, outer color = GreenLighter2!50, inner color = white ] { $\mathbf{ 0 }_{ m }$ }; &
					\node [ shade, outer color = GreenLighter2!50, inner color = white ] { \dots }; &
					\node [ shade, outer color = GreenLighter2!50, inner color = white ] { $\mathbf{ 0 }_{ m }$ }; &
					\node [ shade, outer color = WordBlueVeryLight, inner color = white ] { $\mathbf{ s }_{ 0 }$ }; \\
				};
				\matrix
				[
				column sep = 0.000 mm,
				right = 1.350 cm of Charlie,
				above = -0.350 cm of IR_0, 
				nodes = { minimum size = 8 mm, font = \footnotesize, semithick }, 
				label = { [ label distance = 0.000 cm, align = center, yshift = -0.350 cm ] north: \textbf{Blocks} }, 
				]
				(LabelMatrix)
				{
					\node { $n - 1$ }; &
					\node { $n - 2$ }; &
					\node { \dots }; &
					\node { $1$ }; &
					\node { $0$ }; \\
				};
				\node
				[
				dave,
				below = 1.250 cm of Charlie,
				scale = 1.250,
				anchor = center,
				label = { [ label distance = 0.000 cm ] north: $IB_{ 1 }$ }
				]
				(Dave)
				{ };
				\matrix
				[
				column sep = 0.000 mm,
				right = 1.350 cm of Dave,
				nodes = { draw = black, fill = white, minimum size = 8 mm, semithick, font = \small },
				label = { [ label distance = - 0.150 cm ] west: $\mathbf{ a }_{ 1 } \colon$ },
				]
				(IR_1)
				{
					\node [ shade, outer color = GreenLighter2!50, inner color = white ] { $\mathbf{ 0 }_{ m }$ }; &
					\node [ shade, outer color = GreenLighter2!50, inner color = white ] { \dots }; &
					\node [ shade, outer color = GreenLighter2!50, inner color = white ] { $\mathbf{ 0 }_{ m }$ }; &
					\node [ shade, outer color = WordBlueVeryLight, inner color = white ] { $\mathbf{ s }_{ 1 }$ }; &
					\node [ shade, outer color = GreenLighter2!50, inner color = white ] { $\mathbf{ 0 }_{ m }$ }; \\
				};
				\node
				[
				below = 0.950 cm of Dave,
				scale = 1.250,
				anchor = center,
				]
				(VDots)
				{ \Large \vdots };
				\matrix
				[
				column sep = 0.000 mm,
				right = 1.350 cm of VDots,
				nodes = { minimum size = 8 mm, font = \small, semithick }, 
				]
				(DotsMatrix)
				{
					\node { \vdots }; &
					\node { \vdots }; &
					\node { \vdots }; &
					\node { \vdots }; &
					\node { \vdots }; \\
				};
				\node
				[
				businessman,
				below = 2.500 cm of Dave,
				scale = 1.250,
				anchor = center,
				label = { [ label distance = 0.000 cm ] north: $IB_{ n - 2 }$ }
				]
				(Businessman)
				{ };
				\matrix
				[
				column sep = 0.000 mm,
				right = 1.350 cm of Businessman,
				nodes = { draw = black, fill = white, minimum size = 8 mm, semithick, font = \small },
				label = { [ label distance = - 0.150 cm ] west: $\mathbf{ a }_{ n - 2 } \colon$ },
				]
				(IR_n_2)
				{
					\node [ shade, outer color = GreenLighter2!50, inner color = white ] { $\mathbf{ 0 }_{ m }$ }; &
					\node [ shade, outer color = WordBlueVeryLight, inner color = white ] { $\mathbf{ s }_{ n - 2 }$ }; &
					\node [ shade, outer color = GreenLighter2!50, inner color = white ] { $\mathbf{ 0 }_{ m }$ }; &
					\node [ shade, outer color = GreenLighter2!50, inner color = white ] { \dots }; &
					\node [ shade, outer color = GreenLighter2!50, inner color = white ] { $\mathbf{ 0 }_{ m }$ }; \\
				};
				\node
				[
				bob,
				below = 1.250 cm of Businessman,
				scale = 1.250,
				anchor = center,
				label = { [ label distance = 0.000 cm ] north: $IB_{ n - 1 }$ }
				]
				(Bob)
				{ };
				\matrix
				[
				column sep = 0.000 mm,
				right = 1.350 cm of Bob,
				nodes = { draw = black, fill = white, minimum size = 8 mm, semithick, font = \small },
				label = { [ label distance = - 0.150 cm ] west: $\mathbf{ a }_{ n - 1 } \colon$ },
				]
				(IR_n_1)
				{
					\node [ shade, outer color = WordBlueVeryLight, inner color = white ] { $\mathbf{ s }_{ n - 1 }$ }; &
					\node [ shade, outer color = GreenLighter2!50, inner color = white ] { $\mathbf{ 0 }_{ m }$ }; &
					\node [ shade, outer color = GreenLighter2!50, inner color = white ] { \dots }; &
					\node [ shade, outer color = GreenLighter2!50, inner color = white ] { $\mathbf{ 0 }_{ m }$ }; &
					\node [ shade, outer color = GreenLighter2!50, inner color = white ] { $\mathbf{ 0 }_{ m }$ }; \\
				};
			\end{tikzpicture}
			\vspace*{ 0.250 cm }
			\caption{This figure depicts the construction of the auxiliary segments $\mathbf{ a }_{ 0 }$, $\dots$, $\mathbf{ a }_{ n - 1 }$.}
			\label{fig: The Auxiliary Segments of the Information Brokers}
		\end{figure}
	\end{minipage}
\end{tcolorbox}

As outlined in Definition \ref{def: Primary & Auxiliary Segments}, each segment consists of $n$ blocks, collectively comprising $n m$ bits. By leveraging their primary and auxiliary segments, the information brokers can systematically and symmetrically construct their extended secret vectors $\widetilde { \mathbf{ s } }_{ 0 }, \dots, \widetilde { \mathbf{ s } }_{ n - 1 }$, as specified in Definition \ref{def: Extended Secret Vectors}. This structured approach ensures that the information is organized in a manner that supports both the anonymity and the parallel many-to-many communication objectives of the QDIBP.

\begin{definition} {Extended Secret Vectors} { Extended Secret Vectors}
	Each information broker $IB_{ i }$, $0 \leq i \leq n - 1$, constructs her \emph{extended secret information vector} $\widetilde { \mathbf{ s } }_{ i }$ as depicted in Figure \ref{fig: The Extended Secret Information Vectors of the Information Brokers}.
\end{definition}

Every extended secret vector $\widetilde { \mathbf{ s } }_{ i }$, $0 \leq i \leq n - 1$, is composed of $n$ segments, labeled from right to left as $0, \dots, n - 1$, following the structure illustrated in Figure \ref{fig: The Extended Secret Information Vectors of the Information Brokers}. Collectively, these segments contain a total of $n^{ 2 } m$ bits. As established in Definition \ref{def: Primary & Auxiliary Segments}, the extended secret information vectors can be articulated in a more precise and streamlined manner, as depicted in Figure \ref{fig: Explicit Form of the Extended Secret Information Vectors of the Information Brokers}. This refined representation enhances the clarity and efficiency of the QDIBP by providing a structured framework for organizing complex data while preserving anonymity and supporting seamless many-to-many communication.

\begin{tcolorbox}
	[
		enhanced,
		breakable,
		grow to left by = 0.000 cm,
		grow to right by = 0.000 cm,
		colback = white,
		enhanced jigsaw,			
		frame hidden,
		sharp corners,
	]
	\begin{figure}[H]
		\centering
		\vspace*{ - 0.500 cm }
		\begin{tikzpicture} [ scale = 0.800, transform shape ]
			\node
			[
			charlie,
			scale = 1.250,
			anchor = center,
			label = { [ label distance = 0.000 cm ] north: $IB_{ 0 }$ }
			]
			(Charlie)
			{ };
			\matrix
			[
			column sep = 0.000 mm,
			right = 1.350 cm of Charlie,
			nodes = { draw = black, fill = white, minimum size = 10 mm, semithick, font = \small },
			label = { [ label distance = - 0.150 cm ] west: $\widetilde { \mathbf{ s } }_{ 0 } \colon$ },
			]
			(IR_0)
			{
				\node [ shade, outer color = WordAquaLighter40, inner color = white ] { $\mathbf{ a }_{ 0 }$ }; &
				\node [ shade, outer color = WordAquaLighter40, inner color = white ] { $\mathbf{ a }_{ 0 }$ }; &
				\node [ shade, outer color = WordAquaLighter40, inner color = white ] { \dots }; &
				\node [ shade, outer color = WordAquaLighter40, inner color = white ] { $\mathbf{ a }_{ 0 }$ }; &
				\node [ shade, outer color = RedPurple!50, inner color = white ] { $\mathbf{ p }_{ 0 }$ }; \\
			};
			\matrix
			[
			column sep = 0.000 mm,
			right = 1.350 cm of Charlie,
			above = -0.350 cm of IR_0, 
			nodes = { minimum size = 10 mm, font = \footnotesize, semithick }, 
			label = { [ label distance = 0.000 cm, align = center, yshift = -0.350 cm ] north: \textbf{Segments} }, 
			]
			(SegmentLabelMatrix)
			{
				\node { $n - 1$ }; &
				\node { $n - 2$ }; &
				\node { \dots }; &
				\node { $1$ }; &
				\node { $0$ }; \\
			};
			\node
			[
			dave,
			below = 1.250 cm of Charlie,
			scale = 1.250,
			anchor = center,
			label = { [ label distance = 0.000 cm ] north: $IB_{ 1 }$ }
			]
			(Dave)
			{ };
			\matrix
			[
			column sep = 0.000 mm,
			right = 1.350 cm of Dave,
			nodes = { draw = black, fill = white, minimum size = 10 mm, semithick, font = \small },
			label = { [ label distance = - 0.150 cm ] west: $\widetilde { \mathbf{ s } }_{ 1 } \colon$ },
			]
			(IR_1)
			{
				\node [ shade, outer color = WordAquaLighter40, inner color = white ] { $\mathbf{ a }_{ 1 }$ }; &
				\node [ shade, outer color = WordAquaLighter40, inner color = white ] { \dots }; &
				\node [ shade, outer color = WordAquaLighter40, inner color = white ] { $\mathbf{ a }_{ 1 }$ }; &
				\node [ shade, outer color = RedPurple!50, inner color = white ] { $\mathbf{ p }_{ 1 }$ }; &
				\node [ shade, outer color = WordAquaLighter40, inner color = white ] { $\mathbf{ a }_{ 1 }$ }; \\
			};
			\node
			[
			below = 0.950 cm of Dave,
			scale = 1.250,
			anchor = center,
			]
			(VDots)
			{ \Large \vdots };
			\matrix
			[
			column sep = 0.000 mm,
			right = 1.350 cm of VDots,
			nodes = { minimum size = 10 mm, font = \small, semithick }, 
			]
			(DotsMatrix)
			{
				\node { \vdots }; &
				\node { \vdots }; &
				\node { \vdots }; &
				\node { \vdots }; &
				\node { \vdots }; \\
			};
			\node
			[
			businessman,
			below = 2.500 cm of Dave,
			scale = 1.250,
			anchor = center,
			label = { [ label distance = 0.000 cm ] north: $IB_{ n - 2 }$ }
			]
			(Businessman)
			{ };
			\matrix
			[
			column sep = 0.000 mm,
			right = 1.350 cm of Businessman,
			nodes = { draw = black, fill = white, minimum size = 10 mm, semithick, font = \small },
			label = { [ label distance = - 0.150 cm ] west: $\widetilde { \mathbf{ s } }_{ n - 2 } \colon$ },
			]
			(IR_n_2)
			{
				\node [ shade, outer color = WordAquaLighter40, inner color = white ] { $\mathbf{ a }_{ n - 2 }$ }; &
				\node [ shade, outer color = RedPurple!50, inner color = white ] { $\mathbf{ p }_{ n - 2 }$ }; &
				\node [ shade, outer color = WordAquaLighter40, inner color = white ] { $\mathbf{ a }_{ n - 2 }$ }; &
				\node [ shade, outer color = WordAquaLighter40, inner color = white ] { \dots }; &
				\node [ shade, outer color = WordAquaLighter40, inner color = white ] { $\mathbf{ a }_{ n - 2 }$ }; \\
			};
			\node
			[
			bob,
			below = 1.250 cm of Businessman,
			scale = 1.250,
			anchor = center,
			label = { [ label distance = 0.000 cm ] north: $IB_{ n - 1 }$ }
			]
			(Bob)
			{ };
			\matrix
			[
			column sep = 0.000 mm,
			right = 1.350 cm of Bob,
			nodes = { draw = black, fill = white, minimum size = 10 mm, semithick, font = \small },
			label = { [ label distance = - 0.150 cm ] west: $\widetilde { \mathbf{ s } }_{ n - 1 } \colon$ },
			]
			(IR_n_1)
			{
				\node [ shade, outer color = RedPurple!50, inner color = white ] { $\mathbf{ p }_{ n - 1 }$ }; &
				\node [ shade, outer color = WordAquaLighter40, inner color = white ] { $\mathbf{ a }_{ n - 1 }$ }; &
				\node [ shade, outer color = WordAquaLighter40, inner color = white ] { \dots }; &
				\node [ shade, outer color = WordAquaLighter40, inner color = white ] { $\mathbf{ a }_{ n - 1 }$ }; &
				\node [ shade, outer color = WordAquaLighter40, inner color = white ] { $\mathbf{ a }_{ n - 1 }$ }; \\
			};
		\end{tikzpicture}
		\caption{This figure gives a pictorial representation of the structure of the extended secret information vectors $\widetilde { \mathbf{ s } }_{ 0 }$, $\dots$, $\widetilde { \mathbf{ s } }_{ n - 1 }$.}
		\label{fig: The Extended Secret Information Vectors of the Information Brokers}
	\end{figure}
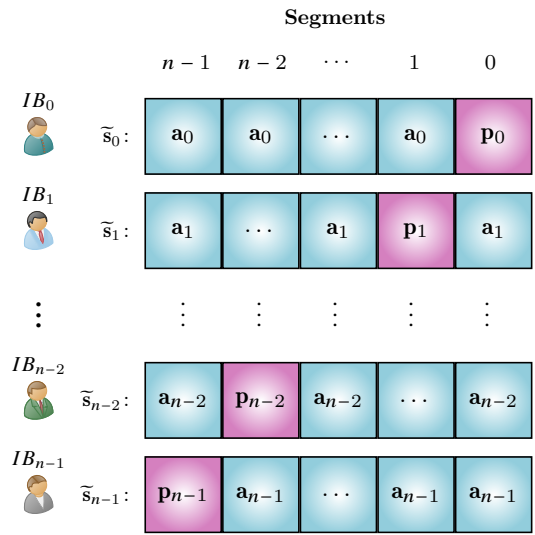
\end{tcolorbox}

\begin{tcolorbox}
	[
		enhanced,
		breakable,
		grow to left by = 1.000 cm,
		grow to right by = 1.000 cm,
		colback = white,
		enhanced jigsaw,			
		frame hidden,
		sharp corners,
	]
	\begin{figure}[H]
		\centering
		\begin{tikzpicture} [ scale = 0.750, transform shape ]
			\node
			[
			charlie,
			scale = 1.250,
			anchor = center,
			label = { [ label distance = 0.000 cm ] north: $IB_{ 0 }$ }
			]
			(Charlie)
			{ };
			\matrix
			[
			column sep = 0.000 mm,
			right = 1.350 cm of Charlie,
			nodes = { draw = black, fill = white, minimum size = 9.000 mm, line width = 0.005 pt, font = \scriptsize },
			label = { [ label distance = - 0.150 cm ] west: $\widetilde { \mathbf{ s } }_{ 0 } \colon$ },
			]
			(IR_0)
			{
				\node [ shade, outer color = GreenLighter2!50, inner color = white ] { $\mathbf{ 0 }_{ m }$ }; &
				\node [ shade, outer color = GreenLighter2!50, inner color = white ] { \tiny \dots }; &
				\node [ shade, outer color = GreenLighter2!50, inner color = white ] { $\mathbf{ 0 }_{ m }$ }; &
				\node [ shade, outer color = WordBlueVeryLight, inner color = white ] { $\mathbf{ s }_{ 0 }$ }; &
				\node
				[
				shade, outer color = GreenLighter2!50, inner color = white,
				append after command =
				{
					(\tikzlastnode.west) edge [ line width = 1.500 pt ] (\tikzlastnode.north west)
					(\tikzlastnode.west) edge [ line width = 1.500 pt ] (\tikzlastnode.south west)
				}
				]
				{ $\mathbf{ 0 }_{ m }$ }; &
				\node [ shade, outer color = GreenLighter2!50, inner color = white ] { \tiny \dots }; &
				\node [ shade, outer color = GreenLighter2!50, inner color = white ] { $\mathbf{ 0 }_{ m }$ }; &
				\node
				[
				shade, outer color = WordBlueVeryLight, inner color = white,
				append after command =
				{
					(\tikzlastnode.east) edge [ line width = 1.500 pt ] (\tikzlastnode.north east)
					(\tikzlastnode.east) edge [ line width = 1.500 pt ] (\tikzlastnode.south east)
				}
				]
				{ $\mathbf{ s }_{ 0 }$ }; &
				\node { \dots }; &
				\node
				[
				shade, outer color = GreenLighter2!50, inner color = white,
				append after command =
				{
					(\tikzlastnode.west) edge [ line width = 1.500 pt ] (\tikzlastnode.north west)
					(\tikzlastnode.west) edge [ line width = 1.500 pt ] (\tikzlastnode.south west)
				}
				]
				{ $\mathbf{ 0 }_{ m }$ }; &
				\node [ shade, outer color = GreenLighter2!50, inner color = white ] { \tiny \dots }; &
				\node [ shade, outer color = GreenLighter2!50, inner color = white ] { $\mathbf{ 0 }_{ m }$ }; &
				\node [ shade, outer color = WordBlueVeryLight, inner color = white ] { $\mathbf{ s }_{ 0 }$ }; &
				\node
				[
				shade, outer color = WordBlueVeryLight, inner color = white,
				append after command =
				{
					(\tikzlastnode.west) edge [ line width = 1.500 pt ] (\tikzlastnode.north west)
					(\tikzlastnode.west) edge [ line width = 1.500 pt ] (\tikzlastnode.south west)
				}
				]
				{ $\mathbf{ s }_{ 0 }$ }; &
				\node [ shade, outer color = WordBlueVeryLight, inner color = white ] { \tiny \dots }; &
				\node [ shade, outer color = WordBlueVeryLight, inner color = white ] { $\mathbf{ s }_{ 0 }$ }; &
				\node [ shade, outer color = GreenLighter2!50, inner color = white ] { $\mathbf{ 0 }_{ m }$ }; &
				\\
			};
			\node
			[
			above = 0.250 cm of IR_0,
			anchor = center,
			label = { [ label distance = 0.000 cm, align = center, yshift = 0.000 cm ] north: \textbf{Blocks} },
			]
			(Blocks)
			{ };
			\node
			[
			dave,
			below = 1.500 cm of Charlie,
			scale = 1.250,
			anchor = center,
			label = { [ label distance = 0.000 cm ] north: $IB_{ 1 }$ }
			]
			(Dave)
			{ };
			\matrix
			[
			column sep = 0.000 mm,
			right = 1.350 cm of Dave,
			nodes = { draw = black, fill = white, minimum size = 9.000 mm, line width = 0.005 pt, font = \scriptsize },
			label = { [ label distance = - 0.150 cm ] west: $\widetilde { \mathbf{ s } }_{ 1 } \colon$ },
			]
			(IR_1)
			{
				\node [ shade, outer color = GreenLighter2!50, inner color = white ] { \tiny \dots }; &
				\node [ shade, outer color = GreenLighter2!50, inner color = white ] { $\mathbf{ 0 }_{ m }$ }; &
				\node [ shade, outer color = WordBlueVeryLight, inner color = white ] { $\mathbf{ s }_{ 1 }$ }; &
				\node [ shade, outer color = GreenLighter2!50, inner color = white ] { $\mathbf{ 0 }_{ m }$ }; &
				\node
				[
				shade, outer color = GreenLighter2!50, inner color = white,
				append after command =
				{
					(\tikzlastnode.west) edge [ line width = 1.500 pt ] (\tikzlastnode.north west)
					(\tikzlastnode.west) edge [ line width = 1.500 pt ] (\tikzlastnode.south west)
				}
				]
				{ \tiny \dots }; &
				\node [ shade, outer color = GreenLighter2!50, inner color = white ] { $\mathbf{ 0 }_{ m }$ }; &
				\node [ shade, outer color = WordBlueVeryLight, inner color = white ] { $\mathbf{ s }_{ 1 }$ }; &
				\node
				[
				shade, outer color = GreenLighter2!50, inner color = white,
				append after command =
				{
					(\tikzlastnode.east) edge [ line width = 1.500 pt ] (\tikzlastnode.north east)
					(\tikzlastnode.east) edge [ line width = 1.500 pt ] (\tikzlastnode.south east)
				}
				]
				{ $\mathbf{ 0 }_{ m }$ }; &
				\node { \dots }; &
				\node
				[
				shade, outer color = WordBlueVeryLight, inner color = white,
				append after command =
				{
					(\tikzlastnode.west) edge [ line width = 1.500 pt ] (\tikzlastnode.north west)
					(\tikzlastnode.west) edge [ line width = 1.500 pt ] (\tikzlastnode.south west)
				}
				]
				{ \tiny \dots }; &
				\node [ shade, outer color = WordBlueVeryLight, inner color = white ] { $\mathbf{ s }_{ 1 }$ }; &
				\node [ shade, outer color = GreenLighter2!50, inner color = white ] { $\mathbf{ 0 }_{ m }$ }; &
				\node [ shade, outer color = WordBlueVeryLight, inner color = white ] { $\mathbf{ s }_{ 1 }$ }; &
				\node
				[
				shade, outer color = GreenLighter2!50, inner color = white,
				append after command =
				{
					(\tikzlastnode.west) edge [ line width = 1.500 pt ] (\tikzlastnode.north west)
					(\tikzlastnode.west) edge [ line width = 1.500 pt ] (\tikzlastnode.south west)
				}
				]
				{ \tiny \dots }; &
				\node [ shade, outer color = GreenLighter2!50, inner color = white ] { $\mathbf{ 0 }_{ m }$ }; &
				\node [ shade, outer color = WordBlueVeryLight, inner color = white ] { $\mathbf{ s }_{ 1 }$ }; &
				\node [ shade, outer color = GreenLighter2!50, inner color = white ] { $\mathbf{ 0 }_{ m }$ }; &
				\\
			};
			\node
			[
			below = 1.250 cm of Dave,
			scale = 1.250,
			anchor = center,
			]
			(VDots)
			{ \Large \vdots };
			\matrix
			[
			column sep = 0.000 mm,
			below = - 0.050 cm of IR_1,
			nodes = { minimum size = 9.000 mm, font = \small }, 
			]
			(DotsMatrix)
			{
				\node [ anchor = center ] { \vdots }; &
				\node [ anchor = center ] { \vdots }; &
				\node [ anchor = center ] { \vdots }; &
				\node [ anchor = center ] { \vdots }; &
				\node [ anchor = center ] { \vdots }; &
				\node [ anchor = center ] { \vdots }; &
				\node [ anchor = center ] { \vdots }; &
				\node [ anchor = center ] { \vdots }; &
				\node [ yshift = 0.375 cm ] { \dots }; &
				\node [ anchor = center ] { \vdots }; &
				\node [ anchor = center ] { \vdots }; &
				\node [ anchor = center ] { \vdots }; &
				\node [ anchor = center ] { \vdots }; &
				\node [ anchor = center ] { \vdots }; &
				\node [ anchor = center ] { \vdots }; &
				\node [ anchor = center ] { \vdots }; &
				\node [ anchor = center ] { \vdots }; \\
			};
			\node
			[
			bob,
			below = 3.000 cm of Dave,
			scale = 1.250,
			anchor = center,
			label = { [ label distance = 0.000 cm ] north: $IB_{ n - 1 }$ }
			]
			(Bob)
			{ };
			\matrix
			[
			column sep = 0.000 mm,
			right = 1.350 cm of Bob,
			nodes = { draw = black, fill = white, minimum size = 9.000 mm, line width = 0.005, font = \scriptsize },
			label = { [ label distance = - 0.150 cm ] west: $\widetilde { \mathbf{ s } }_{ n - 1 } \colon$ },
			]
			(IR_n_1)
			{
				\node [ shade, outer color = GreenLighter2!50, inner color = white ] { $\mathbf{ 0 }_{ m }$ }; &
				\node [ shade, outer color = WordBlueVeryLight, inner color = white ] { $\mathbf{ s }_{ n - 1 }$ }; &
				\node [ shade, outer color = WordBlueVeryLight, inner color = white ] { \tiny \dots }; &
				\node [ shade, outer color = WordBlueVeryLight, inner color = white ] { $\mathbf{ s }_{ n - 1 }$ }; &
				\node
				[
				shade, outer color = WordBlueVeryLight, inner color = white,
				append after command =
				{
					(\tikzlastnode.west) edge [ line width = 1.500 pt ] (\tikzlastnode.north west)
					(\tikzlastnode.west) edge [ line width = 1.500 pt ] (\tikzlastnode.south west)
				}
				]
				{ $\mathbf{ s }_{ n - 1 }$ }; &
				\node [ shade, outer color = GreenLighter2!50, inner color = white ] { $\mathbf{ 0 }_{ m }$ }; &
				\node [ shade, outer color = GreenLighter2!50, inner color = white ] { \tiny \dots }; &
				\node
				[
				shade, outer color = GreenLighter2!50, inner color = white,
				append after command =
				{
					(\tikzlastnode.east) edge [ line width = 1.500 pt ] (\tikzlastnode.north east)
					(\tikzlastnode.east) edge [ line width = 1.500 pt ] (\tikzlastnode.south east)
				}
				]
				{ $\mathbf{ 0 }_{ m }$ }; &
				\node { \dots }; &
				\node
				[
				shade, outer color = WordBlueVeryLight, inner color = white,
				append after command =
				{
					(\tikzlastnode.west) edge [ line width = 1.500 pt ] (\tikzlastnode.north west)
					(\tikzlastnode.west) edge [ line width = 1.500 pt ] (\tikzlastnode.south west)
				}
				]
				{ $\mathbf{ s }_{ n - 1 }$ }; &
				\node [ shade, outer color = GreenLighter2!50, inner color = white ] { $\mathbf{ 0 }_{ m }$ }; &
				\node [ shade, outer color = GreenLighter2!50, inner color = white ] { \tiny \dots }; &
				\node [ shade, outer color = GreenLighter2!50, inner color = white ] { $\mathbf{ 0 }_{ m }$ }; &
				\node
				[
				shade, outer color = WordBlueVeryLight, inner color = white,
				append after command =
				{
					(\tikzlastnode.west) edge [ line width = 1.500 pt ] (\tikzlastnode.north west)
					(\tikzlastnode.west) edge [ line width = 1.500 pt ] (\tikzlastnode.south west)
				}
				]
				{ $\mathbf{ s }_{ n - 1 }$ }; &
				\node [ shade, outer color = GreenLighter2!50, inner color = white ] { $\mathbf{ 0 }_{ m }$ }; &
				\node [ shade, outer color = GreenLighter2!50, inner color = white ] { \tiny \dots }; &
				\node [ shade, outer color = GreenLighter2!50, inner color = white ] { $\mathbf{ 0 }_{ m }$ }; &
				\\
			};
		\end{tikzpicture}
		\caption{This figure provides a detailed and analytical depiction of the extended secret information vectors $\widetilde { \mathbf{ s } }_{ 0 }$, $\dots$, $\widetilde { \mathbf{ s } }_{ n - 1 }$, expressed in terms of their constituent blocks. We clarify that the blocks drawn in green contain the zero vector $\mathbf{ 0 }_{ m }$, while blocks drawn in blue contain secret vectors.}
		\label{fig: Explicit Form of the Extended Secret Information Vectors of the Information Brokers}
	\end{figure}
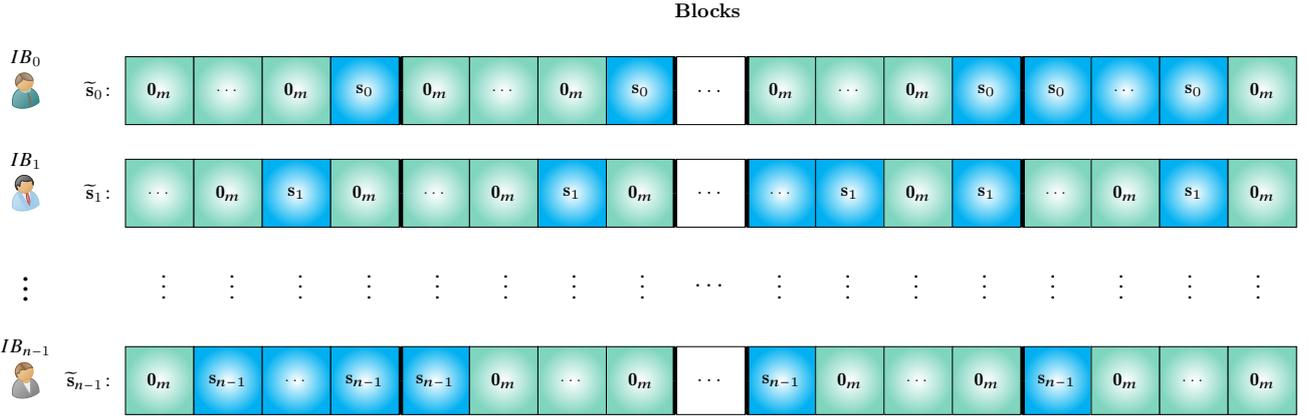
\end{tcolorbox}

The above Figure \ref{fig: Explicit Form of the Extended Secret Information Vectors of the Information Brokers} offers a clear and structured visualization of the block-based composition of the extended secret vectors, enhancing the understanding of how these vectors are organized within the QDIBP. This representation facilitates precise analysis of the vector structure, facilitating the reader's understanding of how the protocol achieves the objectives of maintaining anonymity while accomplishing secure many-to-many communication.

\begin{definition} {Aggregated Secret Vector} { Aggregated Secret Vector}
	Given the extended secret information vectors $\widetilde { \mathbf{ s } }_{ 0 }$, $\dots$, $\widetilde { \mathbf{ s } }_{ n - 1 }$, the \emph{aggregated secret vector} $\mathbf{ t }$ is defined as their sum modulo $2$.
	\begin{align}
		\label{eq: Aggregated Secret Vector}
		\mathbf{ t }
		\coloneq
		\bigoplus_{ i = 0 }^{ n - 1 }
		\
		\widetilde { \mathbf{ s } }_{ i }
		\ .
	\end{align}
	The aggregated secret vector $\mathbf{ t }$ consists of the $n$ aggregated segments $\mathbf{ t }_{ 0 }, \dots, \mathbf{ t }_{ n - 1 }$, enumerated from right to left. Therefore, it can conveniently be expressed as shown below.
	\begin{align}
		\label{eq: Segment Form of Aggregated Secret Vector}
		\mathbf{ t }
		=
		\mathbf{ t }_{ n - 1 }
		\
		\mathbf{ t }_{ n - 2 }
		\dots
		\mathbf{ t }_{ 1 }
		\
		\mathbf{ t }_{ 0 }
		\ .
	\end{align}
\end{definition}

The segments $\mathbf{ t }_{ 0 }, \dots, \mathbf{ t }_{ n - 1 }$ play a capital role in the realization of the QDIBP. Their precise structure in shown in great detail in the next Figure \ref{fig: The Degments of the Aggregated Secret Vector}.

\begin{tcolorbox}
	[
		enhanced,
		breakable,
		grow to left by = 0.000 cm,
		grow to right by = 0.000 cm,
		colback = white,
		enhanced jigsaw,			
		frame hidden,
		sharp corners,
	]
	\begin{figure}[H]
		\centering
		\vspace*{ - 0.500 cm }
		\begin{tikzpicture} [ scale = 0.900, transform shape ]
			\node
			[
			charlie,
			scale = 1.500,
			anchor = center,
			label = { [ label distance = 0.000 cm ] north: $IB_{ 0 }$ }
			]
			(Charlie)
			{ };
			\node
			[
			right = 0.825 cm of Charlie,
			anchor = center,
			]
			(DataToCharlie)
			{ \rotatebox { 90 } { \faWifi[regular] } };
			\matrix
			[
			column sep = 0.000 mm,
			right = 2.000 cm of Charlie,
			nodes = { draw = black, fill = white, minimum width = 17 mm, minimum height = 9 mm, line width = 0.250, font = \footnotesize },
			label = { [ label distance = - 0.150 cm ] west: $\mathbf{ t }_{ 0 } \colon$ },
			]
			(IR_0)
			{
				\node [ shade, outer color = RedPurple!50, inner color = white ] { $\mathbf{ s }_{ 0 } \oplus \mathbf{ s }_{ n - 1 }$ }; &
				\node [ shade, outer color = RedPurple!50, inner color = white ] { $\mathbf{ s }_{ 0 } \oplus \mathbf{ s }_{ n - 2 }$ }; &
				\node [ shade, outer color = RedPurple!50, inner color = white ] { \dots }; &
				\node [ shade, outer color = RedPurple!50, inner color = white ] { $\mathbf{ s }_{ 0 } \oplus \mathbf{ s }_{ 1 }$ }; &
				\node [ shade, outer color = GreenLighter2!50, inner color = white ] { $\mathbf{ 0 }_{ m }$ }; \\
			};
			\matrix
			[
			column sep = 0.000 mm,
			right = 1.350 cm of Charlie,
			above = -0.450 cm of IR_0, 
			nodes = { minimum width = 17 mm, minimum height = 9 mm, font = \footnotesize }, 
			label = { [ label distance = 0.000 cm, align = center, yshift = -0.350 cm ] north: \textbf{Blocks} }, 
			]
			(SegmentLabelMatrix)
			{
				\node { $n - 1$ }; &
				\node { $n - 2$ }; &
				\node { \dots }; &
				\node { $1$ }; &
				\node { $0$ }; \\
			};
			\node
			[
			dave,
			below = 1.500 cm of Charlie,
			scale = 1.500,
			anchor = center,
			label = { [ label distance = 0.000 cm ] north: $IB_{ 1 }$ }
			]
			(Dave)
			{ };
			\node
			[
			right = 0.825 cm of Dave,
			anchor = center,
			]
			(DataToDave)
			{ \rotatebox { 90 } { \faWifi[regular] } };
			\matrix
			[
			column sep = 0.000 mm,
			right = 2.000 cm of Dave,
			nodes = { draw = black, fill = white, minimum width = 17 mm, minimum height = 9 mm, line width = 0.250, font = \footnotesize },
			label = { [ label distance = - 0.150 cm ] west: $\mathbf{ t }_{ 1 } \colon$ },
			]
			(IR_1)
			{
				\node [ shade, outer color = RedPurple!50, inner color = white ] { $\mathbf{ s }_{ 1 } \oplus \mathbf{ s }_{ n - 1 }$ }; &
				\node [ shade, outer color = RedPurple!50, inner color = white ] { \dots }; &
				\node [ shade, outer color = RedPurple!50, inner color = white ] { $\mathbf{ s }_{ 1 } \oplus \mathbf{ s }_{ 2 }$ }; &
				\node [ shade, outer color = GreenLighter2!50, inner color = white ] { $\mathbf{ 0 }_{ m }$ }; &
				\node [ shade, outer color = RedPurple!50, inner color = white ] { $\mathbf{ s }_{ 1 } \oplus \mathbf{ s }_{ 0 }$ }; \\
			};
			\node
			[
			below = 1.000 cm of Dave,
			scale = 1.500,
			anchor = center,
			]
			(VDots)
			{ \Large \vdots };
			\matrix
			[
			column sep = 0.000 mm,
			right = 2.000 cm of VDots,
			nodes = { minimum width = 17 mm, minimum height = 9 mm, font = \footnotesize }, 
			]
			(DotsMatrix)
			{
				\node { \vdots }; &
				\node { \vdots }; &
				\node { \vdots }; &
				\node { \vdots }; &
				\node { \vdots }; \\
			};
			\node
			[
			businessman,
			below = 2.700 cm of Dave,
			scale = 1.500,
			anchor = center,
			label = { [ label distance = 0.000 cm ] north: $IB_{ n - 2 }$ }
			]
			(Businessman)
			{ };
			\node
			[
			right = 0.825 cm of Businessman,
			anchor = center,
			]
			(DataToBusinessman)
			{ \rotatebox { 90 } { \faWifi[regular] } };
			\matrix
			[
			column sep = 0.000 mm,
			right = 2.000 cm of Businessman,
			nodes = { draw = black, fill = white, minimum width = 17 mm, minimum height = 9 mm, line width = 0.250, font = \footnotesize },
			label = { [ label distance = - 0.150 cm ] west: $\mathbf{ t }_{ n - 2 } \colon$ },
			]
			(IR_n_2)
			{
				\node [ shade, outer color = RedPurple!50, inner color = white ] { $\mathbf{ s }_{ n - 2 } \oplus \mathbf{ s }_{ n - 1 }$ }; &
				\node [ shade, outer color = GreenLighter2!50, inner color = white ] { $\mathbf{ 0 }_{ m }$ }; &
				\node [ shade, outer color = RedPurple!50, inner color = white ] { $\mathbf{ s }_{ n - 2 } \oplus \mathbf{ s }_{ n - 3 }$ }; &
				\node [ shade, outer color = RedPurple!50, inner color = white ] { \dots }; &
				\node [ shade, outer color = RedPurple!50, inner color = white ] { $\mathbf{ s }_{ n - 2 } \oplus \mathbf{ s }_{ 0 }$ }; \\
			};
			\node
			[
			bob,
			below = 1.500 cm of Businessman,
			scale = 1.500,
			anchor = center,
			label = { [ label distance = 0.000 cm ] north: $IB_{ n - 1 }$ }
			]
			(Bob)
			{ };
			\node
			[
			right = 0.825 cm of Bob,
			anchor = center,
			]
			(DataToBob)
			{ \rotatebox { 90 } { \faWifi[regular] } };
			\matrix
			[
			column sep = 0.000 mm,
			right = 2.000 cm of Bob,
			nodes = { draw = black, fill = white, minimum width = 17 mm, minimum height = 9 mm, line width = 0.250, font = \footnotesize },
			label = { [ label distance = - 0.150 cm ] west: $\mathbf{ t }_{ n - 1 } \colon$ },
			]
			(IR_n_1)
			{
				\node [ shade, outer color = GreenLighter2!50, inner color = white ] { $\mathbf{ 0 }_{ m }$ }; &
				\node [ shade, outer color = RedPurple!50, inner color = white ] { $\mathbf{ s }_{ n - 1 } \oplus \mathbf{ s }_{ n - 2 }$ }; &
				\node [ shade, outer color = RedPurple!50, inner color = white ] { \dots }; &
				\node [ shade, outer color = RedPurple!50, inner color = white ] { $\mathbf{ s }_{ n - 1 } \oplus \mathbf{ s }_{ 1 }$ }; &
				\node [ shade, outer color = RedPurple!50, inner color = white ] { $\mathbf{ s }_{ n - 1 } \oplus \mathbf{ s }_{ 0 }$ }; \\
			};
		\end{tikzpicture}
		\vspace*{ 0.150 cm }
		\caption{This figure contains a detailed representation of the structure of the segments $\mathbf{ t }_{ 0 }, \dots, \mathbf{ t }_{ n - 1 }$. As in previous figures, the blocks drawn in cyan contain the zero vector $\mathbf{ 0 }_{ m }$, while blocks drawn in red contain encoded information in the form $\mathbf{ s }_{ i } \oplus \mathbf{ s }_{ j }$, $i \neq j$.}
		\label{fig: The Degments of the Aggregated Secret Vector}
	\end{figure}
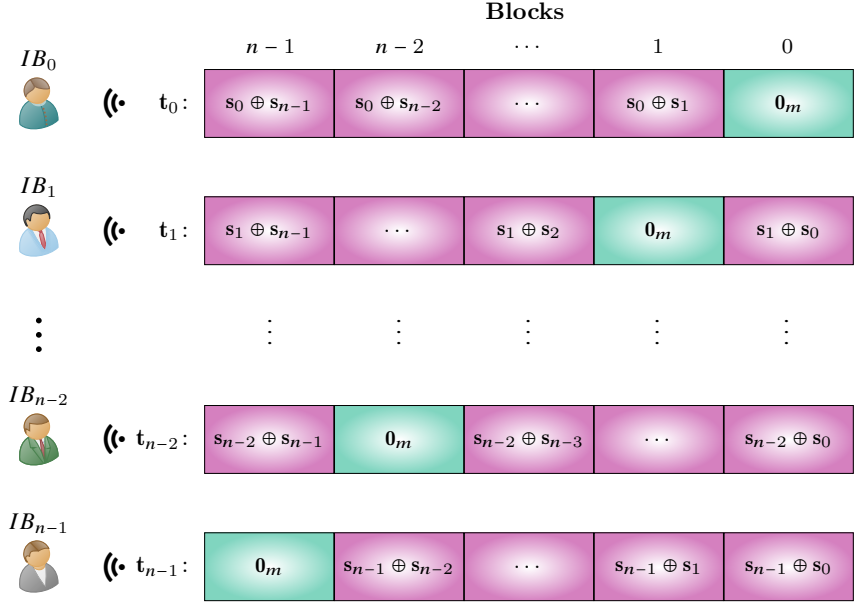
\end{tcolorbox}


The aggregated segments $\mathbf{ t }_{ 0 }, \dots, \mathbf{ t }_{ n - 1 }$ are the primary carriers of information in the QDIBP. Specifically, the segment $\mathbf{ t }_{ i }$ is designed to be delivered to information broker $IB_{ i }$, where $0 \leq i \leq n - 1$, at the conclusion of the protocol. Each $\mathbf{ t }_{ i }$, $0 \leq i \leq n - 1$, contains the secret vector $\mathbf{ s }_{ j }$ of every other broker $IB_{ j }$, where $0 \leq j \neq i \leq n - 1$, obfuscated as $\mathbf{ s }_{ i } \oplus \mathbf{ s }_{ j }$. This encoding ensures that neither Trent nor any external party can decipher the secrets, thereby maintaining the confidentiality and anonymity of the communication. By integrating Definitions \ref{def: Primary & Auxiliary Segments}, \ref{def: Extended Secret Vectors}, and \ref{def: Aggregated Secret Vector}, we may also express the precise structural form of $\mathbf{ t }_{ i }$ by equations \eqref{eq: Aggregated Segment i Decomposition Into Block} and \eqref{eq: Aggregated Block i, k Form}

\begin{align}
	\label{eq: Aggregated Segment i Decomposition Into Block}
	\mathbf{ t }_{ i }
	=
	\mathbf{ b }_{ i, n - 1 }
	\
	\mathbf{ b }_{ i, n - 2 }
	\dots
	\mathbf{ b }_{ i, 1 }
	\
	\mathbf{ b }_{ i, 0 }
	\ ,
	\ 0 \leq i \leq n - 1
	\ ,
\end{align}

where

\begin{align}
	\label{eq: Aggregated Block i, k Form}
	\left
	\{
	\
	\begin{aligned}
		\mathbf{ b }_{ i, i }
		&=
		\mathbf{ 0 }_{ m }
		\\
		\mathbf{ b }_{ i, j }
		&=
		\mathbf{ s }_{ i }
		\oplus
		\mathbf{ s }_{ j }
		\ ,
		\ 0 \leq i \leq n - 1
	\end{aligned}
	\
	\right
	\}
	\ .
\end{align}

\subsection{The $r$-uniform entanglement distribution scheme} \label{subsec: The $r$-Uniform Entanglement Distribution Scheme}

The physical implementation of the QDIBP is based on a composite system comprising multiple local quantum circuits, with no fixed limit on their number. The protocol's functionality hinges on the maximal entanglement of corresponding qubits across all quantum registers. This entanglement is achieved through the $r$-Uniform Entanglement Distribution Scheme, taken from \cite{Andronikos2024b}. The scheme is formally defined in Definition \ref{def: The $r$-Uniform Entanglement Distribution Scheme}.

\begin{definition} {The $r$-Uniform Entanglement Distribution Scheme} { The $r$-Uniform Entanglement Distribution Scheme}
	The $r$-Uniform Distribution Scheme stipulates the following:
	\begin{itemize}
		\item	
		There are $r$ players and each player is endowed with a quantum register consisting of $p$ qubits, and
		\item	
		for each position $k$, where $0 \leq k \leq p - 1$, the qubits in the $k^{ th }$ position across all registers are entangled in the $\ket{ GHZ_{ r } }$ state.
	\end{itemize}
\end{definition}

This entanglement scheme establishes a robust correlation among the quantum registers by ensuring that their corresponding qubits are maximally entangled in the $\ket{ GHZ_{ r } }$ state. A visual representation of this configuration is provided in Figure \ref{fig: The $r$-Uniform Entanglement Distribution Scheme}.

\begin{tcolorbox}
	[
	enhanced,
	breakable,
	grow to left by = 0.000 cm,
	grow to right by = 0.000 cm,
	colback = SkyBlue1!08,
	enhanced jigsaw,			
	frame hidden,
	sharp corners,
	]
	\begin{figure}[H]
		\centering
		\begin{tikzpicture} [ scale = 0.750, transform shape ]
			\node
			[
			anchor = center,
			]
			(Alice)
			{ $QR_{ r - 1 } \colon$ };
			\matrix
			[
			matrix of nodes,
			nodes in empty cells,
			column sep = 4.000 mm,
			right = 0.500 of Alice,
			nodes = { circle, minimum size = 12.000 mm, semithick, font = \footnotesize },
			]
			{
				\node [ shade, outer color = RedPurple!50, inner color = white ] (A_p-1^r-1) { $q_{ p - 1 }^{ r - 1 }$ }; &
				\node [ shade, outer color = WordAquaLighter60, inner color = white ] { \large \dots }; &
				\node [ shade, outer color = GreenLighter2!50, inner color = white ] (A_1^r-1) { $q_{ 1 }^{ r - 1 }$ }; &
				\node [ shade, outer color = WordBlueVeryLight, inner color = white ] (A_0^r-1) { $q_{ 0 }^{ r - 1 }$ };
				\\
			};
			\node
			[
			anchor = center,
			above = 1.500 cm of Alice,
			]
			(Bob)
			{ $QR_{ r - 2 } \colon$ };
			\matrix
			[
			matrix of nodes,
			nodes in empty cells,
			column sep = 4.000 mm,
			right = 0.500 of Bob,
			nodes = { circle, minimum size = 12.000 mm, semithick, font = \footnotesize },
			]
			{
				\node [ shade, outer color = RedPurple!50, inner color = white ] (B_p-1^r-2) { $q_{ p - 1 }^{ r - 2 }$ }; &
				\node [ shade, outer color = WordAquaLighter60, inner color = white ] { \large \dots }; &
				\node [ shade, outer color = GreenLighter2!50, inner color = white ] (B_1^r-2) { $q_{ 1 }^{ r - 2 }$ }; &
				\node [ shade, outer color = WordBlueVeryLight, inner color = white ] (B_0^r_2) { $q_{ 0 }^{ r - 2 }$ };
				\\
			};
			\node
			[
			anchor = center,
			above = 1.000 cm of Bob,
			]
			(Dots) { \Large \vdots };
			\matrix
			[
			right = 4.175 cm of Dots,
			anchor = center,
			column sep = 0.000 mm,
			row sep = 0.000 mm,
			nodes = { minimum size = 14.000 mm, semithick }
			]
			{
				\node { }; &
				\node { }; &
				\node { \Large \vdots }; &
				\node { }; &
				\node { };
				\\
			};
			\node
			[
			anchor = center,
			above = 0.800 cm of Dots,
			]
			(Charlie) { $QR_{ 1 } \colon$ };
			\matrix
			[
			matrix of nodes,
			nodes in empty cells,
			column sep = 4.000 mm, right = 0.700 of Charlie,
			nodes = { circle, minimum size = 12.000 mm, semithick, font = \footnotesize },
			]
			{
				\node [ shade, outer color = RedPurple!50, inner color = white ] (C_p-1^1) { $q_{ p - 1 }^{ 1 }$ }; &
				\node [ shade, outer color = WordAquaLighter60, inner color = white ] { \large \dots }; &
				\node [ shade, outer color = GreenLighter2!50, inner color = white ] (C_1^1) { $q_{ 1 }^{ 1 }$ }; &
				\node [ shade, outer color = WordBlueVeryLight, inner color = white ] (C_0^1) { $q_{ 0 }^{ 1 }$ };
				\\
			};
			\node
			[
			anchor = center,
			above = 1.500 cm of Charlie,
			]
			(Dave)
			{ $QR_{ 0 } \colon$ };
			\matrix
			[
			matrix of nodes,
			nodes in empty cells,
			column sep = 4.000 mm,
			right = 0.700 of Dave,
			nodes = { circle, minimum size = 12.000 mm, semithick, font = \footnotesize },
			]
			{
				\node [ shade, outer color = RedPurple!50, inner color = white ] (D_p-1^0) { $q_{ p - 1 }^{ 0 }$ }; &
				\node [ shade, outer color = WordAquaLighter60, inner color = white ] { \large \dots }; &
				\node [ shade, outer color = GreenLighter2!50, inner color = white ] (D_1^0) { $q_{ 1 }^{ 0 }$ }; &
				\node [ shade, outer color = WordBlueVeryLight, inner color = white ] (D_0^0) { $q_{ 0 }^{ 0 }$ };
				\\
			};
			\node
			[
			above right = 3.250 cm and 3.500 cm of Dave,
			anchor = center,
			shade,
			top color = GreenTeal,
			bottom color = black,
			rectangle,
			text width = 12.000 cm,
			align = center
			]
			(Label)
			{ \color{white} \textbf{A distributed system of $r$ quantum registers $QR_{ 0 }, \dots, QR_{ r - 1 }$, each with $p$ qubits. The characteristic property of this system is that the qubits in the corresponding positions form a $\ket{ GHZ_{ r } }$ tuple.} };
			\begin{scope}[on background layer]
				\node
				[
				above = - 1.500 cm of A_p-1^r-1,
				rectangle,
				rounded corners = 8.000 pt,
				fill = RedPurple!15,
				minimum width = 6.000 mm,
				minimum height = 90.000 mm
				]
				( ) { };
				\node
				[
				above = 0.750 cm of D_p-1^0
				]
				( )
				{\colorbox{RedPurple!50} {$\ket{ GHZ_{ r } }$}};
				\node
				[
				above = - 1.500 cm of A_1^r-1,
				rectangle,
				rounded corners = 8.000 pt,
				fill = GreenLighter2!15,
				minimum width = 6.000 mm,
				minimum height = 90.000 mm
				]
				( )
				{ };
				\node
				[
				above = 0.750 cm of D_1^0
				]
				( )
				{\colorbox{GreenLighter2!50} {$\ket{ GHZ_{ r } }$}};
				\node
				[
				above = - 1.500 cm of A_0^r-1,
				rectangle,
				rounded corners = 8.000 pt,
				fill = WordBlueVeryLight!30,
				minimum width = 6.000 mm,
				minimum height = 90.000 mm
				]
				( )
				{ };
				\node
				[
				above = 0.750 cm of D_0^0
				]
				( )
				{\colorbox{WordBlueVeryLight} {$\ket{ GHZ_{ r } }$}};
			\end{scope}
		\end{tikzpicture}
		\caption{This figure draws the $r$ qubits that populate the same position in the $QR_{ 0 }, \dots, QR_{ r - 1 }$ registers with the same color so as to emphasize that they belong to the same $\ket{ GHZ_{ r } }$ $r$-tuple.}
		\label{fig: The $r$-Uniform Entanglement Distribution Scheme}
	\end{figure}
\end{tcolorbox}

For the practical realization of the QDIBP, an in-game participant, such as Trent or one of the $n$ information brokers, must generate and distribute the necessary $\ket{ GHZ_{ r } }$ tuples through secure quantum channels. Notably, the physical arrangement of the quantum registers—whether they are co-located within a single facility or distributed across geographically distant locations—does not affect the protocol's efficacy. The entanglement-induced correlations, facilitated by the $p$ $\ket{ GHZ_{ r } }$ tuples, remain intact regardless of spatial distribution. This unique property of quantum entanglement allows the entire system to function as a unified, cohesive entity, enabling seamless information broadcasting across the network.

\section{Detailed analysis of the QDIBP} \label{sec: Detailed Analysis Of The QDIBP}

This Section provides an in-depth explanation of the execution of QDIBP that evolves in three phases.

\subsection{Phase 1: Distributing \& obfuscating the secret information} \label{subsec: Phase 1: Distributing & Obfuscating The Secret Information}

In the first phase of the Quantum Dining Information Brokers Protocol (QDIBP), each of the $n$ information brokers employs a private quantum circuit tailored to their specific role. These circuits are identical in structure, with the exception of the unitary transformations $U_{ \widetilde { \mathbf{ s } }_{ i } }$, $0 \leq i \leq n - 1$, which are uniquely determined to encode the extended secret vectors $\widetilde { \mathbf{ s } }_{ i }$ into the relative phase of the entangled distributed system. This phase is realized by the quantum circuit IBtoTQC depicted in Figure \ref{fig: The Quantum Circuit IBtoTQC of the QDIBP}.

\begin{tcolorbox}
	[
		enhanced,
		breakable,
		grow to left by = 0.000 cm,
		grow to right by = 0.000 cm,
		colback = SkyBlue1!08,
		enhanced jigsaw,			
		frame hidden,
		sharp corners,
	]
	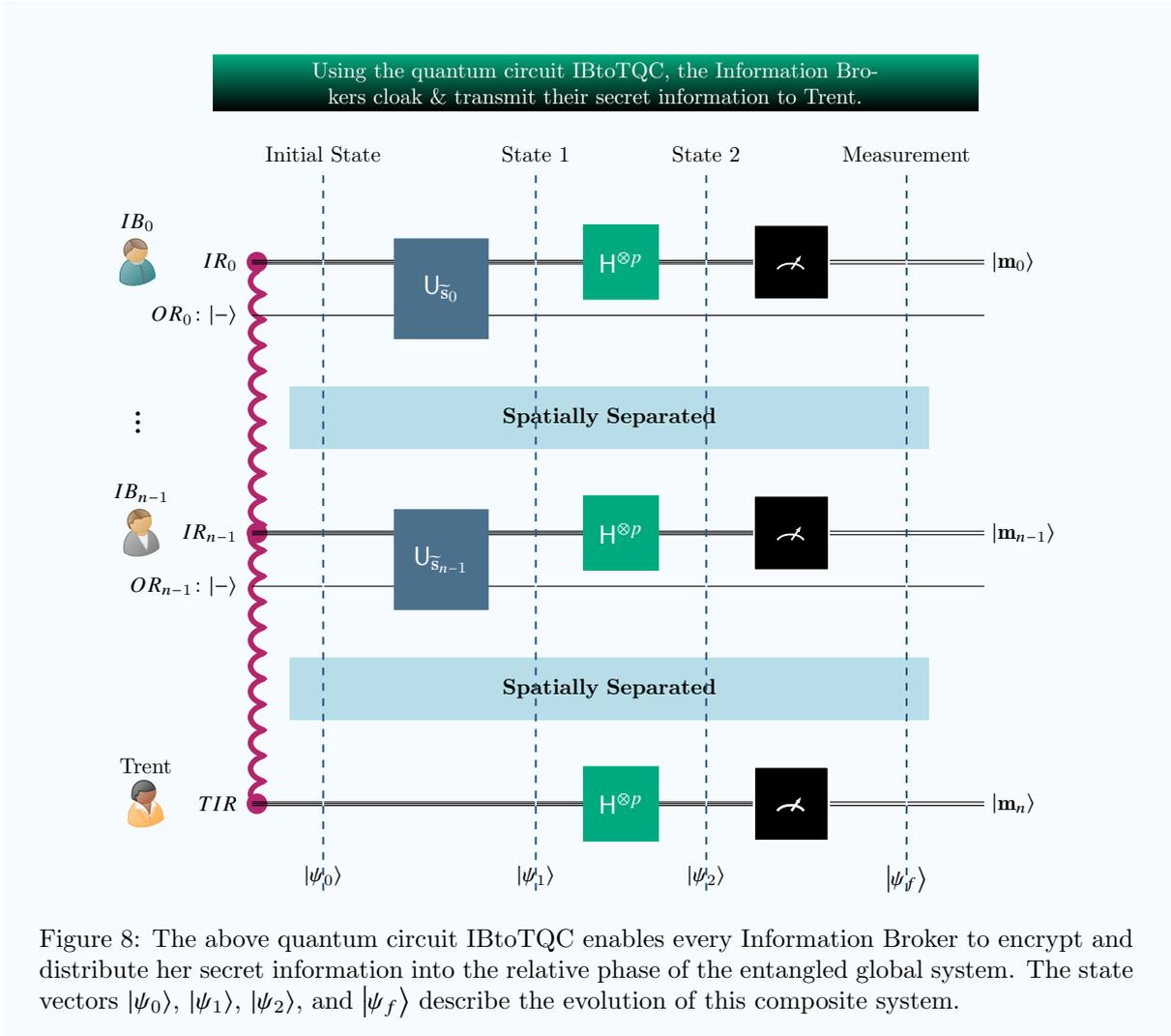
\begin{figure}[H]
		\centering
		\begin{tikzpicture} [ scale = 0.850, transform shape ]
			\begin{yquant}
				nobit AUX_IB_0_0;
				nobit AUX_IB_0_1;
				[ name = Charlie ] qubits { $IR_{ 0 }$ } IR_0;
				qubit { $OR_{ 0 } \colon \ket{ - }$ } OR_0;
				nobit AUX_IB_0_2;
				nobit AUX_IB_0_3;
				[ name = space_0, register/minimum height = 8.000 mm ] nobit space_0;
				nobit AUX_IB_n-1_0;
				nobit AUX_IB_n-1_1;
				[ name = Bob ] qubits { $IR_{ n - 1 }$ } IR_n_1;
				qubit { $OR_{ n - 1 } \colon \ket{ - }$ } OR_n_1;
				nobit AUX_B_n-1_2;
				nobit AUX_B_n-1_3;
				[ name = space_n_2, register/minimum height = 8.000 mm ] nobit space_n_2;
				nobit AUX_T_0;
				nobit AUX_T_1;
				[ name = Alice ] qubits { $TIR$ } TIR;
				nobit AUX_A_2;
				nobit AUX_A_3;
				[ name = Ph0, WordBlueDarker, line width = 0.250 mm, label = { [ label distance = 0.200 cm ] north: Initial State } ]
				barrier ( - ) ;
				[ draw = SkyBlue4, fill = SkyBlue4, radius = 0.750 cm ] box {\color{white} \large \sf{U}$_{ \widetilde { \mathbf{ s } }_{ 0 } }$} (IR_0 - OR_0);
				[ draw = SkyBlue4, fill = SkyBlue4, radius = 0.750 cm ] box {\color{white} \large \sf{U}$_{ \widetilde { \mathbf{ s } }_{ n - 1 } }$} (IR_n_1 - OR_n_1);
				[ name = Ph1, WordBlueDarker, line width = 0.250 mm, label = { [ label distance = 0.200 cm ] north: State 1 } ]
				barrier ( - ) ;
				[ draw = GreenLighter2, fill = GreenLighter2, radius = 0.600 cm ] box {\color{white} \large \sf{H}$^{ \otimes p }$} IR_0;
				[ draw = GreenLighter2, fill = GreenLighter2, radius = 0.600 cm ] box {\color{white} \large \sf{H}$^{ \otimes p }$} IR_n_1;
				[ draw = GreenLighter2, fill = GreenLighter2, radius = 0.600 cm ] box {\color{white} \large \sf{H}$^{ \otimes p }$} TIR;
				[ name = Ph2, WordBlueDarker, line width = 0.250 mm, label = { [ label distance = 0.200 cm ] north: State 2 } ]
				barrier ( - ) ;
				[ line width = .250 mm, draw = white, fill = black, radius = 0.600 cm ] measure IR_0;
				[ line width = .250 mm, draw = white, fill = black, radius = 0.600 cm ] measure IR_n_1;
				[ line width = .350 mm, draw = white, fill = black, radius = 0.600 cm ] measure TIR;
				[ name = Ph3, WordBlueDarker, line width = 0.250 mm, label = { [ label distance = 0.200 cm ] north: Measurement } ]
				barrier ( - ) ;
				output { $\ket{ \mathbf{ m }_{ 0 } }$ } IR_0;
				output { $\ket{ \mathbf{ m }_{ n - 1 } }$ } IR_n_1;
				output { $\ket{ \mathbf{ m }_{ n } }$ } TIR;
				\node [ below = 5.250 cm ] at (Ph0) { $\ket{ \psi_{ 0 } }$ };
				\node [ below = 5.250 cm ] at (Ph1) { $\ket{ \psi_{ 1 } }$ };
				\node [ below = 5.250 cm ] at (Ph2) { $\ket{ \psi_{ 2 } }$ };
				\node [ below = 5.250 cm ] at (Ph3) { $\ket{ \psi_{ f } }$ };
				\node
				[
				charlie,
				scale = 1.500,
				anchor = center,
				left = 0.700 cm of Charlie,
				label = { [ label distance = 0.000 cm ] north: $IB_{ 0 }$ }
				]
				() { };
				\node
				[
				bob,
				scale = 1.500,
				anchor = center,
				left = 0.300 cm of Bob,
				label = { [ label distance = 0.000 cm ] north: $IB_{ n - 1 }$ }
				]
				() { };
				\node
				[
				alice,
				scale = 1.500,
				anchor = center,
				left = 0.500 cm of Alice,
				label = { [ label distance = 0.000 cm ] north: Trent }
				]
				() { };
				\begin{scope} [ on background layer ]
					\node
					[
					above left = 1.350 cm and 0.425 cm of Bob
					]
					{ \LARGE \vdots };
					\node
					[
					above right = - 0.300 cm and 0.600 cm of space_0, rectangle, fill = WordAquaLighter60, text width = 10.000 cm, align = center, minimum height = 10.000 mm
					]
					{ \bf Spatially Separated };
					\node
					[
					above right = - 0.300 cm and 0.600 cm of space_n_2, rectangle, fill = WordAquaLighter60, text width = 10.000 cm, align = center, minimum height = 10.000 mm
					]
					{ \bf Spatially Separated };
				\end{scope}
			\end{yquant}
			\node
			[
			above right = 2.750 cm and 5.500 cm of Charlie,
			anchor = center,
			shade,
			top color = GreenLighter2, bottom color = black,
			rectangle,
			text width = 12.000 cm,
			align = center,
			]
			(Label)
			{ \color{white}
				Using the quantum circuit IBtoTQC, the Information Brokers cloak \& transmit their secret information to Trent.
			};
			\scoped [ on background layer ]
			\draw
			[
			purple4, -, >=stealth, line width = 0.750 mm, decoration = coil, decorate
			]
			( $ (Alice.east) + ( 0.800 mm, 0.000 mm ) $ ) node [ circle, fill, minimum size = 1.500 mm ] () {} -- ( $ (Bob.east) + ( 0.800 mm, 0.000 mm ) $ ) node [ circle, fill, minimum size = 1.5 mm ] () {} -- ( $ (Charlie.east) + ( 0.800 mm, 0.000 mm ) $ ) node [ circle, fill, minimum size = 1.500 mm ] () {};
		\end{tikzpicture}
		\caption{The above quantum circuit IBtoTQC enables every Information Broker to encrypt and distribute her secret information into the relative phase of the entangled global system. The state vectors $\ket{ \psi_{ 0 } }$, $\ket{ \psi_{ 1 } }$, $\ket{ \psi_{ 2 } }$, and $\ket{ \psi_{ f } }$ describe the evolution of this composite system.}
		\label{fig: The Quantum Circuit IBtoTQC of the QDIBP}
	\end{figure}
\end{tcolorbox}

\noindent	Upon completion of this encoding process, all $n + 1$ participants, including the $n$ information brokers and Trent, perform measurements on their respective quantum registers. The information brokers then transmit their measurement outcomes to Trent via secure, pairwise-authenticated classical channels. By combining these measurements with his own, Trent computes the aggregated secret vector $\mathbf{ t }$. It is critical to underscore that, despite having access to $\mathbf{ t }$, Trent is unable to deduce any of the individual secret vectors $\mathbf{ s }_{ i }$, as elaborated in subsection \ref{subsec: Phase 1: Distributing & Obfuscating The Secret Information}. This ensures the confidentiality of each broker's contribution, preserving the security of the protocol through the inherent properties of quantum entanglement and the carefully designed obfuscation mechanism. The quantum circuit denoted as IBtoTQC, consistent with all quantum circuits described in this work, adheres to a set of standardized conventions to ensure clarity and compatibility with established quantum computing frameworks:

\begin{itemize}
	\item	
	Qubits are organized following the Qiskit convention \cite{Qiskit2025}, employing little-endian qubit indexing. In this scheme, the least significant qubit is positioned at the top of the circuit diagram, while the most significant qubit is placed at the bottom.
	\item	
	For each information broker $IB_{ i }$, where $0 \leq i \leq n - 1$, the quantum input register, denoted $IR_{ i }$, consists of $p = n^{ 2 } m$ qubits, sufficient to encode the required information for the protocol.
	\item	
	The output register for each information broker $IB_{ i }$, denoted $IR_{ i }$ for $0 \leq i \leq n - 1$, is a single-qubit register initialized to the state $\ket{ - }$
	\item	
	The unitary transformation $U_{ \widetilde { \mathbf{ s } }_{ i } }$, $0 \leq i \leq n - 1$, is specific to each information broker $IB_{ i }$. Its precise form is determined by the extended secret vector $\widetilde { \mathbf{ s } }_{ i }$ and satisfies the relation specified in equation \eqref{eq: Information Brokers Unitary Transforms}.
	\item	
	The operator $H^{ p }$ represents the $p$-fold Hadamard transform, where $p = n^{ 2 } m$, applied to the input register to create a superposition of states critical to the protocol's operation.
\end{itemize}

\noindent	The information brokers achieve secure and anonymous information exchange by operating on their private, yet entangled, quantum circuits through their respective secret unitary transformations $U_{ \widetilde { \mathbf{ s } }_{ i } }$, $0 \leq i \leq n - 1$. These transformations encode the secret information vectors $\mathbf{ s }_{ i }$, which are embedded in the form of extended secret bit vectors $\widetilde { \mathbf{ s } }_{ i }$, into the relative phases of the entangled composite quantum system. The unitary transformations $U_{ \widetilde { \mathbf{ s } }_{ i } }$ follow the standard form $U_{ \widetilde { \mathbf{ s } }_{ i } } \colon \ket{ y } \ \ket{ \mathbf{ x } }$ $\rightarrow$ $\ket{ y \oplus \left(  \widetilde { \mathbf{ s } }_{ i } \bullet \mathbf{ x } \right) } \ \ket{ \mathbf{ x } }$, where $\oplus$ denotes the bitwise XOR operation and $\bullet$ represents the inner product modulo $2$. This can be expressed more concisely as a phase shift conditional on the inner product of the extended secret vector and the input state. This mechanism ensures that the secret information is securely integrated into the entangled system, preserving anonymity and enabling the protocol's distributed computation objectives.

\begin{align}
	\label{eq: Information Brokers Unitary Transforms}
	U_{ \widetilde { \mathbf{ s } }_{ i } }
	&\colon
	\ket{ - }
	\
	\ket{ \mathbf{ x } }
	\rightarrow
	( - 1 )^{ \widetilde { \mathbf{ s } }_{ i } \bullet \mathbf{ x } }
	\
	\ket{ - }
	\
	\ket{ \mathbf{ x } }
	\ ,
	\ 0 \leq i \leq n - 1
\end{align}

\noindent	Invoking \eqref{eq: p-Fold Extended General GHZ_r State}, where in our case $r$ stands for $n + 1$ and $p$ stands for $n^{ 2 } m$, we can express the initial state $\ket{ \psi_{ 0 } }$ of the IBtoTQC quantum circuit as shown below. To enhance clarity, we use the subscript $T$ to signify Trent, and the subscripts $0 \leq i \leq n - 1$, to designate the information brokers $IB_{ 0 }, \dots, IB_{ n - 1 }$, respectively.

\begin{align}
	\label{eq: Initial State of IBtoTQC}
	\ket{ \psi_{ 0 } }
	=
	2^{ - \frac { p } { 2 } }
	\sum_{ \mathbf{ x } \in \mathbb{ B }^{ p } }
	\
	\ket{ \mathbf{ x } }_{ T }
	\
	\ket{ - }_{ n - 1 }
	\ket{ \mathbf{ x } }_{ n - 1 }
	\
	\dots
	\
	\ket{ - }_{ 0 }
	\ket{ \mathbf{ x } }_{ 0 }
\end{align}

\noindent	The anonymous information exchange begins in earnest when the information brokers act on their private quantum circuits via their secret unitary transforms $U_{ \widetilde { \mathbf{ s } }_{ i } }$, $0 \leq i \leq n - 1$. Their cumulative effect drives the quantum circuit IBtoTQC into the next state $\ket{ \psi_{ 1 } }$.

\begin{align} \label{eq: QDIBP Phase 1 State 1}
	\ket{ \psi_{ 1 } }
	&=
	2^{ - \frac { p } { 2 } }
	\sum_{ \mathbf{ x } \in \mathbb{ B }^{ p } }
	\
	\ket{ \mathbf{ x } }_{ T }
	\
	\left(
	U_{ \widetilde { \mathbf{ s } }_{ n - 1 } }
	\
	\ket{ - }_{ n - 1 }
	\
	\ket{ \mathbf{ x } }_{ n - 1 }
	\right)
	\dots
	\left(
	U_{ \widetilde { \mathbf{ s } }_{ 0 } }
	\
	\ket{ - }_{ 0 }
	\
	\ket{ \mathbf{ x } }_{ 0 }
	\right)
	\nonumber \\
	&\hspace{-0.100 cm} \overset { \eqref{eq: Information Brokers Unitary Transforms} } { = }
	2^{ - \frac { p } { 2 } }
	\sum_{ \mathbf{ x } \in \mathbb{ B }^{ p } }
	\
	\ket{ \mathbf{ x } }_{ T }
	\
	( - 1 )^{ \widetilde { \mathbf{ s } }_{ n - 1 } \bullet \mathbf{ x } }
	\
	\ket{ - }_{ n - 1 }
	\ket{ \mathbf{x} }_{ n - 1 }
	\dots
	( - 1 )^{ \widetilde { \mathbf{ s } }_{ 0 } \bullet \mathbf{ x } }
	\
	\ket{ - }_{ 0 }
	\
	\ket{ \mathbf{ x } }_{ 0 }
	\nonumber \\
	&=
	2^{ - \frac { p } { 2 } }
	\sum_{ \mathbf{ x } \in \mathbb{ B }^{ p } }
	\
	( - 1 )^{ ( \widetilde { \mathbf{ s } }_{ n - 1 } \oplus \dots \oplus \widetilde { \mathbf{ s } }_{ 0 } ) \bullet \mathbf{ x } }
	\
	\ket{ \mathbf{ x } }_{ T }
	\
	\ket{ - }_{ n - 1 }
	\
	\ket{ \mathbf{x} }_{ n - 1 }
	\dots
	\ket{ - }_{ 0 }
	\
	\ket{ \mathbf{ x } }_{ 0 }
	\nonumber \\
	&\hspace{-0.100 cm} \overset { \eqref{eq: Aggregated Secret Vector} } { = }
	2^{ - \frac { p } { 2 } }
	\sum_{ \mathbf{ x } \in \mathbb{ B }^{ p } }
	\
	( - 1 )^{ \mathbf{ t } \bullet \mathbf{ x } }
	\
	\ket{ \mathbf{ x } }_{ T }
	\
	\ket{ - }_{ n - 1 }
	\
	\ket{ \mathbf{ x } }_{ n - 1 }
	\dots
	\ket{ - }_{ 0 }
	\
	\ket{ \mathbf{ x } }_{ 0 }
\end{align}

\noindent	The quantum state $\ket{ \psi_{ 1 } }$, as given by \eqref{eq: QDIBP Phase 1 State 1}, emerges directly from the entanglement phenomenon inherent in QDIBP. In this protocol, each of the $n$ information brokers independently and untraceably embed their secret information into the quantum system by applying their respective unitary transformations. These transformations, ensure that the secret information is encoded securely without revealing individual contributions. The collective effect of these $n$ unitary operations results in the encoding of the aggregated secret vector $\mathbf{ t }$ into the relative phase structure of the distributed quantum circuit. This phase encoding leverages the quantum superposition and entanglement properties to protect the information while enabling its distributed nature. To extract the aggregated secret vector $\mathbf{ t }$, all $n$ information brokers, along with the semi-honest third party, Trent, perform a coordinated quantum operation. Specifically, they apply the $p$-fold Hadamard transform, where $p = n^{ 2 } m$, to their respective input registers, as illustrated in Figure \ref{fig: The Quantum Circuit IBtoTQC of the QDIBP}. This transformation disentangles the system in a controlled manner, allowing the aggregated secret to be reconstructed. As a result of this process, the quantum state of the system transitions from $\ket{ \psi_{ 2 } }$ to $\ket{ \psi_{ 2 } }$. This state transition highlights the power of quantum entanglement and multi-party quantum protocols in secure information processing.

\begin{align}
	\label{eq: QDIBP Phase 1 State 2 - I}
	\ket{ \psi_{ 2 } }
	=
	2^{ - \frac { p } { 2 } }
	\sum_{ \mathbf{ x } \in \mathbb{ B }^{ p } }
	\
	( - 1 )^{ \mathbf{ t } \bullet \mathbf{ x } }
	\
	\left(
	H^{ \otimes p }
	\ket{ \mathbf{ x } }_{ T }
	\right)
	\
	\ket{ - }_{ n - 1 }
	\
	\left(
	H^{ \otimes p }
	\ket{ \mathbf{ x } }_{ n - 1 }
	\right)
	\dots
	\ket{ - }_{ 0 }
	\
	\left(
	H^{ \otimes p }
	\ket{ \mathbf{ x } }_{ 0 }
	\right)
\end{align}

\noindent	At this point, equation \eqref{eq: Hadamard p-Fold Ket x} allows to further analyze $H^{ \otimes p } \ket{ \mathbf{ x } }_{ T }$, $H^{ \otimes p } \ket{ \mathbf{ x } }_{ n - 1 }$, \dots, $H^{ \otimes p } \ket{ \mathbf{ x } }_{ 0 }$, using the expansions shown below. These transformations, which act on the input registers of Trent and the $n$ information brokers, leverage the Hadamard gate’s ability to create superpositions, facilitating the extraction of encoded information from the entangled quantum state.

\begin{align}
	\label{eq: Hadamard p-Fold Transform Auxiliary Expansions}
	\left
	\{
	\
	\begin{aligned}
		H^{ \otimes p }
		\ket{ \mathbf{ x } }_{ T }
		&=
		2^{ - \frac { p } { 2 } }
		\sum_{ \mathbf{ y }_{ n } \in \mathbb{ B }^{ p } }
		\
		( - 1 )^{ \mathbf{ y }_{ n } \bullet \mathbf{ x } }
		\ket{ \mathbf{ y }_{ n } }_{ T }
		\\
		H^{ \otimes p }
		\ket{ \mathbf{ x } }_{ n - 1 }
		&=
		2^{ - \frac { p } { 2 } }
		\sum_{ \mathbf{ y }_{ n - 1 } \in \mathbb{ B }^{ p } }
		\
		( - 1 )^{ \mathbf{ y }_{ n - 1 } \bullet \mathbf{ x } }
		\ket{ \mathbf{ y }_{ n - 1 } }_{ n - 1 }
		\\[ 10.000 pt ]
		\multicolumn{ 2 } { c } { \dots }			
		\\[ 10.000 pt ]
		H^{ \otimes p }
		\ket{ \mathbf{ x } }_{ 0 }
		&=
		2^{ - \frac { p } { 2 } }
		\sum_{ \mathbf{ y }_{ 0 } \in \mathbb{ B }^{ p } }
		\
		( - 1 )^{ \mathbf{ y }_{ 0 } \bullet \mathbf{ x } }
		\ket{ \mathbf{ y }_{ 0 } }_{ 0 }
	\end{aligned}
	\
	\right
	\}
\end{align}

\noindent	By applying the substitutions outlined above, the quantum state $\ket{ \psi_{ 2 } }$ can be reformulated into a more explicit expression, as presented below.

\begin{align}
	\label{eq: QDIBP Phase 1 State 2 - II}
	\hspace* { - 2.000 cm }
	{\small
		\ket{ \psi_{ 2 } }
		=
		2^{ ( - \frac { p } { 2 } )^{ n + 1 } }
		\sum_{ \mathbf{ x } \in \mathbb{ B }^{ p } }
		\sum_{ \mathbf{ y }_{ n } \in \mathbb{ B }^{ p } }
		\sum_{ \mathbf{ y }_{ n - 1 } \in \mathbb{ B }^{ p } }
		\dots
		\sum_{ \mathbf{ y }_{ 0 } \in \mathbb{ B }^{ p } }
		\
		( - 1 )^{ ( \mathbf{ t } \oplus \mathbf{ y }_{ n } \oplus \mathbf{ y }_{ n - 1 } \oplus \mathbf{ y }_{ 0 } ) \bullet \mathbf{ x } }
		\
		\ket{ \mathbf{ y }_{ n } }_{ T }
		\
		\ket{ - }_{ n - 1 }
		\
		\ket{ \mathbf{ y }_{ n - 1 } }_{ n - 1 }
		\dots
		\ket{ - }_{ 0 }
		\
		\ket{ \mathbf{ y }_{ 0 } }_{ 0 }
	}
\end{align}

\noindent	Although this expression may initially appear complex due to its multi-register structure and phase factors, it can be significantly simplified by exploiting the characteristic inner product properties defined in \eqref{eq: Inner Product Modulo $2$ Property For Zero} and \eqref{eq: Inner Product Modulo $2$ Property For NonZero}. To understand the simplification, it is essential to revisit the implications of these inner product properties in the context of the QDIBP.

\begin{itemize}
	\item	
	If $\mathbf{ t } \oplus \mathbf{ y }_{ n } \oplus \mathbf{ y }_{ n - 1 } \oplus \dots \oplus \mathbf{ y }_{ 0 } \neq \mathbf{ 0 }$, or, equivalently, $\mathbf{ t }$ $\neq$ $\mathbf{ y }_{ n } \oplus \mathbf{ y }_{ n - 1 } \oplus \dots \oplus \mathbf{ y }_{ 0 }$, the summation $\sum_{ \mathbf{ x } \in \mathbb{ B }^{ p } }$ $( - 1 )^{ ( \mathbf{ t } \oplus \mathbf{ y }_{ n } \oplus \mathbf{ y }_{ n - 1 } \oplus \mathbf{ y }_{ 0 } ) \bullet \mathbf{ x } }$ $\ket{ \mathbf{ y }_{ n } }_{ T }$ $\ket{ - }_{ n - 1 }$ $\ket{ \mathbf{ y }_{ n - 1 } }_{ n - 1 }$ $\dots$ $\ket{ - }_{ 0 }$ $\ket{ \mathbf{ y }_{ 0 } }_{ 0 }$ in \eqref{eq: QDIBP Phase 1 State 2 - II} evaluates to zero. This cancellation occurs due to the destructive interference of phase factors, a hallmark of quantum mechanics that ensures non-matching configurations contribute negligibly to the final state.
	\item	
	Conversely, if $\mathbf{ t } \oplus \mathbf{ y }_{ n } \oplus \mathbf{ y }_{ n - 1 } \oplus \dots \oplus \mathbf{ y }_{ 0 } = \mathbf{ 0 }$, or, equivalently, $\mathbf{ t }$ $=$ $\mathbf{ y }_{ n } \oplus \mathbf{ y }_{ n - 1 } \oplus \dots \oplus \mathbf{ y }_{ 0 }$, the summation $\sum_{ \mathbf{ x } \in \mathbb{ B }^{ p } }$ $( - 1 )^{ ( \mathbf{ t } \oplus \mathbf{ y }_{ n } \oplus \mathbf{ y }_{ n - 1 } \oplus \mathbf{ y }_{ 0 } ) \bullet \mathbf{ x } }$ $\ket{ \mathbf{ y }_{ n } }_{ T }$ $\ket{ - }_{ n - 1 }$ $\ket{ \mathbf{ y }_{ n - 1 } }_{ n - 1 }$ $\dots$ $\ket{ - }_{ 0 }$ $\ket{ \mathbf{ y }_{ 0 } }_{ 0 }$ simplifies to $2^{ p }$ $\ket{ \mathbf{ y }_{ n } }_{ T }$ $\ket{ - }_{ n - 1 }$ $\ket{ \mathbf{ y }_{ n - 1 } }_{ n - 1 }$ $\dots$ $\ket{ - }_{ 0 }$ $\ket{ \mathbf{ y }_{ 0 } }_{ 0 }$. This amplification arises from constructive interference, where the phase factors align perfectly, resulting in a significant contribution to the quantum state when the aggregated secret vector $\mathbf{ t }$ matches the XOR of the information brokers’ inputs.
\end{itemize}

\noindent	These properties enable us to express $\ket{ \psi_{ 2 } }$ in a reduced, more manageable form, highlighting only the nonzero contributions to the quantum state. This simplification is critical for understanding the protocol’s behavior and verifying the correct encoding and retrieval of the aggregated secret vector $\mathbf{ t }$.

\begin{align}
	\label{eq: QDIBP Phase 1 State 2 - III}
	\ket{ \psi_{ 2 } }
	=
	2^{ ( - \frac { p } { 2 } )^{ n - 1 } }
	\sum_{ \mathbf{ y }_{ n } \in \mathbb{ B }^{ p } }
	\sum_{ \mathbf{ y }_{ n - 1 } \in \mathbb{ B }^{ p } }
	\dots
	\sum_{ \mathbf{ y }_{ 0 } \in \mathbb{ B }^{ p } }
	\
	\ket{ \mathbf{ y }_{ n } }_{ T }
	\
	\ket{ - }_{ n - 1 }
	\
	\ket{ \mathbf{ y }_{ n - 1 } }_{ n - 1 }
	\dots
	\ket{ - }_{ 0 }
	\
	\ket{ \mathbf{ y }_{ 0 } }_{ 0 }
	\ ,
\end{align}

where

\begin{align}
	\label{eq: QDIBP Phase 1 Hadamard Entanglement Property}
	\mathbf{ y }_{ n }
	\oplus
	\mathbf{ y }_{ n - 1 }
	\oplus
	\dots
	\oplus
	\mathbf{ y }_{ 0 }
	=
	\mathbf{ t }
	\ .
\end{align}

\noindent	Following the terminology established in \cite{Ampatzis2023} and \cite{Andronikos2023}, we denote the relation \eqref{eq: QDIBP Phase 1 Hadamard Entanglement Property} as the \textbf{Hadamard Entanglement Property}. This property captures the intricate entanglement among the input registers of Trent and the $n$ information brokers, which is established at the outset of the protocol. The collective action of the $n$ brokers embeds their private information into the global quantum state of the composite circuit. This embedding manifests as a constraint on the input registers’ contents, ensuring that the aggregated secret vector $\mathbf{ t }$ is encoded in the relative phase of the entangled state. The \textbf{Hadamard Entanglement Property} underscores the protocol’s reliance on quantum entanglement to achieve secure and distributed information processing.

The final step of the quantum part of the initial phase of the protocol, all involved parties—Trent and the $n$ information brokers—perform measurements on their respective input registers using the computational basis. This measurement process causes the composite quantum system to collapse into its final state, denoted as $\ket{ \psi_{ f } }$. The collapse reflects the resolution of the entangled state into a classical outcome. This measurement step manifests the \textbf{Hadamard Entanglement Property}, which becomes evident in the classical information now encoded within the input registers of the $n + 1$ participants. By bridging the quantum and classical domains, this transition paves the way for subsequent classical computations, enabling the protocol to proceed with the processing of the resulting classical data.

\begin{align}
	\label{eq: QDIBP Phase 1 Final Measurement}
	\ket{ \psi_{ f } }
	&=
	\ket{ \mathbf{ y }_{ n } }_{ T }
	\
	\ket{ - }_{ n - 1 }
	\
	\ket{ \mathbf{ y }_{ n - 1 } }_{ n - 1 }
	\dots
	\ket{ - }_{ 0 }
	\
	\ket{ \mathbf{ y }_{ 0 } }_{ 0 }
	\ ,
	\text{ where }
	\\
	\label{eq: QDIBP Phase 1 Final Sum}
	&\phantom{==}
	\mathbf{ y }_{ n }
	\oplus
	\mathbf{ y }_{ n - 1 }
	\oplus
	\dots
	\oplus
	\mathbf{ y }_{ 0 }
	=
	\mathbf{ t }
\end{align}

\noindent	The validity of equation \eqref{eq: QDIBP Phase 1 Final Sum} does not depend on all participants—Trent and the information brokers—measuring their input registers at precisely the same instant. The temporal sequence of these measurements does not alter the fundamental entanglement constraint, which ensures that, upon measurement, the qubits collapse into correlated states as dictated by the quantum system’s design. In the context of the QDIBP, although the entanglement structure is significantly more complex and the resulting constraint, as expressed in \eqref{eq: QDIBP Phase 1 Final Sum}, is more intricate, the underlying physical principle remains identical to that of simpler entangled systems, such as a two-qubit Bell state. Specifically, the measurement outcomes of the input registers, denoted $\mathbf{ y }_{ n }, \mathbf{ y }_{ n - 1 }, \dots, \mathbf{ y }_{ 0 }$, obtained by Trent and the information brokers $IB_{ 0 }$, \dots, $IB_{ n - 1 }$, respectively, will adhere to the entanglement constraint specified in \eqref{eq: QDIBP Phase 1 Final Sum}. This constraint ensures that the aggregated secret vector $\mathbf{ t }$ is correctly encoded and recoverable from the collective measurement outcomes, leveraging the non-local correlations inherent in quantum entanglement.

\noindent	The completion of Phase 1 takes place in the classical domain through the final actions of the $n + 1$ players, as ordained below.

\begin{enumerate}
	[ left = 0.300 cm, labelsep = 0.500 cm, start = 1 ]
	\renewcommand \labelenumi { (\theenumi) }
	\item
	Every information broker $IB_{ i }$, $0 \leq i \leq n - 1$, transmits the measured contents of her input register $\mathbf{ y }_{ i }$ to Trent via a secure, pairwise authenticated classical channel. This classical communication ensures that the measurement outcomes are shared reliably, preventing unauthorized tampering.
	\item
	Upon receiving these transmissions, Trent possesses not only the measurement outcome $\mathbf{ y }_{ n }$ of his own input register but also the outcomes $\mathbf{ y }_{ n - 1 }, \dots, \mathbf{ y }_{ 0 }$ from all $n$ information brokers. With this complete set of measurement results, Trent can compute the aggregated secret vector $\mathbf{ t }$ as prescribed by \eqref{eq: QDIBP Phase 1 Final Sum}. This computation reconstructs the secret vector by combining the individual contributions in a manner consistent with the entanglement constraint.
\end{enumerate}

\noindent	Thus, the first phase of the QDIBP successfully enables Trent to compute the aggregated secret vector $\mathbf{ t }$, fulfilling the protocol’s first primary objective. However, it is crucial to emphasize that, despite having access to $\mathbf{ t }$, Trent cannot infer the individual secret vectors $\mathbf{ s }_{ i }$ contributed by each information broker $IB_{ i }$, as detailed in subsection \ref{subsec: Blocks & Segments}. This security feature is a direct consequence of the protocol’s design, which leverages quantum entanglement to distribute information across multiple parties and incorporates a sophisticated obfuscation mechanism to protect individual contributions. The entanglement ensures that the global state encodes the aggregated secret without revealing the individual inputs, while the classical communication phase maintains confidentiality through authenticated channels. This combination of quantum and classical techniques underscores the protocol’s robustness.

\subsection{Phase 2: Permuting the blocks within every segment} \label{subsec: Phase 2: Permuting The Blocks Within Every Segment}

At the conclusion of Phase 1 of the QDIBP, Trent computes the aggregated secret vector $\mathbf{ t }$ by executing a bitwise XOR operation on the bit vectors contributed by all $n$ information brokers, combined with the measurement outcome of his own input register. It is imperative to emphasize that Trent is unable to extract any individual secret vector $\mathbf{ s }_{ i }$ from information broker $IB_{ i }$, as this would breach the stringent confidentiality guarantees of the QDIBP. The protocol is meticulously designed to ensure that the aggregated vector $\mathbf{ t }$ encapsulates the collective secret while concealing individual contributions. This is achieved by leveraging the intrinsic properties of quantum entanglement and the secure classical communication channels established during Phase 1, which together provide a robust framework for privacy-preserving data aggregation.

However, a significant challenge persists: directly transmitting the aggregated secret vector $\mathbf{ t }$ to the information brokers would entail compromising the anonymity of the senders. In the QDIBP, anonymity and privacy are paramount, and any mechanism that could allow an information broker to infer the identity of a sender must be prevented. To address this, during the second phase of the protocol Trent employs a probabilistic strategy to obfuscate the information, ensuring that it is computationally infeasible for any broker to deduce the sender’s identity. This is achieved through a permutation-based shuffling mechanism applied to the internal structure of the data. As described in subsection \ref{subsec: Blocks & Segments}, the aggregated secret vector $\mathbf{ t }$ is organized into segments, each containing $n$ blocks of information. To maintain the protocol’s functionality, the sequence of segments must remain unchanged, as the information brokers rely on this fixed order to correctly interpret the data. However, within each segment, Trent applies a randomly selected permutation to the $n$ blocks. This permutation shuffles the blocks in a way that preserves the information content while breaking any direct correlation between the block positions and the identities of the contributing brokers. By introducing this randomness, the protocol ensures that no broker can trace a specific block back to its sender, thereby guaranteeing anonymity. The permutation mechanism draws on fundamental concepts from group theory, adhering to standard definitions and notations as found in accessible texts such as \cite{Gallian2021, Artin2011, Dummit2004, Matsuura2022}. By introducing controlled randomness through permutations, the QDIBP achieves a balance between maintaining data integrity and ensuring anonymity.

\begin{definition} {Permutation} { Permutation}
	A \emph{permutation} $\sigma$ of the set $\{ 0, 1, \dots, n - 1 \}$ is a function
	\begin{align}
		\label{eq: Permutation}
		\sigma
		\colon
		\{ 0, 1, \dots, n - 1 \}
		\to
		\{ 0, 1, \dots, n - 1 \}
	\end{align}
	that is both one-to-one and onto, i.e., a bijection.
	
	The set of all permutations of $\{ 0, 1, \dots, n - 1 \}$ is termed the \emph{symmetric group} of degree $n$ and is denoted by $S_{ n }$.
\end{definition}

It is a well-established result that $S_{ n }$ is a group that contains $n !$ distinct permutations, providing a vast pool of possible rearrangements for obfuscation purposes.

\begin{definition} {Shuffled Aggregated Secret Vector} { Shuffled Aggregated Secret Vector}
	Given the aggregated secret vector $\mathbf{ t } = \mathbf{ t }_{ n - 1 } \ \mathbf{ t }_{ n - 2 } \dots \mathbf{ t }_{ 1 } \ \mathbf{ t }_{ 0 }$, which consists of $n$ aggregated segments $\mathbf{ t }_{ i } = \mathbf{ b }_{ i, n - 1 } \ \mathbf{ b }_{ i, n - 2 } \dots \mathbf{ b }_{ i, 1 } \ \mathbf{ b }_{ i, 0 }$, $0 \leq i \leq n - 1$, we define the \emph{shuffled} aggregated secret vector
	\begin{align}
		\label{eq: Segment Form of the Shuffled Aggregated Secret Vector}
		\Tilde { \Tilde { \mathbf{ t } } }
		=
		\Tilde { \Tilde { \mathbf{ t } } }_{ n - 1 }
		\
		\Tilde { \Tilde { \mathbf{ t } } }_{ n - 2 }
		\dots
		\Tilde { \Tilde { \mathbf{ t } } }_{ 1 }
		\
		\Tilde { \Tilde { \mathbf{ t } } }_{ 0 }
		\ ,
	\end{align}
	comprising $n$ \emph{shuffled} aggregated segments
	\begin{align}
		\label{eq: Shuffled Aggregated Segment i}
		\Tilde { \Tilde { \mathbf{ t } } }_{ i }
		\coloneq
		\mathbf{ b }_{ i, \sigma_{ i } ( n - 1 ) }
		\
		\mathbf{ b }_{ i, \sigma_{ i } ( n - 2 ) }
		\dots
		\mathbf{ b }_{ i, \sigma_{ i } ( 1 ) }
		\
		\mathbf{ b }_{ i, \sigma_{ i } ( 0 ) }
		\ ,
	\end{align}
	where $\sigma_{ i }$, $0 \leq i \leq n - 1$, is a permutation from $S_{ n }$ chosen randomly by Trent.
\end{definition}

Each shuffled $\Tilde { \Tilde { \mathbf{ t } } }_{ i }$, $0 \leq i \leq n - 1$, contains precisely the same information as the original aggregated segment $\mathbf{ t }_{ i }$, ensuring no loss of information. However, the randomized permutation of the $n$ constituent blocks within each segment effectively disrupts any traceable connection between the blocks and the identities of the contributing brokers. This ensures that the recipient, information broker $IB_{ i }$, cannot infer the sender of any specific block, thereby preserving the anonymity guaranteed by the QDIBP. The permutation-based shuffling mechanism not only enhances anonymity but also strengthens the protocol’s resilience against potential attacks aimed at de-anonymizing contributors. By leveraging the vast combinatorial space of $S_{ n }$, the protocol introduces a high degree of randomness, making it computationally infeasible for an adversary to reverse-engineer the permutation without access to Trent’s random selection process. Furthermore, the use of quantum entanglement in Phase 1, combined with the classical permutation strategy in Phase 2, creates a hybrid quantum-classical framework that maximizes both security and anonymity.

\subsection{Phase 3: Information dissemination} \label{subsec: Phase 3: Information Dissemination}

In the third and final phase of the Quantum Dining Information Brokers Protocol (QDIBP), each of the $n$ information brokers employs identical private quantum circuits, as no unitary transformations are applied by the brokers during this phase, and consequently, single-qubit quantum output registers are not utilized. In this phase, information flows unidirectionally from Trent to the information brokers. Trent applies the unitary transformation $U_{ \Tilde { \Tilde { \mathbf{ t } } } }$ to encode the shuffled aggregated secret vector $\Tilde { \Tilde { \mathbf{ t } } }$ into the relative phase of the entangled distributed quantum system, enabling each information broker to access all secret vectors while maintaining anonymity. This process is implemented through the quantum circuit TtoIBQC, as depicted in Figure \ref{fig: The Quantum Circuit TtoIBQC of the QDIBP}.

Following the methodology established in subsection \ref{subsec: Phase 1: Distributing & Obfuscating The Secret Information}, the analysis begins by defining the initial state $\ket{ \psi_{ 0 } }$ of the TtoIBQC quantum circuit. This state is described using the $p$-fold extended generalized GHZ state, as given in \eqref{eq: p-Fold Extended General GHZ_r State}, where $r = n + 1$ represents the total number of parties (Trent plus the $n$ information brokers), and $p = n^{ 2 } m$ corresponds to the number of entangled qubits. For clarity, the subscript $T$ denotes Trent, while subscripts $0 \leq i \leq n - 1$ correspond to the information brokers $IB_{ 0 }, \dots, IB_{ n - 1 }$, respectively.

\begin{align}
	\label{eq: Initial State of TtoIBQC}
	\ket{ \psi_{ 0 } }
	=
	2^{ - \frac { p } { 2 } }
	\sum_{ \mathbf{ x } \in \mathbb{ B }^{ p } }
	\
	\ket{ - }_{ T }
	\
	\ket{ \mathbf{ x } }_{ T }
	\
	\ket{ \mathbf{ x } }_{ n - 1 }
	\
	\dots
	\
	\ket{ \mathbf{ x } }_{ 0 }
\end{align}

\noindent	Trent achieves secure and anonymous information exchange by operating on his private quantum registers via his secret unitary transformation $U_{ \Tilde { \Tilde { \mathbf{ t } } } }$. Nonetheless, the fact that his input register is entangled with the $n$ input registers of the information brokers, ensures that the aggregated secret vector $\Tilde { \Tilde { \mathbf{ t } } }$ is securely embedded into the entangled system, preserving anonymity and supporting the protocol's distributed computation objectives. The unitary transformation $U_{ \Tilde { \Tilde { \mathbf{ t } } } }$ also follows the typical form $U_{ \Tilde { \Tilde { \mathbf{ t } } } } \colon \ket{ y } \ \ket{ \mathbf{ x } }$ $\rightarrow$ $\ket*{ y \oplus \left(  \Tilde { \Tilde { \mathbf{ t } } } \bullet \mathbf{ x } \right) } \ \ket{ \mathbf{ x } }$, where $\oplus$ denotes the bitwise XOR operation and $\bullet$ stands for the inner product modulo $2$. This can be expressed more conveniently as

\begin{align}
	\label{eq: Trent's Unitary Transform}
	U_{ \Tilde { \Tilde { \mathbf{ t } } } }
	&\colon
	\ket{ - }
	\
	\ket{ \mathbf{ x } }
	\rightarrow
	( - 1 )^{ \Tilde { \Tilde { \mathbf{ t } } } \bullet \mathbf{ x } }
	\
	\ket{ - }
	\
	\ket{ \mathbf{ x } }
	\ .
\end{align}

\begin{tcolorbox}
	[
	enhanced,
	breakable,
	grow to left by = 0.000 cm,
	grow to right by = 0.000 cm,
	colback = SkyBlue1!08,
	enhanced jigsaw,			
	frame hidden,
	sharp corners,
	]
	\begin{figure}[H]
		\centering
		\begin{tikzpicture} [ scale = 0.850, transform shape ]
			\begin{yquant}
				nobit AUX_IB_0_0;
				nobit AUX_IB_0_1;
				[ name = Charlie ] qubits { $IR_{ 0 }$ } IR_0;
				nobit AUX_IB_0_2;
				[ name = space_0, register/minimum height = 8.000 mm ] nobit space_0;
				nobit AUX_IB_n-1_0;
				nobit AUX_IB_n-1_1;
				[ name = Bob ] qubits { $IR_{ n - 1 }$ } IR_n_1;
				nobit AUX_B_n-1_2;
				[ name = space_n_2, register/minimum height = 8.000 mm ] nobit space_n_2;
				nobit AUX_T_0;
				nobit AUX_T_1;
				[ name = Alice ] qubits { $TIR$ } TIR;
				qubit { $TOR \colon \ket{ - }$ } TOR;
				nobit AUX_A_2;
				nobit AUX_A_3;
				[ name = Ph0, WordBlueDarker, line width = 0.250 mm, label = { [ label distance = 0.200 cm ] north: Initial State } ]
				barrier ( - ) ;
				[ draw = SkyBlue4, fill = SkyBlue4, radius = 0.750 cm ] box {\color{white} \large \sf{U}$_{ \Tilde { \Tilde { \mathbf{ t } } } }$} (TIR - TOR);
				[ name = Ph1, WordBlueDarker, line width = 0.250 mm, label = { [ label distance = 0.200 cm ] north: State 1 } ]
				barrier ( - ) ;
				[ draw = GreenLighter2, fill = GreenLighter2, radius = 0.600 cm ] box {\color{white} \large \sf{H}$^{ \otimes r }$} IR_0;
				[ draw = GreenLighter2, fill = GreenLighter2, radius = 0.600 cm ] box {\color{white} \large \sf{H}$^{ \otimes r }$} IR_n_1;
				[ draw = GreenLighter2, fill = GreenLighter2, radius = 0.600 cm ] box {\color{white} \large \sf{H}$^{ \otimes r }$} TIR;
				[ name = Ph2, WordBlueDarker, line width = 0.250 mm, label = { [ label distance = 0.200 cm ] north: State 2 } ]
				barrier ( - ) ;
				[ line width = .250 mm, draw = white, fill = black, radius = 0.600 cm ] measure IR_0;
				[ line width = .250 mm, draw = white, fill = black, radius = 0.600 cm ] measure IR_n_1;
				[ line width = .350 mm, draw = white, fill = black, radius = 0.600 cm ] measure TIR;
				[ name = Ph3, WordBlueDarker, line width = 0.250 mm, label = { [ label distance = 0.200 cm ] north: Measurement } ]
				barrier ( - ) ;
				output { $\ket{ \mathbf{ m }_{ 0 } }$ } IR_0;
				output { $\ket{ \mathbf{ m }_{ n - 1 } }$ } IR_n_1;
				output { $\ket{ \mathbf{ m }_{ n } }$ } TIR;
				\node [ below = 5.250 cm ] at (Ph0) { $\ket{ \psi_{ 0 } }$ };
				\node [ below = 5.250 cm ] at (Ph1) { $\ket{ \psi_{ 1 } }$ };
				\node [ below = 5.250 cm ] at (Ph2) { $\ket{ \psi_{ 2 } }$ };
				\node [ below = 5.250 cm ] at (Ph3) { $\ket{ \psi_{ f } }$ };
				\node
				[
				charlie,
				scale = 1.500,
				anchor = center,
				left = 0.700 cm of Charlie,
				label = { [ label distance = 0.000 cm ] north: $IB_{ 0 }$ }
				]
				()
				{ };
				\node
				[
				bob,
				scale = 1.500,
				anchor = center,
				left = 0.300 cm of Bob,
				label = { [ label distance = 0.000 cm ] north: $IB_{ n - 1 }$ }
				]
				()
				{ };
				\node
				[
				alice,
				scale = 1.500,
				anchor = center,
				left = 0.500 cm of Alice,
				label = { [ label distance = 0.000 cm ] north: Trent }
				]
				()
				{ };
				\begin{scope} [ on background layer ]
					\node
					[
					above left = 1.350 cm and 0.425 cm of Bob
					]
					{ \LARGE \vdots };
					\node
					[
					above right = - 0.300 cm and 0.600 cm of space_0, rectangle, fill = WordAquaLighter60, text width = 10.000 cm, align = center, minimum height = 10.000 mm
					]
					{ \bf Spatially Separated };
					\node
					[
					above right = - 0.300 cm and 0.600 cm of space_n_2, rectangle, fill = WordAquaLighter60, text width = 10.000 cm, align = center, minimum height = 10.000 mm
					]
					{ \bf Spatially Separated };
				\end{scope}
			\end{yquant}
			\node
			[
			above right = 2.750 cm and 5.500 cm of Charlie,
			anchor = center,
			shade,
			top color = RedPurple, bottom color = black,
			rectangle,
			text width = 12.000 cm,
			align = center,
			]
			(Label)
			{ \color{white}
				The quantum circuit TtoIBQC enables Trent to transmit anonymously to the Information Brokers the aggregated secret information.
			};
			\scoped [ on background layer ]
			\draw
			[
			purple4, -, >=stealth, line width = 0.750 mm, decoration = coil, decorate
			]
			( $ (Alice.east) + ( 0.800 mm, 0.000 mm ) $ ) node [ circle, fill, minimum size = 1.500 mm ] () {} -- ( $ (Bob.east) + ( 0.800 mm, 0.000 mm ) $ ) node [ circle, fill, minimum size = 1.5 mm ] () {} -- ( $ (Charlie.east) + ( 0.800 mm, 0.000 mm ) $ ) node [ circle, fill, minimum size = 1.500 mm ] () {};
		\end{tikzpicture}
		\caption{The above quantum circuit TtoIBQC allows Trent to relay the aggregated secret information anonymously to every Information Broker.}
		\label{fig: The Quantum Circuit TtoIBQC of the QDIBP}
	\end{figure}
\end{tcolorbox}

\noindent	Trent's action through $U_{ \Tilde { \Tilde { \mathbf{ t } } } }$ sends the quantum circuit TtoIBQC to the next state $\ket{ \psi_{ 1 } }$.

\begin{align} \label{eq: QDIBP Phase 3 State 1}
	\ket{ \psi_{ 1 } }
	&=
	2^{ - \frac { p } { 2 } }
	\sum_{ \mathbf{ x } \in \mathbb{ B }^{ p } }
	\
	\left(
	U_{ \Tilde { \Tilde { \mathbf{ t } } } }
	\
	\ket{ - }_{ n - 1 }
	\
	\ket{ \mathbf{ x } }_{ T }
	\right)
	\ket{ \mathbf{ x } }_{ n - 1 }
	\dots
	\ket{ \mathbf{ x } }_{ 0 }
	\nonumber \\
	&\hspace{-0.100 cm} \overset { \eqref{eq: Trent's Unitary Transform} } { = }
	2^{ - \frac { p } { 2 } }
	\sum_{ \mathbf{ x } \in \mathbb{ B }^{ p } }
	\
	( - 1 )^{ \Tilde { \Tilde { \mathbf{ t } } } \bullet \mathbf{ x } }
	\
	\ket{ \mathbf{ x } }_{ T }
	\
	\ket{ \mathbf{x} }_{ n - 1 }
	\dots
	\ket{ \mathbf{ x } }_{ 0 }
\end{align}

\noindent	The quantum state $\ket{ \psi_{ 1 } }$, as given by \eqref{eq: QDIBP Phase 3 State 1}, emerges directly from the entanglement properties inherent to the QDIBP. Through $U_{ \Tilde { \Tilde { \mathbf{ t } } } }$ Trent embeds the shuffled aggregated secret vector $\Tilde { \Tilde { \mathbf{ t } } }$ into the relative phase of the distributed quantum circuit. To extract $\Tilde { \Tilde { \mathbf{ t } } }$, all $n$ information brokers, in coordination with the semi-honest third party, Trent, perform a coordinated quantum operation. They apply the $p$-fold Hadamard transform, where $p = n^{ 2 } m$, to their respective input registers, as illustrated in Figure \ref{fig: The Quantum Circuit TtoIBQC of the QDIBP}. This transformation disentangles the system in a controlled manner, enabling the reconstruction of the aggregated secret vector. As a result of this process, the quantum state of the system transitions from $\ket{ \psi_{ 1 } }$ to $\ket{ \psi_{ 2 } }$. This state transition underscores the power of quantum entanglement and multi-party quantum protocols in achieving secure and anonymous.

\begin{align}
	\label{eq: QDIBP Phase 3 State 2 - I}
	\ket{ \psi_{ 2 } }
	=
	2^{ - \frac { p } { 2 } }
	\sum_{ \mathbf{ x } \in \mathbb{ B }^{ p } }
	\
	( - 1 )^{ \Tilde { \Tilde { \mathbf{ t } } } \bullet \mathbf{ x } }
	\
	\ket{ - }_{ T }
	\
	\left(
	H^{ \otimes p }
	\ket{ \mathbf{ x } }_{ T }
	\right)
	\
	\left(
	H^{ \otimes p }
	\ket{ \mathbf{ x } }_{ n - 1 }
	\right)
	\dots
	\
	\left(
	H^{ \otimes p }
	\ket{ \mathbf{ x } }_{ 0 }
	\right)
\end{align}

\noindent	Using the relations outlined in \eqref{eq: Hadamard p-Fold Transform Auxiliary Expansions}, the quantum state $\ket{ \psi_{ 2 } }$ can be recast into a more explicit expression, providing a clearer representation of the disentangled system and the extracted secret vector.

\begin{align}
	\label{eq: QDIBP Phase 3 State 2 - II}
	\hspace* { - 2.000 cm }
	{\small
		\ket{ \psi_{ 2 } }
		=
		2^{ ( - \frac { p } { 2 } )^{ n + 1 } }
		\sum_{ \mathbf{ x } \in \mathbb{ B }^{ p } }
		\sum_{ \mathbf{ y }_{ n } \in \mathbb{ B }^{ p } }
		\sum_{ \mathbf{ y }_{ n - 1 } \in \mathbb{ B }^{ p } }
		\dots
		\sum_{ \mathbf{ y }_{ 0 } \in \mathbb{ B }^{ p } }
		\
		( - 1 )^{ ( \Tilde { \Tilde { \mathbf{ t } } } \oplus \mathbf{ y }_{ n } \oplus \mathbf{ y }_{ n - 1 } \oplus \mathbf{ y }_{ 0 } ) \bullet \mathbf{ x } }
		\
		\ket{ - }_{ T }
		\
		\ket{ \mathbf{ y }_{ n } }_{ T }
		\
		\ket{ \mathbf{ y }_{ n - 1 } }_{ n - 1 }
		\dots
		\ket{ \mathbf{ y }_{ 0 } }_{ 0 }
	}
\end{align}

\noindent	Similar to the analysis conducted for Phase 1 of the QDIBP, the expression for the quantum state $\ket{ \psi_{ 2 } }$ may initially appear intricate due to its multi-register structure and the presence of phase factors. However, it can be significantly simplified by leveraging the inner product properties defined in \eqref{eq: Inner Product Modulo $2$ Property For Zero} and \eqref{eq: Inner Product Modulo $2$ Property For NonZero}. To fully appreciate this simplification, it is crucial to examine the implications of these properties within the context of the QDIBP, particularly in how they govern the behavior of the quantum state and facilitate the secure retrieval of the shuffled aggregated secret vector $\Tilde { \Tilde { \mathbf{ t } } }$.

\begin{itemize}
	\item	
	If $\Tilde { \Tilde { \mathbf{ t } } } \oplus \mathbf{ y }_{ n } \oplus \mathbf{ y }_{ n - 1 } \oplus \dots \oplus \mathbf{ y }_{ 0 } \neq \mathbf{ 0 }$, or, equivalently, $\Tilde { \Tilde { \mathbf{ t } } }$ $\neq$ $\mathbf{ y }_{ n } \oplus \mathbf{ y }_{ n - 1 } \oplus \dots \oplus \mathbf{ y }_{ 0 }$, the summation $\sum_{ \mathbf{ x } \in \mathbb{ B }^{ p } }$ $( - 1 )^{ ( \Tilde { \Tilde { \mathbf{ t } } } \oplus \mathbf{ y }_{ n } \oplus \mathbf{ y }_{ n - 1 } \oplus \mathbf{ y }_{ 0 } ) \bullet \mathbf{ x } }$ $\ket{ - }_{ T }$ $\ket{ \mathbf{ y }_{ n } }_{ T }$ $\ket{ \mathbf{ y }_{ n - 1 } }_{ n - 1 }$ $\dots$ $\ket{ \mathbf{ y }_{ 0 } }_{ 0 }$ in \eqref{eq: QDIBP Phase 3 State 2 - II} reduces to zero. This cancellation results from the destructive interference of phase factors, a fundamental quantum mechanical phenomenon. In this case, the non-matching configurations between the shuffled aggregated secret vector and the XOR of the brokers’ inputs lead to phase terms that destructively interfere, effectively nullifying their contribution to the final quantum state. This ensures that only the correct configurations contribute meaningfully to the protocol’s outcome.
	\item	
	In contrast, if $\Tilde { \Tilde { \mathbf{ t } } } \oplus \mathbf{ y }_{ n } \oplus \mathbf{ y }_{ n - 1 } \oplus \dots \oplus \mathbf{ y }_{ 0 } = \mathbf{ 0 }$, or, equivalently, $\Tilde { \Tilde { \mathbf{ t } } }$ $=$ $\mathbf{ y }_{ n } \oplus \mathbf{ y }_{ n - 1 } \oplus \dots \oplus \mathbf{ y }_{ 0 }$, the summation $\sum_{ \mathbf{ x } \in \mathbb{ B }^{ p } } ( - 1 )^{ ( \Tilde { \Tilde { \mathbf{ t } } } \oplus \mathbf{ y }_{ n } \oplus \mathbf{ y }_{ n - 1 } \oplus \mathbf{ y }_{ 0 } ) \bullet \mathbf{ x } } \ket{ - }_{ T } \ket{ \mathbf{ y }_{ n } }_{ T } \ket{ \mathbf{ y }_{ n - 1 } }_{ n - 1 } \dots \ket{ \mathbf{ y }_{ 0 } }_{ 0 }$ equals $2^{ p }$ $\ket{ - }_{ T }$ $\ket{ \mathbf{ y }_{ n } }_{ T }$ $\ket{ \mathbf{ y }_{ n - 1 } }_{ n - 1 }$ $\dots$ $\ket{ \mathbf{ y }_{ 0 } }_{ 0 }$. This is the result of constructive interference, where the phase factors align coherently when the shuffled aggregated secret vector matches the XOR of the information brokers’ inputs. This alignment results in a significant contribution to the quantum state, enabling the precise retrieval of $\Tilde { \Tilde { \mathbf{ t } } }$.
\end{itemize}

\noindent	These inner product properties allow for a streamlined representation of the quantum state $\ket{ \psi_{ 2 } }$, focusing exclusively on the nonzero contributions. This simplification is pivotal for analyzing the protocol’s behavior, as it clarifies how the QDIBP ensures the accurate encoding and retrieval of the shuffled aggregated secret vector $\Tilde { \Tilde { \mathbf{ t } } }$. The destructive interference in the non-matching case ensures that irrelevant configurations do not affect the outcome, while the constructive interference in the matching case amplifies the correct state, facilitating efficient and secure information extraction. This mechanism underscores the power of quantum interference in achieving the protocol’s objectives of anonymity and data security.

\begin{align}
	\label{eq: QDIBP State 2 - III}
	\ket{ \psi_{ 2 } }
	=
	2^{ ( - \frac { p } { 2 } )^{ n - 1 } }
	\sum_{ \mathbf{ y }_{ n } \in \mathbb{ B }^{ p } }
	\sum_{ \mathbf{ y }_{ n - 1 } \in \mathbb{ B }^{ p } }
	\dots
	\sum_{ \mathbf{ y }_{ 0 } \in \mathbb{ B }^{ p } }
	\
	\ket{ - }_{ T }
	\
	\ket{ \mathbf{ y }_{ n } }_{ T }
	\
	\ket{ \mathbf{ y }_{ n - 1 } }_{ n - 1 }
	\dots
	\ket{ \mathbf{ y }_{ 0 } }_{ 0 }
	\ ,
\end{align}

where

\begin{align}
	\label{eq: QDIBP Phase 3 Hadamard Entanglement Property}
	\mathbf{ y }_{ n }
	\oplus
	\mathbf{ y }_{ n - 1 }
	\oplus
	\dots
	\oplus
	\mathbf{ y }_{ 0 }
	=
	\Tilde { \Tilde { \mathbf{ t } } }
	\ .
\end{align}

\noindent	As established in our analysis of Phase 1, the \textbf{Hadamard Entanglement Property} plays the most critical role also in Phase 3 of the QDIBP. This property encapsulates the complex entanglement established at the protocol’s outset among the input registers of Trent and the $n$ information brokers. Through Trent’s application of the unitary transformation, the shuffled aggregated secret vector $\Tilde { \Tilde { \mathbf{ t } } }$ is embedded into the global quantum state of the composite circuit. This embedding imposes a constraint on the contents of the input registers, encoding $\Tilde { \Tilde { \mathbf{ t } } }$ into the relative phase of the entangled state. The \textbf{Hadamard Entanglement Property} thus highlights the QDIBP’s dependence on quantum entanglement to enable secure, anonymous, and distributed information processing, ensuring that the aggregated secret is accessible to all authorized parties without revealing individual contributions.

At the conclusion of the quantum part of Phase 3, mirroring the process observed at the end of the quantum component of Phase 1, all participants, i.e., Trent and the $n$ information brokers, carry out measurements on their respective input registers in the computational basis. This measurement induces the collapse of the composite quantum system into its final state, denoted $\ket{ \psi_{ f } }$. This collapse resolves the entangled quantum state into a definitive classical outcome, marking a crucial transition from the quantum to the classical domain. This quantum-to-classical shift facilitates subsequent classical processing of the measurement outcomes while leveraging the unique properties of the quantum system. The \textbf{Hadamard Entanglement Property}, intrinsic to the distributed quantum circuit, ensures that the encoded information, represented as the shuffled aggregated secret vector $\Tilde { \Tilde { \mathbf{ t } } }$, is faithfully extracted in a classical form suitable for further processing, all while upholding the protocol’s guarantees of security and anonymity.

\begin{align}
	\label{eq: QDIBP Phase 3 Final Measurement}
	\ket{ \psi_{ f } }
	&=
	\ket{ - }_{ T }
	\
	\ket{ \mathbf{ y }_{ n } }_{ T }
	\
	\ket{ \mathbf{ y }_{ n - 1 } }_{ n - 1 }
	\dots
	\ket{ \mathbf{ y }_{ 0 } }_{ 0 }
	\ ,
	\text{ where }
	\\
	\label{eq: QDIBP Phase 3 Final Sum}
	&\phantom{==}
	\mathbf{ y }_{ n }
	\oplus
	\mathbf{ y }_{ n - 1 }
	\oplus
	\dots
	\oplus
	\mathbf{ y }_{ 0 }
	=
	\Tilde { \Tilde { \mathbf{ t } } }
\end{align}

\noindent	Ergo, as a direct consequence of the \textbf{Hadamard Entanglement Property}, the measurement outcomes from the input registers—denoted as $\mathbf{ y }_{ n }, \mathbf{ y }_{ n - 1 }, \dots, \mathbf{ y }_{ 0 }$ for Trent and the information brokers $IB_{ 0 }$, \dots, $IB_{ n - 1 }$, respectively—satisfy the entanglement constraint formalized in Equation \eqref{eq: QDIBP Phase 3 Final Sum}. This constraint ensures that the shuffled aggregated secret vector $\Tilde { \Tilde { \mathbf{ t } } }$ is accurately encoded within the entangled quantum state prior to measurement and can be reliably reconstructed from the collective classical outcomes. By harnessing the non-local correlations inherent in quantum entanglement, the QDIBP guarantees that the aggregated secret is distributed across the participants in a way that safeguards both security and anonymity. This distribution prevents any single participant from accessing or reconstructing individual contributions, thereby preserving the integrity of the protocol.

To contextualize these outcomes, we revisit the virtual hierarchical structure assigned to the quantum registers, as defined in Definition \ref{def: Blocks & Segments}. This structure enables the recasting of the measurement outcomes $\mathbf{ y }_{ n }, \mathbf{ y }_{ n - 1 }, \dots, \mathbf{ y }_{ 0 }$ into their segmented forms. Combined with Equation \eqref{eq: Segment Form of the Shuffled Aggregated Secret Vector}, which we restate here for clarity, we can express the relationships as follows:

\begin{align}
	\label{eq: Segment Form of the Measured Quantum Registers}
	\left
	\{
	\
	\begin{aligned}
		\mathbf{ y }_{ 0 }
		&=
		\mathbf{ y }_{ 0, n - 1 }
		\
		\mathbf{ y }_{ 0, n - 2 }
		\dots
		\mathbf{ y }_{ 0, 1 }
		\
		\mathbf{ y }_{ 0, 0 }
		\\
		\mathbf{ y }_{ n - 1 }
		&=
		\mathbf{ y }_{ n - 1, n - 1 }
		\
		\mathbf{ y }_{ n - 1, n - 2 }
		\dots
		\mathbf{ y }_{ n - 1, 1 }
		\
		\mathbf{ y }_{ n - 1, 0 }
		\\[ 10.000 pt ]
		\multicolumn{ 2 } { c } { \dots }			
		\\[ 10.000 pt ]
		\mathbf{ y }_{ n }
		&=
		\mathbf{ y }_{ n, n - 1 }
		\
		\mathbf{ y }_{ n, n - 2 }
		\dots
		\mathbf{ y }_{ n, 1 }
		\
		\mathbf{ y }_{ n, 0 }
		\\
		\Tilde { \Tilde { \mathbf{ t } } }
		&=
		\Tilde { \Tilde { \mathbf{ t } } }_{ n - 1 }
		\
		\Tilde { \Tilde { \mathbf{ t } } }_{ n - 2 }
		\dots
		\Tilde { \Tilde { \mathbf{ t } } }_{ 1 }
		\
		\Tilde { \Tilde { \mathbf{ t } } }_{ 0 }
	\end{aligned}
	\
	\right
	\}
	\ ,
\end{align}

\noindent	where $\mathbf{ y }_{ i, j }$ is the $j^{ th }$ segment of the measured contents of the input register of information broker $IB_{ i }$, $0 \leq i, j \leq n - 1$, $\mathbf{ y }_{ n, j }$ is the $j^{ th }$ segment of the measured contents of the input register of Trent, $0 \leq j \leq n - 1$, and $\Tilde { \Tilde { \mathbf{ t } } }_{ j }$ is the $j^{ th }$ segment of the measured contents of the input register of shuffled aggregated secret vector $\Tilde { \Tilde { \mathbf{ t } } }$.

In light of equation \eqref{eq: Segment Form of the Measured Quantum Registers}, the entanglement constraint articulated in \eqref{eq: QDIBP Phase 3 Final Sum} can be expressed in a more granular form, incorporating individual segments as follows:

\begin{align}
	\label{eq: Segment Form of the QDIBP Phase 3 Final Sum}
	\mathbf{ y }_{ n, j }
	\oplus
	\mathbf{ y }_{ n - 1, j }
	\oplus
	\dots
	\oplus
	\mathbf{ y }_{ 0, j }
	=
	\Tilde { \Tilde { \mathbf{ t } } }_{ j }
	\ ,
	\ 0 \leq j \leq n - 1
	\ .
\end{align}

This refined expression elucidates the correlations among the measured contents of the quantum registers, a direct consequence of the entanglement present in the initial state of the quantum circuit. Conceptually, this scenario can be understood as follows: the contents of any $n$ out of the $n + 1$ registers can vary independently, but the contents of the remaining register are fully determined by equation \eqref{eq: QDIBP Phase 3 Final Sum} and its bitwise counterpart, equation \eqref{eq: Segment Form of the QDIBP Phase 3 Final Sum}. This relationship encapsulates the \textbf{Bitwise Hadamard Entanglement Property}, which underscores the deterministic interdependence of the measurement outcomes due to the underlying quantum entanglement.

The culmination of Phase 3 signifies the completion of the QDIBP. This final stage transitions fully into the classical domain, where all $n + 1$ participants—Trent and the $n$ information brokers—execute the following prescribed actions:

\begin{enumerate}
	[ left = 0.300 cm, labelsep = 0.500 cm, start = 1 ]
	\renewcommand \labelenumi { (\theenumi) }
	\item
	Each information broker $IB_{ i }$, $0 \leq i \leq n - 1$, securely transmits the  $j^{ th }$ segment of her measured input register, denoted $\mathbf{ y }_{ i, j }$, to every other information broker $IB_{ j }$, $0 \leq j \neq i \leq n - 1$, via a secure, pairwise authenticated classical channel. This controlled communication ensures the reliable exchange of measurement outcomes. Notably, each information broker $IB_{ i }$ keeps their own $i^{ th }$ segment, $\mathbf{ y }_{ i, i }$, private and does not share it with any other participant, thereby preserving the protocol’s security.
	\item
	Trent transmits the $i^{ th }$ segment of his measured input register, $\mathbf{ y }_{ n, i }$, to information broker $IB_{ i }$, $0 \leq i \leq n - 1$, through a secure, pairwise authenticated classical channel. This step ensures that Trent’s measurement outcomes are shared reliably, maintaining the integrity of the data exchange.
	\item
	Upon receiving these $n$ transmissions, each information broker $IB_{ i }$, $0 \leq i \leq n - 1$, possesses a complete set of the $i^{ th }$ segments: her own $\mathbf{ y }_{ i, i }$, Trent’s $\mathbf{ y }_{ n, i }$, and the $i^{ th }$ segments from all other information brokers $IB_{ i }$, for $j \neq i$. With this comprehensive collection of measurement outcomes, $IB_{ i }$ can compute the $i^{ th }$ segment of the shuffled aggregated secret vector, $\Tilde { \Tilde { \mathbf{ t } } }_{ i }$, as prescribed by Equation \eqref{eq: Segment Form of the QDIBP Phase 3 Final Sum}. Subsequently, $IB_{ i }$ performs an XOR operation between each block of  $\Tilde { \Tilde { \mathbf{ t } } }_{ i }$ and her own secret vector $\mathbf{ s }_{ i }$. This computation enables $IB_{ i }$ to retrieve all other secret vectors $\mathbf{ s }_{ j }$, $0 \leq j \neq i \leq n - 1$. Crucially, this revelation of the secret information occurs without compromising the anonymity of the contributors, as the identities of the senders remain remain entirely untraceable.
\end{enumerate}

Thus, the successful completion of the Quantum Dining Information Brokers Protocol (QDIBP) ensures a fully parallel, completely anonymous, and untraceable exchange of information among the $n$ information brokers. This remarkable achievement is enabled by the synergistic interplay of quantum entanglement and the random shuffling facilitated by Trent, who acts as a semi-honest coordinator. The entanglement phenomenon, combined with the protocol’s structured classical communication, guarantees that the aggregated secret is distributed and reconstructed securely, preserving both the privacy of individual contributions and the anonymity of the participants.

\section{A small scale realization of the QDIBP} \label{sec: A Small Scale Realization Of The QDIBP}

This section presents a compact yet comprehensive example illustrating the practical implementation of the QDIBP. This example serves as a definitive proof of the protocol’s validity and its applicability to real-world scenarios, demonstrating its capability to facilitate secure and anonymous information exchange.

\subsection{Implementing Phase 1 of the QDIBP} \label{subsec: Implementing Phase 1 of the QDIBP}

Consider a scenario involving three information brokers—Alice, Bob, and Charlie—who aim to securely exchange their confidential data in a single transaction while preserving their anonymity and leaving no traceable evidence. To accomplish this, they enlist the assistance of a semi-honest intermediary, Trent, who facilitates the process without compromising their privacy. Due to hardware constraints, we simplify the example by assuming each broker exchanges a single bit of information. The secret vectors held by Alice, Bob, and Charlie, denoted as $\mathbf{ s }_{ A }$, $\mathbf{ s }_{ B }$, and $\mathbf{ s }_{ C }$, respectively, are detailed in Table \ref{tbl: Secret Information}. This table also includes their extended secret vectors, $\widetilde { \mathbf{ s } }_{ A }$, $\widetilde { \mathbf{ s } }_{ B }$, $\widetilde { \mathbf{ s } }_{ C }$, as well as the resulting aggregated secret vector derived from the protocol’s execution.

\begin{tcolorbox}
	[
		enhanced,
		breakable,
		center title,
		fonttitle = \bfseries,
		grow to left by = 0.000 cm,
		grow to right by = 0.000 cm,
		colback = SkyBlue1!08,
		enhanced jigsaw,			
		frame hidden,
		sharp corners,
	]
	\vspace* { - 0.500 cm }
	\begin{table}[H]
		\caption{This table shows Alice, Bob, and Charlie's secret vectors, extended secret vectors, and the resulting aggregated secret vector.}
		\label{tbl: Secret Information}
		\centering
		\SetTblrInner { rowsep = 1.000 mm }
		\begin{tblr}
			{
				colspec =
				{
					Q [ r, m, 4.000 cm ]
					| [ 1.000 pt, cyan3 ]
					| [ 1.000 pt, cyan3 ]
					Q [ c, m, 3.000 cm ]
					| [ 0.500 pt, cyan3 ]
					Q [ c, m, 4.000 cm ]
				},
				rowspec =
				{
					|
					[ 3.500 pt, cyan3 ]
					|
					[ 0.750 pt, cyan3 ]
					|
					[ 0.250 pt, white ]
					Q
					|
					[ 0.500 pt, cyan3 ]
					Q
					|
					[ 0.500 pt, cyan3 ]
					Q
					|
					[ 0.500 pt, cyan3 ]
					Q
					|
					[ 0.500 pt, cyan3 ]
					Q
					|
					[ 3.500 pt, cyan3 ]
				}
			}
			\SetCell { bg = cyan4, fg = white }
			&
			\SetCell { bg = cyan4, fg = white } Secret Vectors
			&
			\SetCell { bg = cyan4, fg = white } Extended Secret Vectors
			\\
			Alice
			&
			$\mathbf{ s }_{ A } = 1$
			&
			$\widetilde { \mathbf{ s } }_{ A } = 011 \ 100 \ 100$
			\\
			Bob
			&
			$\mathbf{ s }_{ B } = 0$
			&
			$\widetilde { \mathbf{ s } }_{ B } = 000 \ 000 \ 000$
			\\
			Charlie
			&
			$\mathbf{ s }_{ C } = 1$
			&
			$\widetilde { \mathbf{ s } }_{ C } = 001 \ 001 \ 110$
			\\
			\SetCell { bg = cyan4, fg = white } Aggregated Secret Vector
			&
			&
			$\hspace { 0.300 cm } \mathbf{ t } = 010 \ 101 \ 010$
			\\
		\end{tblr}
	\end{table}
\end{tcolorbox}

\begin{figure}[htp]
	\centering
	\includegraphics [ scale = 0.375, trim = {0.000cm 0.000cm 0.000cm 0.000cm}, clip ]
	{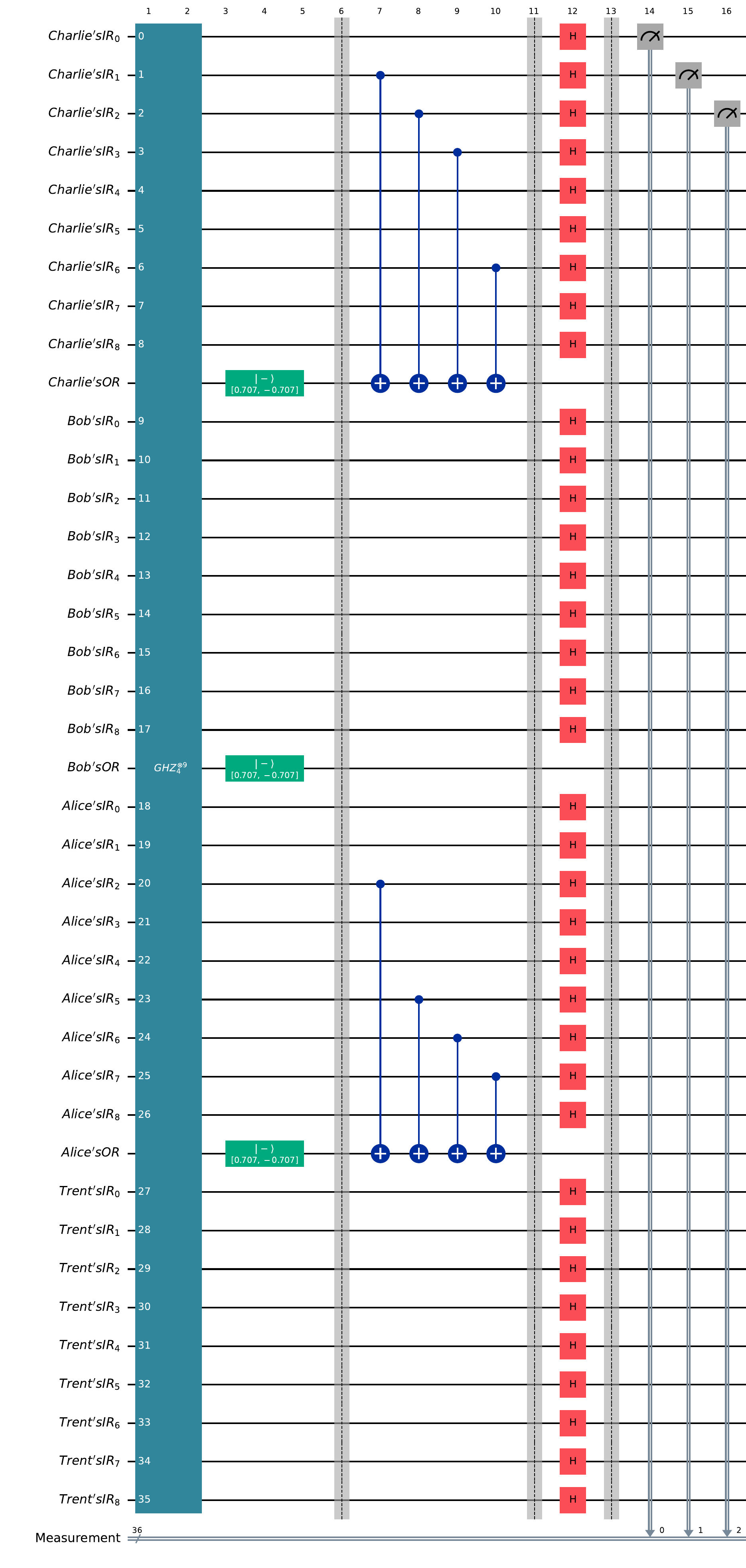}
	\caption{This figure depicts the implementation of the IBtoTQC quantum circuit for this scenario.}
	\label{fig: IBtoTQC}
\end{figure}

The quantum circuit implementing the first phase of this example is constructed using Qiskit \cite{Qiskit2025} and is derived by adapting the abstract quantum circuit presented in Figure \ref{fig: The Quantum Circuit IBtoTQC of the QDIBP} to this specific case. The resulting circuit is depicted in Figure \ref{fig: IBtoTQC}. Given the circuit’s complexity and to improve readability, Figure \ref{fig: IBtoTQC} shows only the left portion of the circuit, which captures the core operations of the QDIBP. The right portion, consisting solely of measurement gates for each qubit in every input register, has been omitted for clarity, as it does not contribute significantly to understanding the protocol’s mechanics.

\begin{figure}[htp]
	\hspace* { - 1.750 cm }
	\centering
	\includegraphics [ scale = 0.475, trim = {0.000cm 0.000cm 1.500cm 0.000cm}, clip ]
	{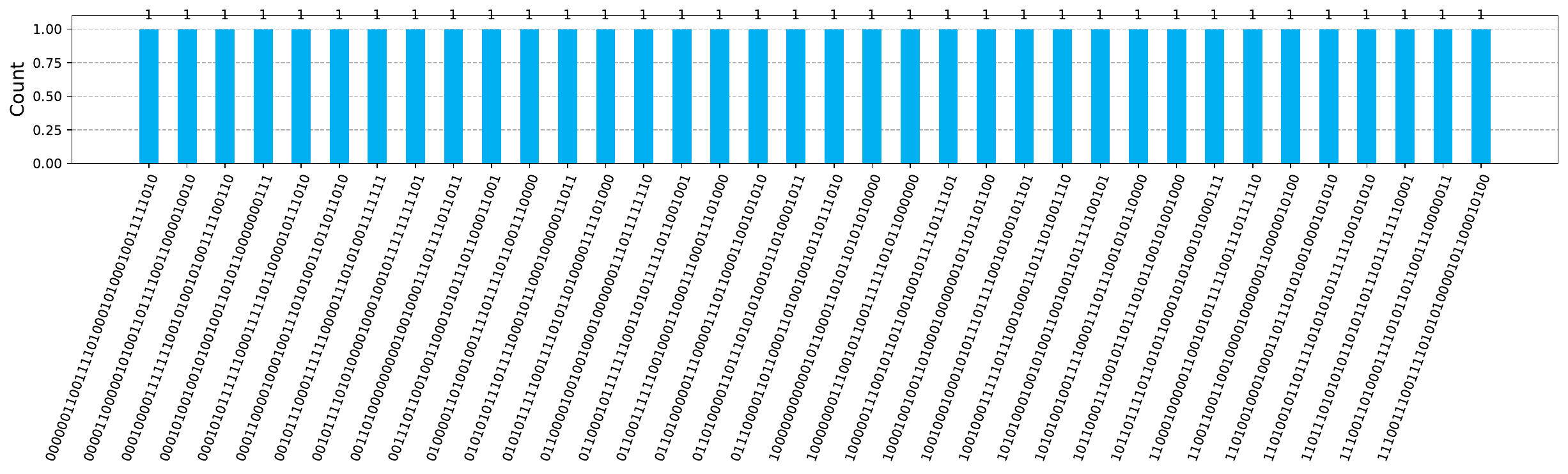}
	\caption{Few of the equiprobable measurements and their corresponding counts for the circuit of Figure \ref{fig: IBtoTQC}.}
	\label{fig: IBtoTQCHistogram}
\end{figure}

Displaying all possible equiprobable outcomes from the measurements performed by Alice, Bob, Charlie, and Trent would result in a cluttered and difficult-to-interpret figure. Therefore, we have chosen to illustrate a representative subset of these outcomes in Figure \ref{fig: IBtoTQCHistogram}, accompanied by the corresponding measurement counts for each outcome. Crucially, every possible outcome adheres to the \textbf{Hadamard Entanglement Property} and satisfies equation \eqref{eq: QDIBP Phase 1 Final Sum}, ensuring the protocol’s correctness. After measuring their respective input registers to obtain $\mathbf{ y }_{ A }$, $\mathbf{ y }_{ B }$ and $\mathbf{ y }_{ C }$, Alice, Bob, and Charlie transmit these measurement results to Trent. Trent then computes the aggregated secret vector by performing an XOR operation: $\mathbf{ t } = \mathbf{ y }_{ A } \oplus \mathbf{ y }_{ B } \oplus \mathbf{ y }_{ C }$. It is straightforward to verify that all outcomes shown in Figure \ref{fig: IBtoTQCHistogram} consistently yield the same aggregated secret vector, $\mathbf{ t } = 010 \ 101 \ 010$, confirming the protocol’s reliability and precision in achieving secure information exchange.

\subsection{Implementing Phase 2 of the QDIBP} \label{subsec: Implementing Phase 2 of the QDIBP}

The use of probabilities is a critical mechanism for ensuring anonymity in the QDIBP, as elaborated in subsection \ref{subsec: Phase 2: Permuting The Blocks Within Every Segment}. Trent, the semi-honest intermediary, plays an indispensable role in this process, as his actions directly safeguard the anonymity of the information brokers. Specifically, Trent is tasked with applying three randomly selected permutations to shuffle the aggregated secret vector, making it probabilistically infeasible for Alice, Bob, or Charlie to trace the origin of any individual piece of information.

Following the protocol’s specifications, Trent selects three random permutations from the symmetric group $S_{ 3 }$ and uses them to construct the shuffled aggregated secret vector, denoted as $\Tilde { \Tilde { \mathbf{ t } } }$, which is presented in Table \ref{tbl: Shuffled Aggregated Secret Vector}. As highlighted in the protocol, $\Tilde { \Tilde { \mathbf{ t } } }$ contains exactly the same information as the original aggregated secret vector $\mathbf{ t }$, but it is reorganized in such a way that identifying the sender of any specific data segment becomes computationally intractable. This shuffling process leverages the randomness of the permutations to obscure the relationship between the input data and its source, thereby ensuring robust anonymity for all participants.

\begin{tcolorbox}
	[
		enhanced,
		breakable,
		center title,
		fonttitle = \bfseries,
		grow to left by = 0.000 cm,
		grow to right by = 0.000 cm,
		colback = SkyBlue1!08,
		enhanced jigsaw,			
		frame hidden,
		sharp corners,
	]
	\vspace* { - 0.500 cm }
	\begin{table}[H]
		\caption{This table shows the original aggregated secret vector and the shuffled aggregated secret vector constructed by Trent.}
		\label{tbl: Shuffled Aggregated Secret Vector}
		\centering
		\SetTblrInner { rowsep = 1.000 mm }
		\begin{tblr}
			{
				colspec =
				{
					Q [ c, m, 6.000 cm ]
					| [ 1.000 pt, cyan3 ]
					| [ 1.000 pt, cyan3 ]
					Q [ c, m, 6.000 cm ]
				},
				rowspec =
				{
					|
					[ 3.500 pt, cyan3 ]
					|
					[ 0.750 pt, cyan3 ]
					|
					[ 0.250 pt, white ]
					Q
					|
					[ 0.500 pt, cyan3 ]
					Q
					|
					[ 0.500 pt, cyan3 ]
					Q
					|
					[ 3.500 pt, cyan3 ]
				}
			}
			\SetCell { bg = cyan4, fg = white } Aggregated Secret Vector
			&
			\SetCell { bg = cyan4, fg = white } Shuffled Aggregated Secret Vector
			\\
			$\mathbf{ t } = 010 \ 101 \ 010$
			&
			$\Tilde { \Tilde { \mathbf{ t } } } = 001 \ 110 \ 100$
			\\
		\end{tblr}
	\end{table}
\end{tcolorbox}

Trent’s careful execution of these permutations is pivotal to the protocol’s success. By introducing controlled randomness, the QDIBP guarantees that no single broker can reverse-engineer the contributions of others, even if they attempt to analyze the shuffled output. This probabilistic approach, combined with the quantum properties of the protocol, establishes a high degree of security and anonymity, making the QDIBP a powerful tool for privacy-preserving information exchange in distributed systems.

\subsection{Implementing Phase 3 of the QDIBP} \label{subsec: Implementing Phase 3 of the QDIBP}

\begin{figure}[htp]
	\centering
	\includegraphics [ scale = 0.375, trim = {0.000cm 0.000cm 0.000cm 0.000cm}, clip ]
	{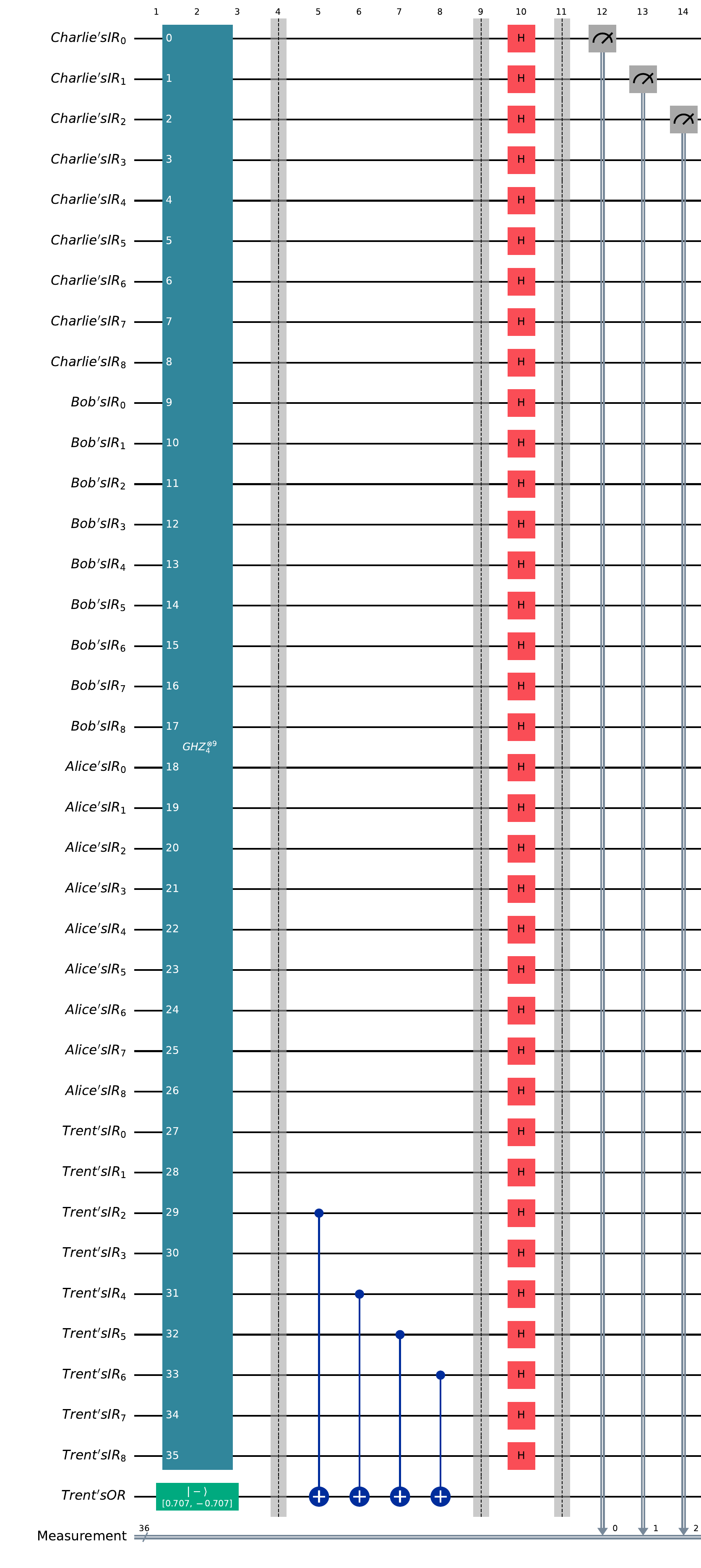}
	\caption{This figure shows the implementation of the TtoIBQC quantum circuit for this example.}
	\label{fig: TtoIBQC}
\end{figure}

The quantum circuit for the third and final phase of the QDIBP, implemented using the Qiskit framework \cite{Qiskit2025}, is constructed by tailoring the abstract quantum circuit shown in Figure \ref{fig: The Quantum Circuit TtoIBQC of the QDIBP} to the specific requirements of this phase. The resulting circuit is illustrated in Figure \ref{fig: TtoIBQC}. To enhance clarity and manage the complexity of the circuit, Figure \ref{fig: TtoIBQC} depicts only the left portion, which encapsulates the core quantum operations of the QDIBP. The right portion, which consists exclusively of measurement gates applied to each qubit in every input register, is omitted to avoid visual clutter, as it contributes minimally to understanding the protocol’s operational mechanics.

\begin{figure}[htp]
	\hspace* { - 1.750 cm }
	\centering
	\includegraphics [ scale = 0.475, trim = {0.000cm 0.000cm 1.500cm 0.000cm}, clip ]
	{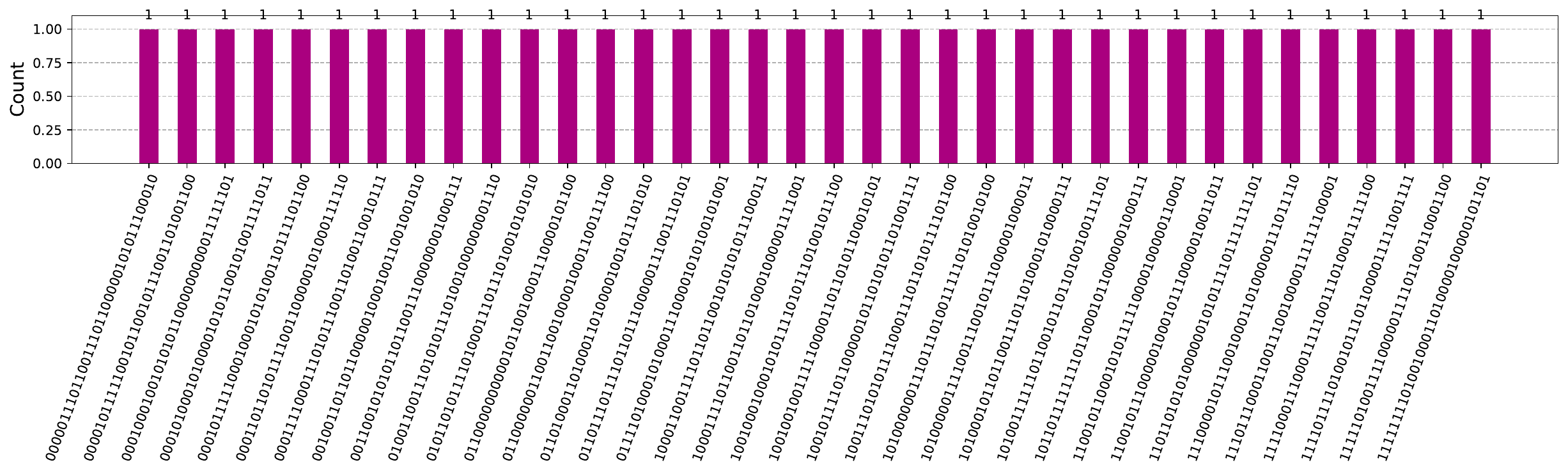}
	\caption{Some of the equiprobable measurements and their corresponding counts for the circuit of Figure \ref{fig: TtoIBQC}.}
	\label{fig: TtoIBQCHistogram}
\end{figure}

As previously discussed, presenting all possible equiprobable measurement outcomes from the participants—Alice, Bob, Charlie, and Trent—would result in an overly complex and challenging-to-interpret diagram. To address this, Figure \ref{fig: TtoIBQCHistogram} illustrates a carefully selected subset of these outcomes, accompanied by their corresponding measurement counts. This selective representation ensures clarity while effectively conveying the protocol’s behavior. Each outcome strictly adheres to the \textbf{Bitwise Hadamard Entanglement Property}, as defined by Equation \eqref{eq: Segment Form of the QDIBP Phase 3 Final Sum}, ensuring the quantum entanglement properties critical to the protocol’s functionality are preserved.

In the QDIBP, each information broker, denoted $IB_{ i }$, $0 \leq i \leq n - 1$, securely transmits the $j^{ th }$ segment of their measured input register, $\mathbf{ y }_{ i, j }$, to every other information broker $IB_{ j }$, where $0 \leq j \neq i \leq n - 1$. This transmission occurs over a secure, pairwise authenticated classical channel, guaranteeing reliable and tamper-proof communication. Importantly, each $IB_{ i }$ keeps their own $i^{ th }$ segment, $\mathbf{ y }_{ i, i }$, private, withholding it from all other participants to safeguard the protocol’s security. Meanwhile, Trent, acting as a semi-honest facilitator, transmits the $i^{ th }$ segment of his measured input register, $\mathbf{ y }_{ n, i }$, to the corresponding information broker $IB_{ i }$ for $0 \leq i \leq n - 1$, also via a secure, pairwise authenticated classical channel. This controlled exchange ensures the integrity and reliability of Trent’s measurement outcomes.

Upon receiving these $n$ transmissions, each information broker $IB_{ i }$ possesses a complete set of the $i^{ th }$ segments: their own $\mathbf{ y }_{ i, i }$, Trent’s $\mathbf{ y }_{ n, i }$, and the $i^{th}$ segments from all other information brokers $IB_j$ for $j \neq i$. With this comprehensive dataset, $IB_{ i }$ can compute the $i^{ th }$ segment of the shuffled aggregated secret vector, $\Tilde { \Tilde { \mathbf{ t } } }_{ i }$, as specified by Equation \eqref{eq: Segment Form of the QDIBP Phase 3 Final Sum}. Subsequently, $IB_{ i }$ performs an XOR operation between each block of $\Tilde { \Tilde { \mathbf{ t } } }_{ i }$ and their own secret vector $\mathbf{ s }_{ i }$. This computation enables $IB_{ i }$ to reconstruct all other secret vectors $\mathbf{ s }_{ j }$ for $0 \leq j \neq i \leq n - 1$, effectively recovering the shared secrets. A critical feature of this process is that it preserves the anonymity of the contributors, as the identities of the senders remain entirely untraceable, ensuring no linkage between the revealed information and its source.

This example underscores the QDIBP’s ability to enable secure, anonymous, and efficient data sharing among multiple parties, with Trent acting as a semi-honest facilitator. The use of quantum entanglement and the \textbf{Hadamard Entanglement Property} ensures that the protocol maintains confidentiality and integrity, making it a robust solution for privacy-preserving applications in distributed systems.

\section{Discussion and conclusions} \label{sec: Discussion and Conclusions}

This work introduces and resolves the novel Quantum Dining Information Brokers Problem, a scenario involving $n$ information brokers, distributed across diverse geographic locations, participating in a virtual, metaphorical dinner. During this interaction, the brokers aim to exchange arbitrarily large volumes of data in a completely anonymous and untraceable manner. To address this challenge, we propose the Quantum Dining Information Brokers Protocol (QDIBP), a pioneering entanglement-based quantum cryptographic protocol. Building upon foundational works that leverage quantum properties to ensure uncompromising privacy and anonymity, our protocol advances the field through three transformative innovations that significantly enhance the landscape of quantum cryptographic protocols.

\begin{itemize}
	\item	[\faPeopleArrows]
	\textbf{Many-to-Many Simultaneous Information Exchange.}

	The QDIBP introduces a groundbreaking capability for simultaneous, fully parallel communication among all participants, regardless of their geographical distribution. Unlike traditional protocols that often rely on sequential or one-to-many communication models, our approach is among the first to enable a true many-to-many exchange in a single operation. This innovation ensures efficient, real-time data sharing, making it particularly suited for large-scale, distributed systems where speed and concurrency are paramount.
	\item	[\faMask]

	By harnessing the unique properties of quantum entanglement, the QDIBP encodes information into the relative phases of a distributed entangled quantum system. This approach renders the exchanged data untraceable and ensures complete anonymity for all participants. Unlike sequential applications of one-to-many protocols, which often compromise sender identity, our protocol guarantees robust anonymity by leveraging entanglement to obscure individual contributions, marking a significant advancement over existing methods.
	\item	[\faGlobe]
	\textbf{Fully Distributed Framework.}

	Traditional formulations such as the Dining Cryptographers Problem typically assume participants are physically co-located, limiting their applicability in modern, globalized contexts. The QDIBP transcends this constraint by designing a fully distributed framework, enabling secure and seamless communication among information brokers situated across vast geographical distances. By exploiting quantum entanglement, the protocol ensures that data exchange remains secure and efficient, regardless of physical separation, thus redefining the scope of quantum cryptographic applications.
\end{itemize}

\noindent	The QDIBP leverages the intricate interplay of quantum entanglement and the effects of constructive and destructive quantum interference to manage complex multi-party interactions. By exploiting the cancellation and amplification properties of quantum phase factors, the protocol ensures that only the intended aggregated information is recovered, while individual contributions remain confidential. Central to the QDIBP is the \textbf{Hadamard Entanglement Property}, which, combined with a carefully designed measurement step in the computational basis, facilitates a seamless quantum-to-classical transition. This approach ensures that no single party can access individual contributions, preserving confidentiality while enabling scalable processing of classical outcomes.

The protocol’s measurement mechanism serves as a controlled method to extract the aggregated secret, aligning with the objectives of secure multi-party interaction. This scalability is further enhanced by the ability to efficiently process and verify classical outcomes, making the QDIBP suitable for large-scale applications. The protocol’s design positions it as a versatile framework for quantum cryptography, distributed quantum computing, and privacy-preserving data aggregation, where the synergy of quantum entanglement and classical processing is critical for achieving both security and anonymity.

As with any quantum protocol, the QDIBP adheres to the principle of ``no free lunch.'' The protocol requires quantum registers comprising $n^{ 2 } m$ qubits, where $n$ represents the number of information brokers and $m$ denotes the number of bits needed to encode each piece of secret information. This resource demand reflects the complexity of enabling simultaneous, many-to-many, and anonymous information exchange. Recognizing that qubits remain a scarce and valuable resource, we are committed to exploring more efficient coding schemes to reduce the qubit overhead while preserving the protocol’s security and anonymity guarantees. Future research will focus on optimizing quantum resource utilization and extending the QDIBP’s applicability to even larger and more diverse distributed systems.

\bibliographystyle{ieeetr}
\bibliography{QSTDIB}

\end{document}